\documentclass[aps,amsmath,amssymb,showkeys]{revtex4}
\usepackage[dvips]{graphicx}
\usepackage{palatino}
\usepackage{amsmath}
\usepackage{mathrsfs}
\usepackage{graphicx}
\usepackage{epsfig}
\usepackage{dcolumn}
\usepackage{xcolor}
\usepackage{hyperref}
\hypersetup{colorlinks=true,linkcolor=red,citecolor=blue, urlcolor=teal}
\newcommand{\orcid}[1]{\href{https://orcid.org/#1}{\includegraphics[width=8pt]{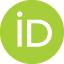}}}
\begin{document}
\title{Galactic dynamics in the presence of scalaron: A perspective from 
$\boldsymbol{f(R)}$ gravity}	
\author{Gayatri Mohan \orcid{0009-0008-0654-5227}}
\email[Email: ]{gayatrimohan704@gmail.com}
\author{Umananda Dev Goswami\footnote{Corresponding author} \orcid{0000-0003-0012-7549}}
\email[Email: ]{umananda2@gmail.com}	
\affiliation{Department of Physics, Dibrugarh University, Dibrugarh 786004, 
	Assam, India}	

\begin{abstract}
We consider $f(R)$ modified gravity theory incorporating the chameleon 
mechanism to address galactic dynamics. By employing the metric formalism and 
utilizing a conformal transformation, we simplify the field equations and 
describe the extra degree of freedom $f_{R}$ via a scalar field (scalaron) 
with chameleonic behavior. A recently proposed $f(R)$ model is analyzed to 
illustrate this behavior effectively. Subsequently, the rotational velocity 
equation including the scalaron's contribution is derived for a test particle 
in a static, spherically symmetric spacetime. Then we generate rotation curves 
and fit them to observational data of thirty seven galaxies using two fitting 
parameters, $M_0$ and $r_c$, the total mass and core radius of a galaxy 
respectively.

\end{abstract}

\keywords{Modified gravity theory; Chameleon mechanism; Scalaron; Conformal 
transformation; Rotation curves.}  

\maketitle

\section{Introduction}
\label{sec1}
The captivating enigma of dark matter (DM) and the accelerated expansion of 
the Universe are two formidable challenges of modern astrophysics and 
cosmology. Behaviors of observed galactic rotation curves 
\cite{1970_Rubin, 1978_Rubin, 1983_Rubin} and gravitational lensing 
\cite{2011_Gar,2010_Massey} in galaxies and their clusters manifest the mass
discrepancy in them and suggest the need for DM as a major matter component in 
the Universe. However, there is currently no experimental
evidence supporting the existence of this elusive missing matter and dark
energy responsible for late-time acceleration. These two so-called greatest puzzles 
of modern times have stimulated the scientific community to think about the 
modifications of Einstein's theory of General relativity (GR) and upshot the 
modified theories of gravity (MTGs) \cite{2022_Shankar, 1983_Milikan, 
	2008_capoz, 2010_harko, 2011_harko, 2011_Capoz,2021_Atayde}.~These theories 
are the consequences of the modification of gravity part of Einstein-Hilbert 
(EH) action and can be treated as a modified approach for explaining the 
issues of DM \cite{2006_Sobouti, 2008_Harko, 2010_Harko, 2011_Gegurgel, 
	2016_Zaregonbadi, 2021_Nashiba, 2023_Nashiba, 2018_Finch,2023_Shabani} as well 
as dark energy (DE), the dominant unknown component of mass-energy that 
is supposed to be responsible for the present accelerated expansion of the 
Universe \cite{2013_Chakraborti, 2008_capozz, 2016_Joyce, 2002_Peebles}.
$f(R)$ gravity \cite{1970_Buchd,1980_starobinsky, 2020_dhruba, 2007_sawiki},
$f(R, T)$ gravity \cite{2011_harko,2013_alva,2024_mohan}, Gauss-Bonnet gravity \cite{1971_David, 2005_nojiri},
Scalar-tensor theories \cite{2003_chiba,2007_martin}, 
Braneworld models \cite{2011_sahidi} 
etc.\ are some widely studied gravity theories as alternatives to GR. Among 
different MTGs, the simplest modification of GR is the $f(R)$ theory of 
gravity, proposed by H.~A.~Buchdahl in $1970$ through the replacement of the 
Ricci scalar $R$ in the EH action with an arbitrary function $f(R)$ of 
it \cite{1970_Buchd}. In this theory, field equations can be derived from the 
action using mainly two approaches: one is the metric formalism in which the
affine connection $\Gamma^{\rho}_{\alpha\beta}$ depends on the metric 
$g_{\mu\nu}$ and the field equations are obtained by varying the action with 
respect to the metric only. 
Another is the Palatini formalism in which the field equations are derived 
from the variation of the action treating affine connection and the metric as 
independent variables \cite{2010_felice}. Such covariant modifications of GR 
always introduce an extra degree of freedom besides the tensor degree of 
freedom. We restrict ourselves to metric $f(R)$ theory in this work. This 
theory can be reformulated as a scalar-tensor theory by describing this 
additional degree of freedom as a scalar field called scalaron 
\cite{2006_Faulkner}. Indeed, it is equivalent to a scalar-tensor theory if 
it is transformed to the Einstein frame via conformal transformation 
\cite{2008_tamaki, 2012_cliftone, 2015_joyce, 2014_terukina, 2016_koyama, 
	2018_burrage, 2007_Chiba}. Under this transformation the scalaron field 
couples with matter minimally and acquires effective mass that depends 
sensitively on the density of the environment. The scalaron field exhibiting 
such a local matter density dependent physical property is referred to as a 
chameleon field and the mechanism involved is called the chameleon mechanism 
\cite{2004_Khoury, 2004_Khoury2}. It is a screening mechanism in which the 
scalaron changes its properties to fit the surroundings where the field is 
allowed to exhibit significant mass in dense environments, such as within the 
solar system, and less mass in low density regions, such as on cosmological 
scales. In high density surroundings, the force induced by the scalaron field 
is suppressed while in low density localities it mediates a force of 
gravitational strength extending its impacts over a long range 
\cite{2013_Khoury, 2010_Hinter}. Such a low-mass scalar field may be a genuine 
substitute for DM.

A range of studies have been undertaken to investigate the scalaron as an 
alternative to DM and DE in MTGs. Especially, the effectiveness of 
chameleonically viable $f(R)$ models in mimicking the DM component 
of the Universe was explored at different scales. Ref.~\cite{2018_verma} has 
explained the DM problem in chameleon $f(R)$ gravity by considering 
a model of the from $f(R) = R^{(1+\delta)}/R^\delta_c$ and suggested 
that behaviour of the scalaron derived from this model makes it suitable to 
mimic as DM. The study of DM taking into account the screening mechanism in 
the Starobinsky model is addressed in Ref.~\cite{2017_katsu}. This study has 
analysed three different possible ways for revealing light scalaron to be a DM 
candidate and also estimated its lifetime by investigating the coupling 
between the scalaron and standard model particles. 
Refs.~\cite{2018_kat,2021_Nashiba} have illustrated DM based on chameleonic 
features of scalaron. It is suggested in Ref.~\cite{2021_Nashiba} that the mass 
of the scalaron can be found to be close to the mass of ultralight axions 
so that it can mimic DM candidate. Refs.~\cite{2012_Burikham} 
investigated the effects of a chameleon scalar field on rotation curves of 
certain late-type low surface brightness (LSB) galaxies and observed a cuspier 
rotation curve for each galaxy. Some other investigations 
\cite{2012_mann,2021_Shtanov,2018_Naik,2020_Li,2006_brown} have been conducted 
by analyzing the chameleon scalar field to address DM issues. In addition to 
the dark matter and dark energy issues, multiple investigations such as 
exploration of black hole properties \cite{2023_Falco}, determination of the 
gravitational wave echo frequencies emitted by ultra-compact objects like 
strange stars \cite{2022_Bora} etc.~have been performed in $f(R)$ gravity that
will enable to test the predictions of this theory with observations. These 
works have motivated us to consider a study for the explanation of this exotic 
matter component by exploring the chameleonic behaviour of a recently 
introduced $f(R)$ gravity model \cite{2020_dhruba,2022_dhruba}. Specifically, 
in this study, we intend to check the chameleonic viability of this 
model
of $f(R)$ gravity by examining the characteristics of its 
scalar 
degree of freedom (scalaron) as no chameleonic study has been performed with this
specific model till now. The principal focus of this work is to 
investigate the impact of scalaron on galactic dynamics taking into 
consideration 
of chameleon mechanism via conformal transformation in metric 
$f(R)$
gravity theory, utilizing this new model within this gravitational theory. With this 
objective, we first present field equations of metric $f(R)$ gravity. Then by 
conformal transformation, the fourth-order  equations generated by the theory
are simplified to second-order equations through the introduction of a scalar
field that shows chameleonic features. We study this feature of the field shown by the model with a correction of the 
singularity
problem usually suffered by $f(R)$ gravity models \cite{2021_Nashiba}. 
The 
singularity correction makes scalaron reasonably light in denser regions,
such as 
in the galactic center, which may mitigate the suppression of the field
due 
to its induced force. Using the singularity corrected model, we obtain rotational
velocity with scalaron contribution term for test particles moving around galaxies in 
stable circular orbits. The rotation curves thus predicted by the theory are then fitted with observations 
of some samples of high surface brightness (HSB), LSB and dwarf galaxies, and 
achieve well-fitted rotation curves almost for all sample galaxies.

The work is organized as follows. In Section \ref{sec2}, we present the field 
equations of the $f(R)$ gravity and then via conformal 
transformation the minimal coupling of a scalar field (scalaron) to 
non-relativistic matter is presented. Also, the equation of motion of the 
scalaron, its effective potential and expression for mass are derived. In 
Section \ref{sec3}, by considering a new f($R$) gravity 
model~\cite{2020_dhruba,2022_dhruba} as mentioned above, the chameleonic 
behavior of the scalaron is outlined.
In Section \ref{sec4}, assuming a 
static, spherically symmetric spacetime the orbital motion of a test particle 
in the presence of the scalaron is presented. Finally, we conclude our work in 
Section \ref{sec5} with some future prospects. In this work, we set 
$c=\hbar =1$ and adopt the metric signature as $(-,+,+,+)$. 

\section{$\boldsymbol{f(R)}$ Gravity Field Equations and Conformal 
	Transformation}
\label{sec2}
\subsection{Field equations in metric formalism}
To derive field equations in metric formalism, we proceed from the action of 
$f(R)$ gravity, which is given as ~\cite{2010_felice,2010_sotiriou}
\begin{equation} 
	S = \frac{1}{2\kappa^2}\int\!d^4x\,\sqrt{-\, g}\,{f(R)}  + S_m\left(g_{\mu\nu}, \psi\right),
	\label{eq1}
\end{equation}
where $g$ is the determinant of the metric $g_{\mu\nu}$, 
$\kappa^2 = 8\pi G = M_{pl}^{-2}$, $M_{pl}$ is the reduced Plank mass 
$\approx 2.4\times 10^{18}$ GeV, and $S_m(g_{\mu\nu}, \psi) = 
\int\!d^4x\,\sqrt{-\, g}\,\mathcal{L}_m(g_{\mu\nu},\psi)$ is the 
matter action with the non-relativistic matter field $\psi$. The Ricci scalar 
$R$ is defined as $R=g^{\mu\nu}R_{\mu\nu}$, where the Ricci tensor 
$R_{\mu\nu}$ is expressed in the from:
\begin{equation} 
	R_{\mu\nu} = R^\sigma_{\mu\sigma\nu}=\partial_\sigma\Gamma^\sigma_{\mu\nu}-\partial_\nu\Gamma^\sigma_{\mu\sigma} +\Gamma^\lambda_{\mu\nu}\Gamma^\sigma_{\lambda\sigma}-\Gamma^\lambda_{\mu\sigma}\Gamma^\sigma_{\lambda\nu}.
	\label{eq2}
\end{equation}
The variation of action \eqref{eq1} with respect to metric $g_{\mu\nu}$ 
results in the field equations of the metric formalism of $f(R)$ gravity as 
given by
\begin{equation} 
	f_R R_{\mu\nu}- \frac{1}{2}\,g_{\mu\nu} f(R)-\big[\nabla_{\mu}\nabla_{\nu}-g_{\mu\nu}\,\square \big]\,f_R=\kappa^2\, T_{\mu\nu}\left(g_{\mu\nu}, \psi\right).
	\label{eq3}
\end{equation}
Here, $f_R=df/dR$, $\nabla_{\mu}$ is the covariant derivative connected with 
the Levi-Civita connection of metric $g_{\mu\nu}$, 
$\square=\nabla^{\mu}\nabla_{\mu}$ is the Laplace operator in four dimensions
and $T_{\mu\nu}$ is the energy-momentum tensor which satisfies the condition: 
$\nabla^\mu T_{\mu\nu}=0$, and usually it is given by  
\begin{equation} 
	T_{\mu\nu}= -\, \frac{2}{\sqrt{-g}}\, \frac{\delta S_m\left(g_{\mu\nu}, \psi\right)}{\delta g^{\mu\nu}}
	\equiv -\, \frac{2}{\sqrt{-g}}\, \frac{\delta (\sqrt{-g}\,\mathcal{L}_m\left(g_{\mu\nu}, \psi\right))}{\delta g^{\mu\nu}}.
	\label{eq4}
\end{equation}
The field equations \eqref{eq3} in the form of Einstein field equations can be 
written as
\begin{equation} 
	G_{\mu\nu}=\frac{\kappa^2\,T_{\mu\nu}}{f_R}+g_{\mu\nu}  \frac{f(R)-Rf_R}{2f_R}+\frac{\nabla_\mu \nabla_\nu f_R-g_{\mu\nu}\,\square f_R}{f_R}=\frac{\kappa^2}{f_R}\left(T_{\mu\nu}+T_{\mu\nu}^{eff}\right),
	\label{eq5}
\end{equation}
where $T_{\mu\nu}^{eff}$ is the effective energy-momentum tensor of the 
following form:
\begin{equation} 
	T_{\mu\nu}^{eff}=\frac{1}{\kappa^2}\left[g_{\mu\nu} \frac{f(R)-Rf_R}{2}+\big(\nabla_\mu \nabla_\nu -g_{\mu\nu}\,\square\big) f_R\right].
	\label{eq6}
\end{equation}
Further, the modified field equations \eqref{eq5} may also take the form: 
\begin{equation} 
	f_R\left[G_{\mu\nu} + G_{\mu\nu}^d\right] =\kappa^2\,T_{\mu\nu},
	\label{eq7}
\end{equation}
where $$G_{\mu\nu}^d=\frac{1}{f_R}\left[g_{\mu\nu}\frac{Rf_R -f(R)}{2}-\big(\nabla_\mu \nabla_\nu -g_{\mu\nu}\square\big)f_R\right].$$ is the deviated part
of the Einstein tensor that appears due to the change in the curvature 
introduced by the $f(R)$ gravity. The effective energy-momentum tensor is also 
emerged due to the same reason, i.e.~the change of the behavior of curvature 
in $f(R)$ gravity.

The existence of the last two terms on the left hand side of equations 
\eqref{eq3} expresses them as fourth-order partial differential equations. 
The trace of equations \eqref{eq3} is obtained as  
\begin{equation} 
	3\, \square f_R + f_R\, R - 2\,f(R)  = \kappa^2 T,
	\label{eq8}
\end{equation}
where $T=T_\mu^\mu$ is the trace of the energy-momentum tensor and 
$\square f_R=\partial_\mu(\sqrt{-g}\,g^{\mu\nu}\partial_\nu f_R)/\sqrt{-g}$.
Equation \eqref{eq8} is the equation of motion for $f_R$ that determines the 
dynamics of $f_R$. It shows that the Ricci scalar $R$ is dynamic. With 
$f(R)$ as a linear function of $R$, $f_R$ is constant and hence 
$\square f_R = 0$. In this case, the theory reduces to GR leading the equation 
\eqref{eq8} as $R=-\,\kappa^2 T$, implying the direct dependence of Ricci 
scalar on matter. Instead, in $f(R)$ theory $\square f_R $ does not vanish. It 
signifies that the theory includes an extra propagating degree of freedom 
$f_R$, called the scalar degree of freedom that characterizes the modification 
of gravity and plays a central role in explaining the galactic dynamics without 
dark matter~\cite{2017_katsu,2018_verma,2012_li}. With this conformity, 
we proceed with the conformal transformation in the following 
subsection to identify $f_{R}$ conveniently in an explicit form.

\subsection{Conformal transformation}
Conformal transformation is a mathematical technique used expediently to 
simplify the equations of motion in the Jordan frame into an amenable form
\cite{1984_whitt}. In fact, it is a tool that physicists generally use for 
a better understanding of the implications of MTGs and their connection to the 
standard framework of GR. Through a conformal transformation, the higher-order 
and non-minimally coupled terms can always be shown as the sum of the Einstein 
gravity and one or more minimally coupled scalar field(s)~\cite{2011_Capoz}. In 
the case of the fourth-order gravity when undergoing this transformation, gives 
one scalar field with the Einstein gravity. Before performing such a 
transformation we recast the action \eqref{eq1} in the following 
form~\cite{2010_felice}:
\begin{equation}
	S = \int\!d^4x\,\sqrt{-\, g}\,\left[\frac{Rf_R}{2\kappa^2}-X(R)\right] + 
	\int\!d^4x\,\sqrt{-\, g}\,\mathcal{L}_m\left(g_{\mu\nu},\psi\right),
	\label{eq9}
\end{equation}
where \begin{equation}X(R)=\frac{1}{2\kappa^2}\big[Rf_R-f(R)\big].
	\label{eq10}
\end{equation}
Equation \eqref{eq10} can be treated as the potential of an auxiliary scalar 
field defined in the Jordan frame~\cite{2021_Nashiba,2010_sotiriou,2020_vel}.
This potential is non-zero only when $f_R \ne$ constant.
In the case $f(R) = {R}$, it becomes zero and hence the theory 
reduces to GR.  
We consider now the following conformal transformation that reduces the 
fourth-order non-minimally coupled field equations \eqref{eq3} into 
a minimally coupled tractable form through the transition of the Jordan 
frame metric $g_{\mu\nu}$ to the Einstein frame metric $\tilde{g}_{\mu\nu}$ 
as \cite{2010_felice,2007_Faraoni,2009_dab,1999_far}
\begin{equation} 
	g_{\mu\nu}\rightarrow\tilde{g}_{\mu\nu}= \Omega^2 g_{\mu\nu},
	\label{eq11}
\end{equation}
where $\Omega^2$ is a non-vanishing regular function called the conformal 
factor. Under transformation rule \eqref{eq11} the Ricci scalars $R$ and 
$\tilde R$ in these two respective frames are related by 
\begin{equation} 
	R= \Omega^2\left[\tilde R+6\,\square\,(\ln\Omega)-6\,\tilde g^{\mu\nu}\nabla_\mu\,(\ln\Omega)\,\nabla_\nu\,(\ln\Omega)\right].
	\label{eq12}
\end{equation}
Also, the determinants of the metric tensors in the two frames hold the 
following relation: 
\begin{equation} 
	\sqrt{-g}= \Omega^{-4}\sqrt{-\tilde g}\,.
	\label{eq13}
\end{equation}    
Using equations \eqref{eq9}, \eqref{eq12} and \eqref{eq13}, and also 
Gauss's theorem the action in Einstein's frame is derived as given below:
\begin{align}
	S_E= \int\!d^4x\,\sqrt{-\tilde g}\,\Big[\frac{1}{2\kappa^2}f_R\,\Omega^{-2} \big\{\tilde{R}&-6\,\tilde g^{\mu\nu}\nabla_\mu(\ln\Omega)\nabla_\nu(\ln\Omega)\big\}-\frac{X(f_R)}{\Omega^4}\Big]\notag \\ & + \int\!d^4x\;\Omega^{-4}\sqrt{-\tilde g}\,\mathcal{L}_m\left(\Omega^{-2}\,\tilde g_{\mu\nu},\psi\right).
	\label{eq14}
\end{align}

At this stage, we are in a position to introduce a new scalar field called 
scalaron $\phi$ defined by~\cite{2013_kob}
\begin{equation} 
	\phi=\frac{1}{\kappa}\,\sqrt\frac{3}{2}\ln f_R.
	\label{eq15}  
\end{equation}
This field corresponds to an extra scalar degree of freedom $f_R$ in $f(R)$ 
theory as
\begin{equation} 
	f_R=e^{\sqrt{\frac{2}{3}}\kappa\phi}.
	\label{eq16}
\end{equation}
Using the scalar field defined in equation \eqref{eq15} and choosing the 
conformal factor $\Omega^2=f_R$, the action \eqref{eq14} in Einstein's frame 
can be written in the form:
\begin{align} 
	S_E=\frac{1}{2\kappa^2}\! \int\!d^4x\,\sqrt{-\tilde g}\,\tilde R & + \int\!d^4x\,\sqrt{-\tilde g}\,\left[-\frac{1}{2}\,\tilde g^{\mu\nu}\partial_\mu \phi\,\partial_\nu\phi-V(\phi)\right]\notag \\ &+ \int\!d^4x\;\frac{\sqrt{-\tilde{g}}}{f_R^{\,2}}\,\mathcal{L}_m\left(f_R(\phi)^{-1}\,\tilde g_{\mu\nu},\psi\right).
	\label{eq17}
\end{align}
Here, we may write $f_R(\phi)^{-1}=e^{2\beta\kappa\phi}$ which gives constant 
coupling factor $\beta=-1/\sqrt{6}$. It indicates that the scalar field is 
directly coupled to matter field $\psi$ with a constant coupling $\beta$ in 
the Einstein frame. Further, $V(\phi)$ represents the potential of the 
scalaron field defined by
\begin{equation} 
	V(\phi)=\frac{X}{f_R^{\,2}}=\frac{1}{2\kappa^2}\left[\frac{Rf_R-f(R)}{f_R^{\,2}}\right].
	\label{eq18}
\end{equation}
Equation \eqref{eq17} shows that $f(R)$ gravity can be expressed by means of 
GR along with the additional scalar field, the scalaron, that holds noteworthy 
significance in the theory.

\subsection{Equation of motion of scalaron and its mass}
Variation of action \eqref{eq17} with respect to $\phi$ gives the following 
equation:
\begin{equation} 
	\tilde \square\,\phi-\dfrac{dV(\phi)}{d\phi}+\frac{1}{\sqrt {-\tilde g}}\frac{\partial(\sqrt{- g}\,\mathcal{L}_m)}{\partial \phi} = 0,
	\label{eq19}
\end{equation}
where $$\tilde \square=\frac{1}{\sqrt {-\tilde g}}\,\partial _\mu(\sqrt{-\tilde g}\,\tilde g^{\mu\nu}\partial_\nu).$$ 
Again, the derivative factor of the last term on the left-hand side of 
equation \eqref{eq19} can be expressed as
\begin{equation} 
	\frac{\partial(\sqrt{- g}\,\mathcal{L}_m)}{\partial \phi} = -\,\frac{\sqrt{- g}}{2}\left(-\,\frac{2}{\sqrt {- g}}\,\frac{\delta (\sqrt{- g}\,\mathcal{L}_m)}{\delta g^{\mu\nu}}\right)\frac{\partial g^{\mu\nu}}{\partial \phi}=-\,\frac{\sqrt {- g}}{2}\,T_{\mu\nu}\,\frac{\partial g^{\mu\nu}}{\partial \phi}.
	\label{eq20}
\end{equation}
Also we have,
\begin{equation} 
	\tilde T_{\mu\nu}=-\,\frac{2}{\sqrt {-\tilde g}}\,\frac{\delta (\sqrt{- g}\,\mathcal{L}_m)}{\delta \tilde g^{\mu\nu}}=\frac{T_{\mu\nu}}{f_R}
	\label{eq21}.
\end{equation}
Use of equations \eqref{eq11} in contravariant form, and equations 
\eqref{eq13} and \eqref{eq21} in equation \eqref{eq20} yields 
\begin{equation} 
	\frac{\partial(\sqrt{- g}\,\mathcal{L}_m)}{\partial \phi} = -\,\frac{\sqrt {-\tilde{g}}}{2f_R}\,\frac{df_R}{d\phi}\,\tilde T_\mu^\mu.
	\label{eq22}
\end{equation}
From $\tilde T_\mu^\mu=T_\mu^\mu/f_R^{\,2}$, the relation between the trace of 
the energy-momentum tensors in two frames and equation \eqref{eq22} together 
with relation \eqref{eq16} we can rewrite scalaron's equation of motion 
\eqref{eq19} as
\begin{equation} 
	\tilde \square\,\phi-\frac{dV_{eff}}{d\phi}= 0,
	\label{eq23}
\end{equation}  
where 
\begin{align} 
	\dfrac{dV_{eff}(\phi)}{d\phi} & =\dfrac{dV(\phi)}{d\phi}+\frac{\kappa}{\sqrt 6}\,\tilde{T}_\mu^\mu\notag\\
	& = \dfrac{dV(\phi)}{d\phi}+\frac{\kappa}{\sqrt 6}\,e^{-2\sqrt{\frac{2}{3}}\kappa\phi}\,T_\mu^\mu. 
	\label{eq24}
\end{align} 
$V_{eff}(\phi)$ refers the effective potential of the scalaron field. Equation 
\eqref{eq23} shows in conjunction with equation \eqref{eq24} that scalaron 
couples to the matter field in the Einstein frame with a constant coupling
strength $1/\sqrt{6}$ as mentioned earlier. In the Jordan frame, this coupling 
strength decreases exponentially with the increasing scalaron field value.  
Also it should be noted that as matter moves on geodesic of Jordan frame 
metric, the Einstein frame energy-momentum tensor $\tilde T_{\mu\nu}$ is not 
covariantly conserved, i.e.~$\tilde{\nabla}_\mu \tilde{T}_{\mu\nu}\neq 0$
\cite{2018_burrage}. By integrating equation \eqref{eq24} one may 
obtain the effective potential $V_{eff}$ of the scalaron as follows:
\begin{equation} 
	V_{eff} = V(\phi)-\frac{1}{4}\,\,e^{-2\sqrt{\frac{2}{3}}\kappa\phi}\,T_\mu^\mu.
	\label{eq25}
\end{equation} 
The presence of the second term in equation \eqref{eq25} indicates that 
the dynamics of the scalaron described by the equation \eqref{eq23} is 
influenced by the surrounding matter content represented by 
$T_\mu^\mu$. Consequently, the effective potential results in the environment 
matter dependent scalaron mass
\,$m_\phi$ as the square of 
$m_\phi$ corresponds to the second order derivative of effective potential 
at the minimum of the field value $\phi_0$.
	So, we obtain the mass of the scalaron by differentiating equation 
	\eqref{eq24} with respect to $\phi$ at $\phi=\phi_0$ as
	\begin{equation} 
		m_\phi^2=\dfrac{d^2V_{eff}(\phi)}{d\phi^2}=\dfrac{d^2V(\phi_0)}{d\phi^2}-
		\frac{2\kappa^2}{3}e^{-2\sqrt {\frac{2}{3}}\kappa\phi_0}\,T_\mu^\mu.
		\label{eq26}
	\end{equation}
	From equation \eqref{eq26} it is found that the potential $V(\phi)$ is not only the determining factor of the scalaron mass, it is strongly 
	affected by \textbf{$T_\mu^\mu$} of environment as mentioned above. 
	To explicitly express the factor contributed
	by the potential $V(\phi)$ to the scalaron mass and also to express the 
	scalaron mass in a suitable form we proceed as follows. From equations 
	\eqref{eq15} and \eqref{eq18} we may write,
	\begin{align} 
		\dfrac{dV(\phi)}{dR} & = \frac{1}{2\kappa^2}\,\frac{2 f(R)f_{RR}-Rf_Rf_{RR}}{f_R^{\,3}}.
		\label{eq27}\\[5pt]
		\dfrac{d\phi}{dR} & =\frac{1}{\kappa}\,\sqrt{\frac{3}{2}}\frac{f_{RR}}{f_R}.
		\label{eq28}
	\end{align}
	These two equations provide us the  derivatives of $V(\phi)$ with respect to 
	$\phi$ as
	\begin{equation} 
		\dfrac{dV(\phi)}{d\phi}=\frac{1}{\sqrt{6}\,\kappa}\,\frac{2f(R)-Rf_R}{f_R^{\,2}}.
		\label{eq29}
	\end{equation}
	From this equation, we may obtain,
	\begin{equation} 
		\dfrac{d^2V(\phi)}{d\phi^2}=\frac{1}{3f_{RR}}\left[1+\frac{Rf_{RR}}{f_R}-\frac{4f(R)f_{RR}}{f_R^{\,2}}\right].
		\label{eq30}
	\end{equation}
	It should be mentioned that at the minimum of the scalaron field value i.e., 
	at $\phi=\phi_0$, the effective potential satisfies the condition 
	$dV_{eff}(\phi_0)/d\phi=0$, which turns the left-hand side of equation 
	\eqref{eq24} to zero. Thus, inserting equations \eqref{eq15} and \eqref{eq29} 
	into \eqref{eq24}, the stationary condition satisfied by the minimum of 
	potential is obtained as   
	\begin{equation} 
		2f(R_0)-R_0\,f_R(R_0)= -\kappa^2\,T_\mu^\mu,
		\label{eq31}
	\end{equation}
	where $R_0$ is the Ricci scalar corresponding to the minimum of the scalaron
	$\phi_0$. Using equations \eqref{eq30} and \eqref{eq31} we can write a general 
	expression from equation \eqref{eq26} for $m_\phi^2$ at minimum field value in 
	the form given below:
	\begin{equation} 
		m_\phi^2=\frac{1}{3f_{RR}(R_0)}\left[1-\frac{R_0f_{RR}(R_0)}{f_R(R_0)}\right].
		\label{eq32}
	\end{equation} 
	This equation is useful to study scalaron mass for a specific $f(R)$ gravity 
	model.
	
	\section{Chameleon Mechanism in a New $\boldsymbol{f(R)}$ Gravity Model}
	\label{sec3}
	In this section, we study the scalaron mass and its chameleonic behavior in a
	model specific case considering a recently proposed $f(R)$ gravity model 
	\cite{2020_dhruba,2022_dhruba}, which is given by
	\begin{equation} 
		f(R)=R-\frac{\alpha}{\pi}\,R_c\,\cot^{-1}\!\left(\frac{R_c^2}{R^2}\right)-\beta R_c\left[1-e^{-\frac{R}{R_c}}\right], 
		\label{eq33}
	\end{equation}
	where $\alpha$, $\beta$ are two dimensionless positive constants and $R_c$ is 
	a characteristic curvature constant having the dimension same as the curvature 
	scalar. For this model \eqref{eq33}, $f_R$ can be obtained as
	\begin{equation} 
		f_R(R) = 1-\frac{2\,\alpha R_c^3}{\pi R^3\left(1+\frac{R_c^4}{R^4}\right)}-\beta e^{-\frac{R}{R_c}}.
		\label{eq34}
	\end{equation}
	Notably, in $f(R)$ gravity, the curvature $R$ and the trace of 
the energy-momentum tensor $T_\mu^\mu$ are related through the modified trace 
equation \eqref{eq8}, which shows how the function $f(R)$ governs the 
spacetime curvature response in the presence of matter and energy. Here, we 
assume a large curvature limit,
i.e.~$R\gg R_c$, we can approximate equation (33) in this
limit as
	\begin{equation} 
		f(R)=R-\frac{\alpha R_c}{2}-\beta R_c. 
		\label{eq35}
	\end{equation}
Similarly, under this 
	consideration equation \eqref{eq34} reduces to
	\begin{equation} 
		f_R(R) = 1-\frac{2\,\alpha R_c^3}{\pi R^3}-\beta\, e^{-\frac{R}{R_c}}.
		\label{eq36}
	\end{equation}
	The derivative of it appears as
	\begin{equation} 
		f_{RR}(R)=\frac{6\,\alpha\, R_c^3}{\pi\,R^4} + \frac{\beta}{R_c}\,e^{-\frac{R}{R_c}}.
		\label{eq37}
	\end{equation}
Assuming the energy-momentum tensor for an ideal non-relativistic 
matter we obtain $T_\mu^\mu = - \rho$, where $\rho$ is the local matter energy 
density. Thus equation \eqref{eq31} which is a consequence of minimum 
effective potential ensues a relation between matter density and curvature in high curvature limit $R >> R_c$ as~\cite{2021_Nashiba,2018_kat}
	\begin{equation} 
		R_0= \rho\kappa^2+\alpha R_c+2\beta R_c.
		\label{eq38}
	\end{equation}
	This relation when used in equation \eqref{eq32}, along with equations 
	\eqref{eq36} and \eqref{eq37} for the respective $f_R(R_0)$ and $f_{RR}(R_0)$ 
	terms, we obtain a local matter density dependent mass of the scalaron for 
	the chosen model as 
	\begin{equation} 
		m_\phi^2=\frac{\pi\left(\rho\kappa^2+\alpha R_c+2\beta R_c \right)^{\!4}}{3}\!\left[\frac{R_c}{6\alpha R_c^4 + \pi \beta\left(\rho\kappa^2+\alpha R_c+2\beta R_c \right)^{\!4}e^{-\frac{R_0}{R_c}}}-
		\frac{1}{\pi\left(\rho\kappa^2+\alpha R_c+2\beta R_c\right)^{\!3}[1 - \beta e^{-\frac{R_0}{R_c}}]-2\alpha R_c^3}\right].
		\label{eq39}
	\end{equation}
	
\begin{figure}[!h]
		\centerline{
			\includegraphics[scale = 0.35]{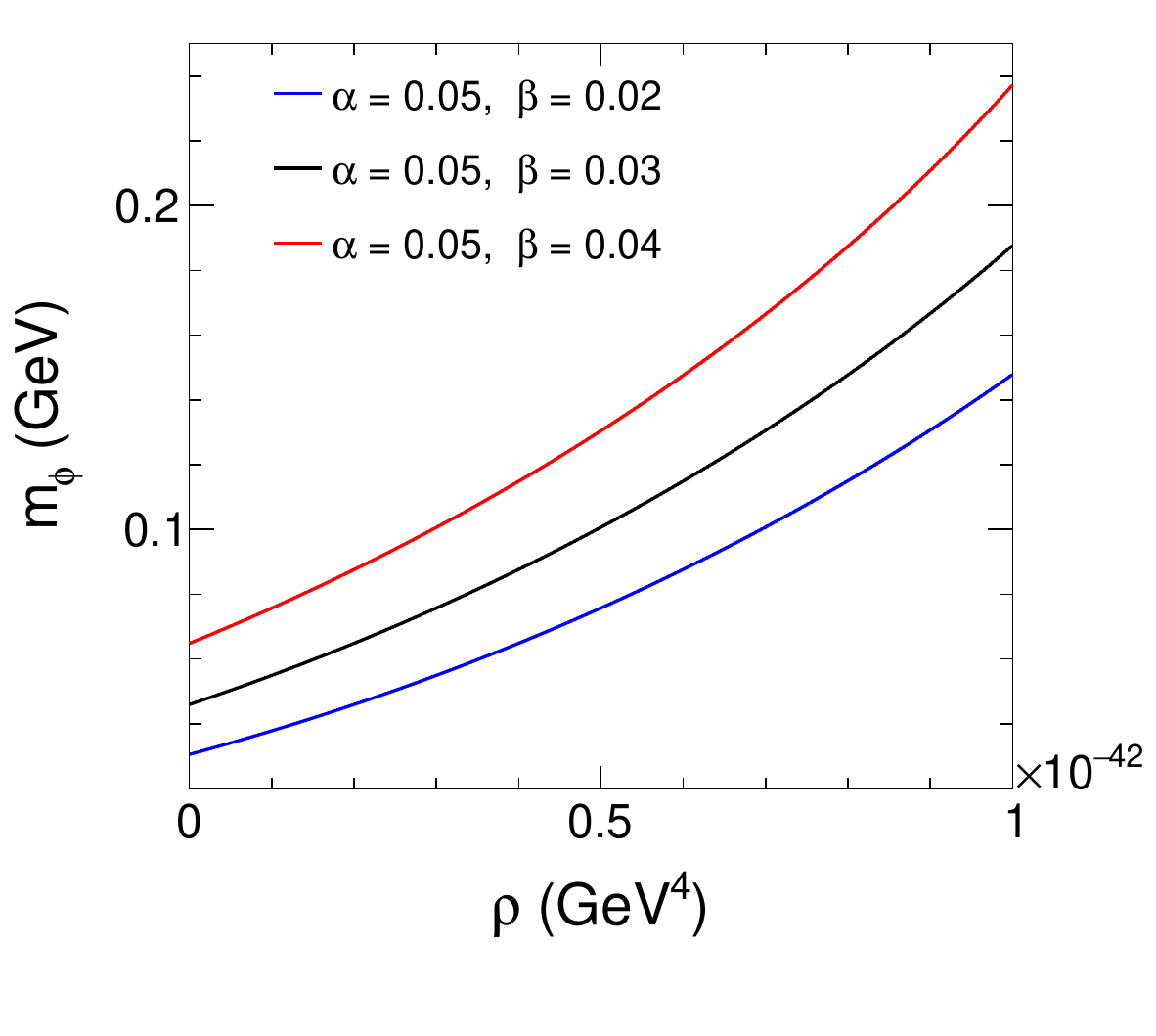}\hspace{0.5cm}
			\includegraphics[scale = 0.35]{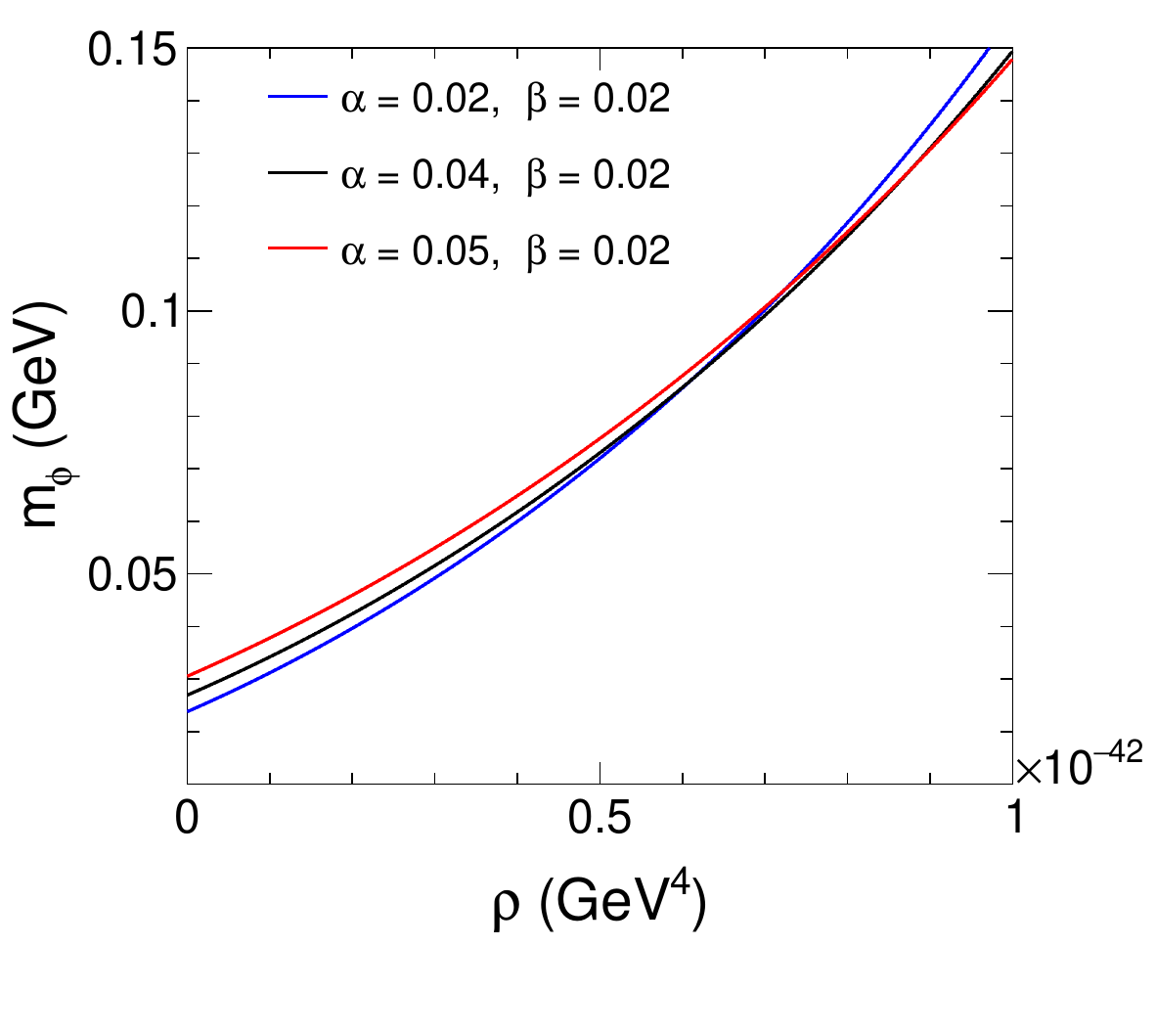}}\vspace{-0.2cm}
		\caption{Relations between scalaron mass and matter density for two sets of model 
			parameters $\alpha$ and $\beta$ with $R_c=1$. The left plot is for different values of $\beta$ by fixing $\alpha$ to 0.05 and the right plot is for different values of $\alpha$ by fixing $\beta$ to a value 0.02.}
		\label{fig1}
	\end{figure}

    It is clear from equation \eqref{eq39} that the scalaron mass depends on 
	matter density $\rho$ of the surrounding environment and it increases as a 
	certain power function of density $\rho$. This behaviour of the scalaron mass 
	with increasing $\rho$ confirms the chameleonic nature of the scalaron in the 
	model we have considered for our study. We depict this particular behavior of 
	scalaron mass in Fig.~\ref{fig1} for different values of model parameters by 
	setting the characteristic curvature constant $R_c=1$ \cite{2020_dhruba}. 
	Of course there is no specific reason for choosing these particular values
	of model parameters. These are considered arbitarily to study the variation of
	scalaron mass $m_\phi$ with environment density. However, it is seen that 
for each set of parameters $\alpha$ and $\beta$, the scalaron mass 
increases as the matter density increases. The plot in the left panel 
illustrates the variation of $m_\phi$ for changing $\beta$, while the right 
panel demonstrates the effect of varying $\alpha$ on $m_\phi$. 
	\\
	It is seen from equations \eqref{eq18} and \eqref{eq33} that the scalaron 
	potential is a function of curvature scalar $R$. By substituting equation 
	\eqref{eq16} into equation \eqref{eq18}, it can also be expressed as a 
	function of the scalaron field $\phi$ where the relation between curvature 
	scalar and scalaron field is given by~\cite{2021_Nashiba,2018_kat} 
	\begin{equation} 
		R=\left(\frac{2\alpha}{\pi}\right)^{1/3}\!\!\!\!R_c\left(1-e^{\sqrt{\frac{2}{3}}\kappa\phi}\right)^{-1/3}\!\!\!,
		\label{eq40}
	\end{equation}
	which is obtained from equations \eqref{eq16} and \eqref{eq36} in high 
	curvature regime approximation consideration. Although, the 
approximation made here is a bit stringent, the result gives us a clear 
indication of the presence of curvature singularity~\cite{2008_frolov} which 
is discussed in Appendix \ref{ap1}. 

\section{Rotation curves and scalaron}
\label{sec4}
\subsection{Galactic rotational velocity}
To derive the equation of motion of a test particle (a star) using conformal 
coupling we consider the motion of the particle in a static gravitational 
field within a constant velocity region described by the following 
spherically symmetric metric: 
\begin{equation}
	ds^2 = - e^{\mu} dt^2 + e^{\nu} dr^2 + r^2d\theta^2 + r^2 \sin^2\theta\,d\varphi^2, 	
	\label{eq40}
\end{equation}
where the metric components $\mu$ and $\nu$ are functions of the radial 
coordinate $r$ only. Since the conformal coupling given by equation 
\eqref{eq11} with the conformal factor expressed by equation \eqref{eq16} 
makes the interaction between scalaron field with matter distribution, we will 
use this coupling in the geodesic equation of the particle. On the other hand, 
the matter field $\psi$ couples to Jordan frame metric $g_{\mu\nu}$ instead of 
Einstein frame metric $\tilde g_{\mu\nu}$~\cite{2018_burrage}. Hence the 
particle follows timelike geodesic of $g_{\mu\nu}$ not that of 
$\tilde g_{\mu\nu}$. The geodesic of the particle is, therefore, given by the 
following equation ~\cite{1992_stefan,2015_Zanzi,2024_mohan}:   
\begin{equation} 
	\dfrac{d^2x^\eta}{d\tau^2}+\Gamma_{\mu\nu}^\eta\dfrac{dx^\mu}{d\tau}\dfrac{dx^\nu}{d\tau}=0,
	\label{eq41}
\end{equation}
where the Christoffel symbol (connection coefficient) which governs the motion 
is defined as
\begin{equation} 
	\Gamma_{\mu\nu}^{\eta} = \,\frac{1}{2}\,g^{\eta\lambda} \left[\frac{\partial g_{\mu\lambda}}{\partial x^\nu} + \frac{\partial g_{\nu\lambda}}{\partial x^\mu}-\frac{\partial g_{\mu\nu}}{\partial x^\lambda}\right].
	\label{eq42}
\end{equation}
From equations \eqref{eq11} and \eqref{eq16}, we have
\begin{equation} 
	g_{\mu\nu}=e^{-\sqrt{\frac{2}{3}}\kappa\phi}\, \tilde g_{\mu\nu},\qquad
	g^{\mu\nu}=e^{\sqrt{\frac{2}{3}}\kappa\phi}\,\tilde g^{\mu\nu}.
	\label{eq43}
\end{equation}
Each term of the right-hand side of equation \eqref{eq42} can be replaced by 
respective conformally transformed equations that are obtained from equation 
\eqref{eq43}. Thus, a relation between the connection coefficients in the 
Jordan and the Einstein frames is established~\cite{2015_Zanzi} which may be written as
\begin{equation} 
	\Gamma_{\mu\nu}^{\eta} =\tilde \Gamma_{\mu\nu}^{\eta}- \,\frac{\kappa}{\sqrt{6}}\,\left[\frac{\partial\phi}{\partial x^\nu}\,\delta _\mu^\eta+\frac{\partial\phi}{\partial x^\mu}\, \delta _\nu^\eta-\frac{\partial\phi}{\partial x^\lambda}\,\tilde g^{\eta\lambda}\,\tilde g_{\mu\nu} \right].
	\label{eq44}
\end{equation}
Substitution of this equation in equation \eqref{eq42} yields
\begin{equation} 
	\dfrac{d^2x^\eta}{d\tau^2}+\tilde \Gamma_{\mu\nu}^\eta\dfrac{dx^\mu}{d\tau}\dfrac{dx^\nu}{d\tau}-\frac{\kappa}{\sqrt{6}}\,\left[2\, \frac{\partial\phi}{\partial x^\mu}\,\dfrac{dx^\mu}{d\tau}\, \dfrac{dx^\eta}{d\tau}\,-\tilde g^{\eta\lambda}\,\frac{\partial\phi}{\partial x^\lambda}\,\tilde g_{\mu\nu}\,\dfrac{dx^\mu}{d\tau}\dfrac{dx^\nu}{d\tau} \right]=0.
	\label{eq45}
\end{equation}

In the non-relativistic or weak gravity limit, the particle's velocity is 
sufficiently slow, i.e. $v\ll c$ and in such a slow velocity situation the 
proper time $\tau$ may be approximated to the coordinate time $t$. Hence, the 
spatial components of four velocity $v^i = dx^i/dt\equiv\left(dx^1/dt, dx^2/dt,
dx^3/dt \right) \ll dx^0/dt$~\cite{1984_wald,2018_sporea,2024_mohan}. Thereby, it is seen 
from the above equation that the first term within the square bracket is 
negligible in comparison with the second term. So, in the weak field limit 
and for static spacetime the equation of motion contains radial component 
only. With the radial component of the geodesic equation, we get from equation 
\eqref{eq45} as given by 
\begin{equation} 
	\dfrac{d^2r}{dt^2}= -\left[\tilde \Gamma_{00}^1 -\frac{\kappa}{\sqrt{6}}\,\frac{d\phi}{dr}\right].
	\label{eq46}
\end{equation}
This is the equation of motion of the particle in the Einstein frame in the 
weak field approximation. It shows that the geodesic in this frame contains a 
term $\kappa/\sqrt{6}\,(d\phi/dr)$ due to the scalaron field in addition to the 
gravitational term. This term can be defined as the acceleration caused by 
chameleon force $F_\phi$ that appears as a result of the chameleonic nature of 
the scalaron field. Obviously, in the weak field limit, chameleon force exists 
and a test particle experiences chameleonic force apart from the gravitation 
force also. So, the dynamics of the test particle will be controlled by both of 
these two forces. Now, for simplicity, if we assume that the orbit of the 
particle is circular, the centripetal acceleration of the particle moving with 
velocity $v$ will be   
\begin{equation} 
	a=-\frac{v^2}{r}.
	\label{eq47}
\end{equation}
These two equations \eqref{eq46} and \eqref{eq47} result in the following 
velocity equation:
\begin{equation} 
	v^2=r\,\left[\tilde \Gamma_{00}^1-\frac{\kappa}{\sqrt{6}}\,\frac{d\phi}{dr}\right].
	\label{eq48}
\end{equation}
As in the weak field limit, the velocity of a particle is vanishingly 
small~\cite{2014_stabile,2007_Capozzie} as discussed before, one may obtain 
\begin{equation}
	\Gamma_{00}^1\simeq-\frac{1}{2}\frac{\partial g_{00}}{\partial r}. 
	\label{eq49} 
\end{equation} 
Hence,  
\begin{equation}
	\tilde \Gamma_{00}^1\simeq-\frac{1}{2}\frac{\partial \tilde g_{00}}{\partial r}   \simeq-\frac{1}{2}\frac{\partial (\Omega^2 g_{00})}{\partial r}.
	\label{eq50} 
\end{equation} 
Next, to compute $d\phi/dr$ we will use equation \eqref{eq28} and the metric 
\eqref{eq40}. It can be expressed as 
\begin{equation}
	\frac{d\phi}{dr}=\frac{d\phi}{dR}\frac{dR}{dr}.
	\label{eq51}
\end{equation}
From the metric \eqref{eq40} Ricci scalar is obtained as 
\begin{equation}
	R=\frac{e^{-\nu}}{2r^2}\Big[-4+4e^\nu-r^2\mu'^2+4r\nu'+r\mu'(-\,4+r\nu')-2r^2\mu''\ \Big],
	\label{eq52}
\end{equation}
where the prime denotes the derivative with respect to the radial 
coordinate $r$. Thus
\begin{align} 
	\dfrac{dR}{dr}=\frac{e^{-\nu}}{2r^3}\Big[8-8e^\nu+r^3\mu'^2\nu+r\mu'\big\{4&+r(4\nu'-r\nu^2+r(-\,2\mu''+\nu''))\big\}\,+ \notag \\ r^2&(-\,4\nu'^2-4\mu''+3r\nu'\mu''+4\nu''-2r\mu''')\Big].
	\label{eq53}
\end{align} 
It is clear from equations \eqref{eq50} and \eqref{eq53} that to have a 
precise velocity equation from equation \eqref{eq48}, we need to derive the 
explicit forms of the $e^\mu$ and $e^\nu$ first. These coefficients will shape 
the velocity profile of the particle by determining $dR/dr$ and 
$\Gamma_{00}^1$ respectively.

In the said context, it is worth mentioning that the generalized field 
equation of an MTG can be recast in the form given 
below ~\cite{2023_Nashiba,2018_sporea,2015_mimoso,2016_Wojnar}:
\begin{equation} 
	\xi(\varphi)\left[G_{\mu\nu}+ X_{\mu\nu}\right]=\kappa^2\,T_{\mu\nu},
	\label{eq54}
\end{equation}
where $X_{\mu\nu}$ is an additional term that appeared due to modification of 
geometry in an MTG, $\xi(\varphi)$ is a coupling factor that couples geometry 
to matter, $\varphi$ may be a curvature invariant or other gravitational field 
such as a scalar field. For $X_{\mu\nu}=0$ and $\xi(\varphi)=1$, one can 
recover GR. In our case, we have equation \eqref{eq7} similar to generalized 
equation \eqref{eq54}, where the coupling factor $\xi(\varphi)=f_R$ and the 
tensor $G_{\mu\nu}^d$ represents the modified term regarding GR. From the 
metric \eqref{eq40} and the generalized field equation \eqref{eq54}, 
derivation of the metric coefficient $e^\nu$ is accomplished in Refs.~\cite{2023_Nashiba}
and is found as
\begin{equation} 
	e^\nu= \left(1-\frac{2G\bar{M}(r)}{r}\right)^{-1}=\left(1-\frac{2GM(r)}{\Omega^2\,r}\right)^{-1}\!\!\!\!,
	\label{eq55}
\end{equation}
where $\bar{M}(r)$ denotes modified mass profile ~\cite{2023_Nashiba,2018_sporea} which is
\begin{equation} 
	\bar{M}(r)= \frac{M(r)}{\Omega^2}=\frac{1}{\Omega^2}\int_{0}^{r}\left(\frac{\kappa^2\rho(r)\, r^2}{2Gf_R(r)}-\Omega^2\,\frac{r^2X_{00}(r)}{2Ge^\mu}\right)dr.
	\label{eq56}
\end{equation}
Equation \eqref{eq55} indicates the resemblance of the form of metric 
coefficient $e^\nu$ to the standard Schrawzchild form. As in the limit 
$r\rightarrow\infty$, i.e.~at very large distance from the source of the 
gravitation field (in our case from the center of the galaxy), the metric 
coefficients can plausibly be assumed in the standard Schrawzchild form. 
Therefore, we consider the weak field limit where the spacetime can be adopted 
as Minkowskian and thereby can be expressed $\tilde g_{\mu\nu}$ in terms of 
Minkowski metric $\eta_{\mu\nu}$ as ~\cite{1992_stefan,2023_Nashiba} 
\begin{equation} 
	\tilde g_{\mu\nu}= \Omega^2\left(\eta_{\mu\nu}+g_{\mu\nu}\right),
	\label{eq57}
\end{equation}
and we obtain $g_{00}=-\,2\phi$ to the first order in $g_{\mu\nu}$, where 
$\phi=G\bar{M}(r)/r=GM(r)/\Omega^2r$. Hence, we may write $e^{\mu}$ from equation 
\eqref{eq57} as
\begin{equation} 
	e^{\mu}\approx1-\frac{2G\bar{M}(r)}{r}=1-\frac{2GM(r)}{\Omega^2\,r}.
	\label{eq58}
\end{equation}
Thus, $\tilde \Gamma_{00}^1$ of equation \eqref{eq50} takes the form: 
\begin{equation} 
	\tilde \Gamma_{00}^1=\frac{GM(r)}{r^2}.
	\label{eq59}
\end{equation}
On the other hand, we require derivatives of metric functions to get $dR/dr$ 
exactly. On that account, the metric functions are written from \eqref{eq55} 
and \eqref{eq58} as
\begin{align} 
	\nu(r) & = \ln\left(1-\frac{2GM(r)}{\Omega^2\,r}\right)^{-1}\!\!\!\!\!,
	\label{eq60}\\[5pt] 
	\mu(r) & = \ln\left(1-\frac{2GM(r)}{\Omega^2\,r}\right).
	\label{eq61}
\end{align}
After performing derivations of these two equations with respect to $r$ we are 
able to find
\begin{align} 
	\nu'& = -\left(\frac{2M(r)}{\Omega^2\,r^2}-\frac{4M^2(r)}{\Omega^4\,r^3}\right)e^{2\nu},\;\;\; \nu''= \frac{4M(r)}{\Omega^2\,r^3}e^\nu+\frac{4M^2(r)}{\Omega^4\,r^4}e^{2\nu}.
	\label{eq62}\\[5pt]  
	\mu' & =\frac{2M(r)}{\Omega^2\,r^2}e^\nu,\;\;\; \mu''=-\frac{4M(r)}{\Omega^2\,r^3}e^\nu-\frac{4M^2(r)}{\Omega^4\,r^4}e^{2\nu},
	\label{eq63}\\[5pt] 
	\mu'''& =\frac{24M^2(r)}{\Omega^4\,r^5}e^{2\nu}+\frac{12M(r)e^\nu}{\Omega^2\,r^4}+\frac{16M^3(r)}{\Omega^6\,r^6}e^{3\nu}.
	\label{eq64}
\end{align}
On substitution of equations \eqref{eq55}, \eqref{eq60} and \eqref{eq62} -- 
\eqref{eq64} into equation \eqref{eq53} we attain the expression of $d\phi/dr$ 
from equation \eqref{eq51} with the help of equation \eqref{eq28} after being 
carried out some algebraic calculations as given by  
\begin{equation} 
	\frac{d\phi}{dr}=\frac{2\sqrt{6}f_{RR}}{r^3\kappa f_R}\left[-1+e^{-\nu}-\frac{2M^3(r)}{\Omega^6\,r^3}e^{2\nu}-\frac{3M^2(r)}{\Omega^4\,r^2}e^\nu+\frac{3M^2(r)}{\Omega^4\,r^3}e^\nu+\frac{2M^3(r)}{\Omega^6\,r^4}e^\nu+\frac{4M(r)}{\Omega^2r}\right].
	\label{eq65}
\end{equation}
Our purpose is to find the rotational velocity of a particle (star) according 
to equation \eqref{eq48}. This is the equation of motion of a particle that 
follows the geodesic of metric $g_{\mu\nu}$ basically under the static and weak 
field approximation of the gravitation field sourced by a galaxy. Thus, we 
obtain the rotational velocity equation for the particle by inserting 
equations \eqref{eq59} and \eqref{eq65} into equation \eqref{eq48}, and also 
writing $\Omega^2=f_R$ as follows:
\begin{equation} 
	v^2=\frac{GM(r)}{r}+\frac{2f_{RR}}{r^2f_R}\left[1-e^{-\nu}+\frac{2M^3(r)}{f_R^3\,r^3}e^{2\nu}+\frac{3M^2(r)}{f_R^2\,r^2}e^\nu-\frac{3M^2(r)}{f_R^2\,r^3}e^\nu-\frac{2M^3(r)}{f_R^3\,r^4}e^\nu-\frac{4M(r)}{rf_R}\right].
	\label{eq66}
\end{equation}
We may write this equation as 
	$$v=\sqrt{{v_{Neff}^2} + v_{\phi}^2},$$
where $v_{Neff}= \sqrt{\frac{GM(r)}{r}}$\, can be considered as the effective Newtonian velocity term and $v_{\phi}$ as the velocity due to the contribution of chameleon field. The $v_{Neff}$ is differ from the standard Newtonian velocity because the mass function $M(r)$, defined by the equation \eqref{eq56}, is linked to the modification of gravity.
Notably, we focus on a model of a spherically symmetric system (galaxy) with 
energy density $\rho(r)$. However, the case of spherical symmetry assumed here 
is only for the sake of simplicity, as all galaxies considered in this study 
may not have this geometry. Moreover, from the derivative of the mass 
distribution 
described by \eqref{eq56} with respect to $r$, we may obtain an expression for the energy density function $\rho(r)$. Since the choice of 
$\rho(r)$ 
influences how mass is distributed within a system, therefore, by selecting an appropriate
function for $\rho(r)$ in equation \eqref{eq56}, the mass distribution can be made 
equivalent to a simpler one. In this context, we can assume the following simple mass 
distribution in a galaxy instead of the complex distribution illustrated by the equation equation \eqref{eq56}~\cite{2018_sporea}:
\begin{equation} 
	M(r)=M_0 \left[\sqrt{\frac{r_0}{r_c}}\left(\frac{r}{r+r_c}\right)\right]^{3b}\!\!\!\!,
	\label{eq67}
\end{equation}
where $r_0$ is the scale length, $M_0$ and $r_c$ are the total mass and core 
radius of the galaxy respectively. $b$ is a parameter that determines the slope of the mass profiles of galaxies. The values of $M_0$ and $r_c$ will be 
estimated from the fitting of theoretical rotation curves with observational 
data of respective galaxies. It should be mentioned that the expression 
\begin{equation}
	\rho(r)=\frac{3\,M(r)}{{4\pi\,r^3}}\left[b\left(\frac{r_c}{r+r_c}\right)+\frac{r^2\,X_{00}(r)}{3\,(1-e^\mu)e^\mu}\right]f_R(r),
	\label{eq68}
\end{equation}
which can be obtained by equating $\Omega^2\bar{M}'(r)=\frac{\kappa^2\rho(r)\, r^2}{2Gf_R(r)}-\,\Omega^2\frac{r^2X_{00}(r)}{2Ge^\mu}$ to $M'(r)=\frac{3b\,M(r)r_c}{r(r+r_c)}$ is suitable one to make mass distributions \eqref{eq56} and \eqref{eq67} identical. 
\begin{table}[!h]
	\centering
	\caption{Different physical properties of a set of fifteen observed LSB
		galaxies and the best-fitted values of total mass $M_0$ and core radius $r_c$
		of each galaxy for the $f(R)$ gravity model \eqref{eq33}.}
	\vspace{0.2cm}
	\scalebox{0.95}{
		\begin{tabular}{c c c c c c c c c c c c c c c c }  \hline\\[-8pt]
			& Galaxy Name& ~~~Scale length  & ~~~Distance $D$ &  ~~~Luminosity $L_{B}$ &  ~~~Mass $M_0$ &~~~Core Radius &~~~ $\chi^2_{red}$~~~&\\& &~~~$r_0$ (kpc)~~~&~~~ (Mpc)~~~&~~~($10^{10}L_{\odot}$)~~~&~~~($10^{10}M_{\odot}$)~~~& $r_c$ (kpc)~~~& ~~~& &  \\[2pt] \hline \hline\\[-8pt]
			& DDO 064  & 0.69  & 6.8   & 0.015 & 4.1    &2.04&  0.53& \\ [7pt]
			& F563-V2  & 2.43  & 59.7  & 0.266 & 2.00   &1.50&  2.20 & \\ [7pt]
			& F 568-3  & 4.99  & 82.4  & 0.351 & 1.20   &2.50&  0.42& \\ [7pt]
			& F 583-1  & 2.36  & 35.4  & 0.064 & 4.10   &2.16&  0.39& \\ [7pt]
			& F 583-4  & 1.93  & 53.3  & 0.096 & 0.50   &1.23&  0.33& \\ [7pt]
			& NGC 300  & 1.75  & 2.08  & 0.271 & 3.30   &1.54&  1.50& \\ [7pt]
			& NGC 2976 & 1.01  & 3.58  & 0.201 & 1.01   &0.84&  0.92& \\ [7pt]
			& NGC 3109 & 1.56  & 1.33  & 0.064 & 1.50   &1.45&  1.50& \\ [7pt]
			& NGC 3917 & 2.63  & 18.0  & 1.334 & 8.00   &2.15&  5.10& \\ [7pt]
			& NGC 4010 & 2.81  & 18.0  & 0.883 & 4.10   &2.00&  0.20& \\ [7pt]
			& NGC 1003 & 1.61  & 11.4  & 1.480 & 5.60   &1.80&  0.09& \\ [7pt]
			& UGC 5750 & 3.46  & 58.7  & 0.472 & 9.60   &3.56&  0.31& \\ [7pt]
			& UGC 6983 & 3.21  & 18.0  & 0.577 & 1.80   &1.90&  1.50& \\ [7pt]
			& UGC 7089 & 2.26  & 18.0  & 0.352 & 11.6   &2.70&  0.24& \\ [7pt]
			& UGC 11557& 2.75  & 24.2  & 1.806 & 3.10   &2.20&  0.59& \\  [2pt] \hline \hline
	\end{tabular}}\\[5pt]
	\label{table1}
\end{table}

\begin{table}[!h]
	\centering
	\caption{Different physical properties of a set of ten observed HSB
		galaxies and the best-fitted values of total mass $M_0$ and core radius $r_c$
		of each galaxy for the $f(R)$ gravity model \eqref{eq33}.}
	\vspace{0.2cm}
	\scalebox{0.95}{
		\begin{tabular}{c c c c c c c c c c c c c c c c }  \hline\\[-8pt]
			& Galaxy Name& ~~~Scale length  & ~~~Distance $D$ &  ~~~Luminosity $L_{B}$ &  ~~~Mass $M_0$ &~~~Core Radius &~~~ $\chi^2_{red}$~~~& \\& &~~~$r_0$ (kpc)~~~&~~~ (Mpc)~~~&~~~($10^{10}L_{\odot})$~~~&~~~($10^{10}M_{\odot}$)~~~& $r_c$ (kpc)~~~& ~~~& &  \\[2pt] \hline \hline\\[-8pt]
			& NGC 2903 & 2.33  & 6.6 & 4.088 &23.00 & 2.50 &  4.50& \\ [7pt]
			& NGC 2998 & 6.20  &68.1 & 5.186 &16.00 & 3.90 &  1.04& \\ [7pt]
			& NGC 3198 & 3.14  &13.8 & 3.241 &102.50& 6.70 &  1.20& \\ [7pt]
			& NGC 3769 & 3.38  &18.0 & 0.684 &22.80 & 4.90 &  0.12& \\ [7pt]
			& NGC 4088 & 2.58  &18.0 & 2.957 &69.40 &5.00  &  0.62& \\ [7pt]
			& NGC 4157 & 2.32  &18.0 & 2.901 &81.60 &4.46  &  1.91& \\ [7pt]
			& NGC 4183 & 2.79  &18.0 & 1.042 &29.00 &5.25  &  1.70& \\ [7pt]
			& NGC 4217 & 2.94  &18.0 & 3.031 &57.20 &4.70  &  0.71& \\ [7pt]
			& NGC 5055 & 3.20  & 9.9 & 3.622 &56.00 & 4.50 &  0.41& \\ [7pt]
			& NGC 7793 & 1.21  & 3.61 & 0.910 &20.00 & 2.80 & 3.06& \\ [2pt] \hline \hline
	\end{tabular}}\\[5pt]
	\label{table2}
\end{table}

\begin{table}[!h]
	\centering
	\caption{Different physical properties of a set of twelve observed
		dwarf galaxies and the best-fitted values of total mass $M_0$ and core radius
		$r_c$ of each galaxy for the $f(R)$ gravity model \eqref{eq33}.}
	\vspace{0.2cm}
	\scalebox{0.95}{
		\begin{tabular}{c c c c c c c c c c c c c c c c }  \hline\\[-8pt]
			& Galaxy Name& ~~~Scale length  & ~~~Distance ($D$) &  ~~~Luminosity ($L_{B})$ &  ~~~Mass ($M_0$) &~~~Core Radius  &~~~ $\chi^2_{red}$~~~ &\\& &~~~$r_0$ (kpc)~~~&~~~ (Mpc)~~~&~~~$10^{10}L_{\odot}$~~~&~~~$10^{10}M_{\odot}$~~~& $r_c$ (kpc)~~~& ~~~& &  \\[2pt] \hline \hline\\[-8pt]
			& NGC 2366 & 0.65 & 3.27 & 0.236 & 5.44  &1.10 &  5.04& \\ [7pt]
			& NGC 3877 & 2.53 & 18.0 & 1.948 & 6.00  &1.72 &  4.60& \\ [7pt]
			& NGC 3972 & 2.18 & 18.0 & 0.978 & 7.40  &1.85 &  1.20& \\ [7pt]
			& NGC 5585 & 1.53 & 7.06 & 0.333 & 5.40  &1.60 &  4.90& \\ [7pt]
			& UGC 4325 & 1.86 & 9.6  & 2.026 & 0.25  &0.84 &  7.10& \\ [7pt]
			& UGC 4499 & 1.73 & 12.5 & 1.552 & 1.40  &1.38 &  2.01& \\ [7pt]
			& UGC 6446 & 1.49 & 12.0 & 0.988 & 2.00  &1.28 &  0.54& \\ [7pt]
			& UGC 6667 & 5.15 & 18.0 & 0.422 & 0.13  &1.60 &  2.20& \\ [7pt]
			& UGC 6818 & 1.39 & 18.0 & 0.352 & 9.00  &1.90 &  0.96& \\ [7pt]
			& UGC 6917 & 2.76 & 18.0 & 6.832 & 0.92  &1.50 &  1.20& \\ [7pt]
			& UGC 7524 & 3.46 & 4.74 & 2.436 & 0.90  &2.00 &  0.77& \\ [7pt]
			& UGC 12632& 2.42 & 9.77 & 1.301 & 1.06  &1.68 &  0.70& \\ [2pt] \hline \hline
	\end{tabular}}\\[5pt]
	\label{table3}
\end{table}

\begin{figure}[p]
	\centerline{
		\includegraphics[scale = 0.26]{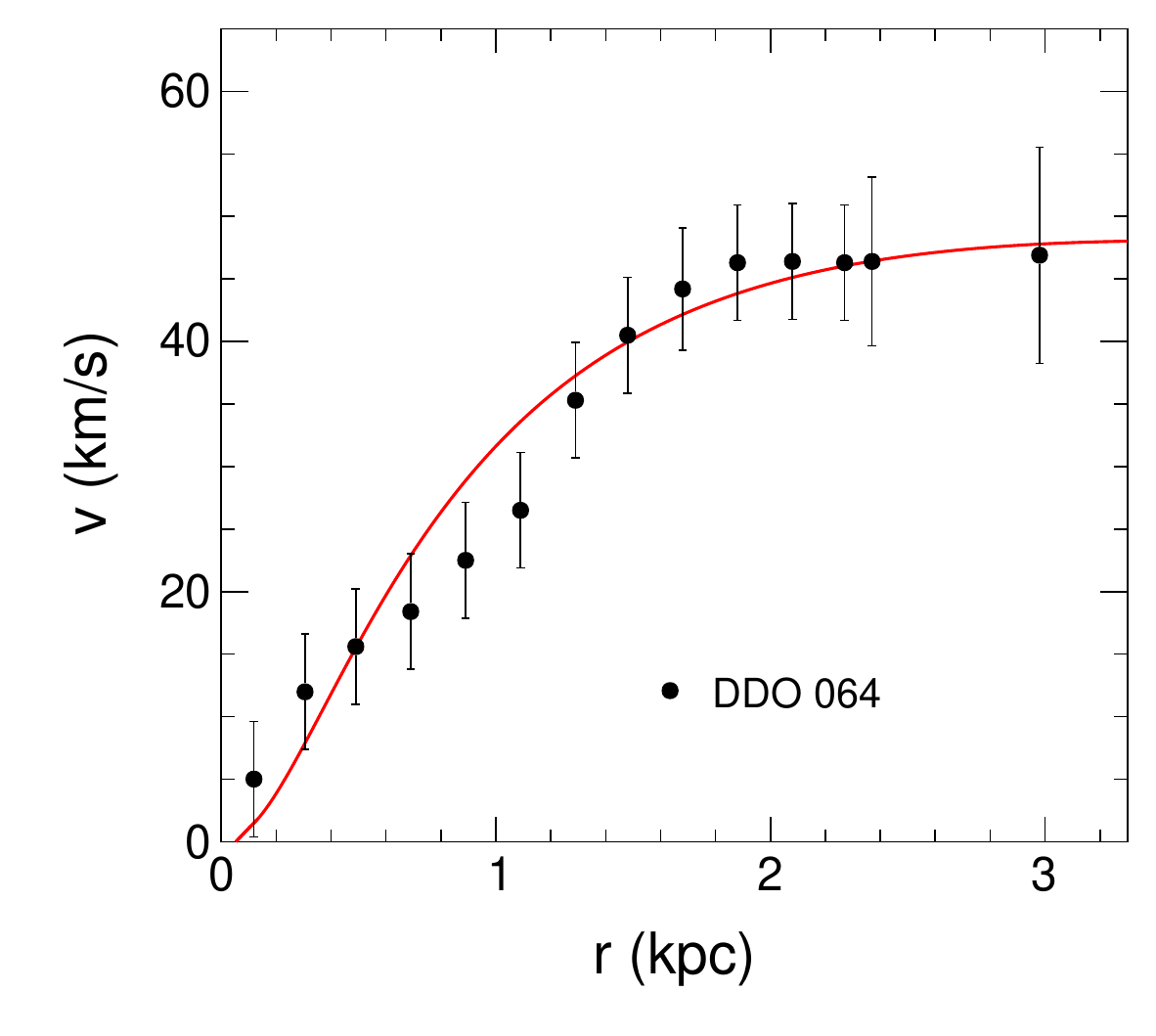}\hspace{0.3cm}
		\includegraphics[scale = 0.26]{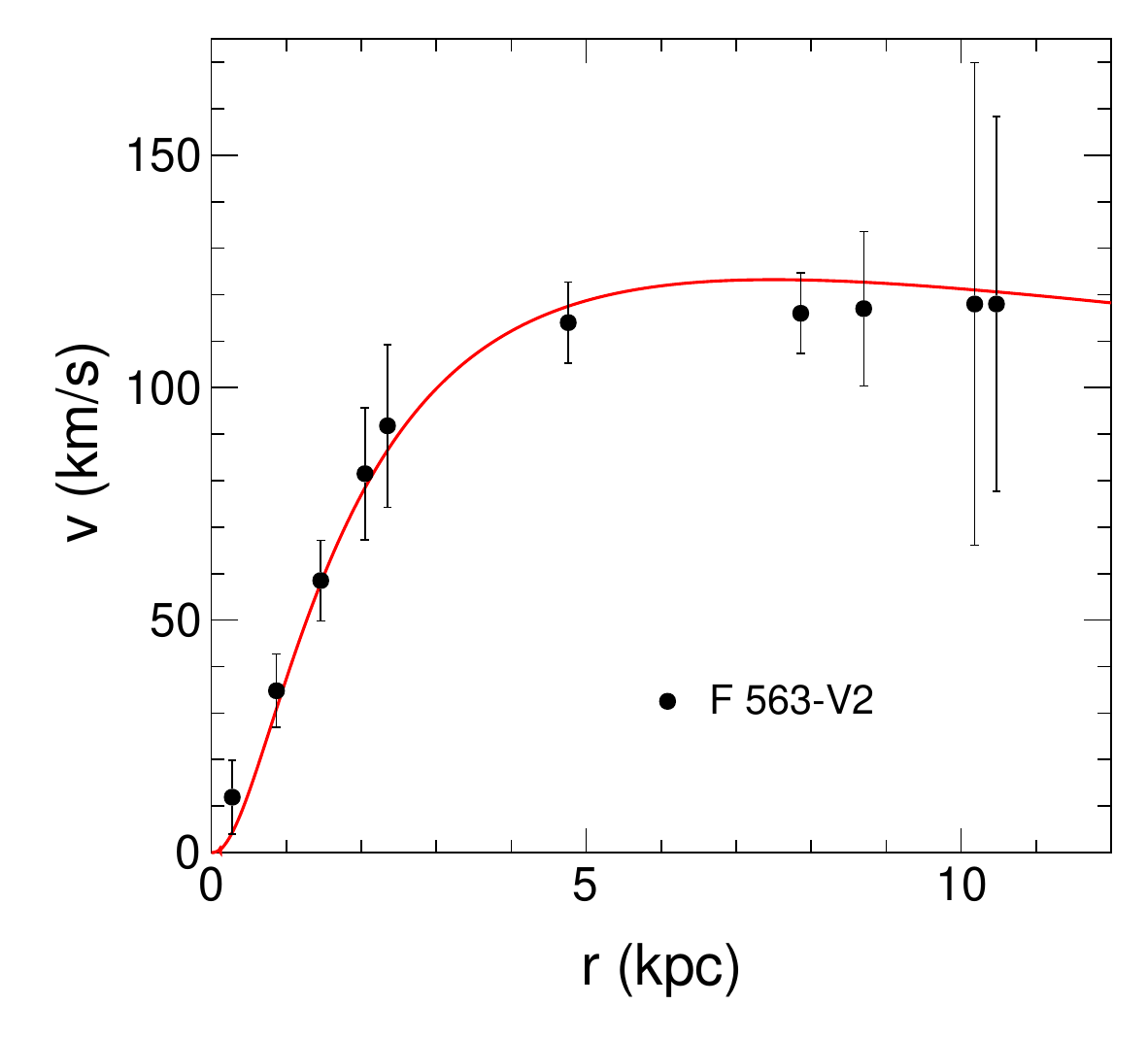}\hspace{0.3cm}
		\includegraphics[scale = 0.26]{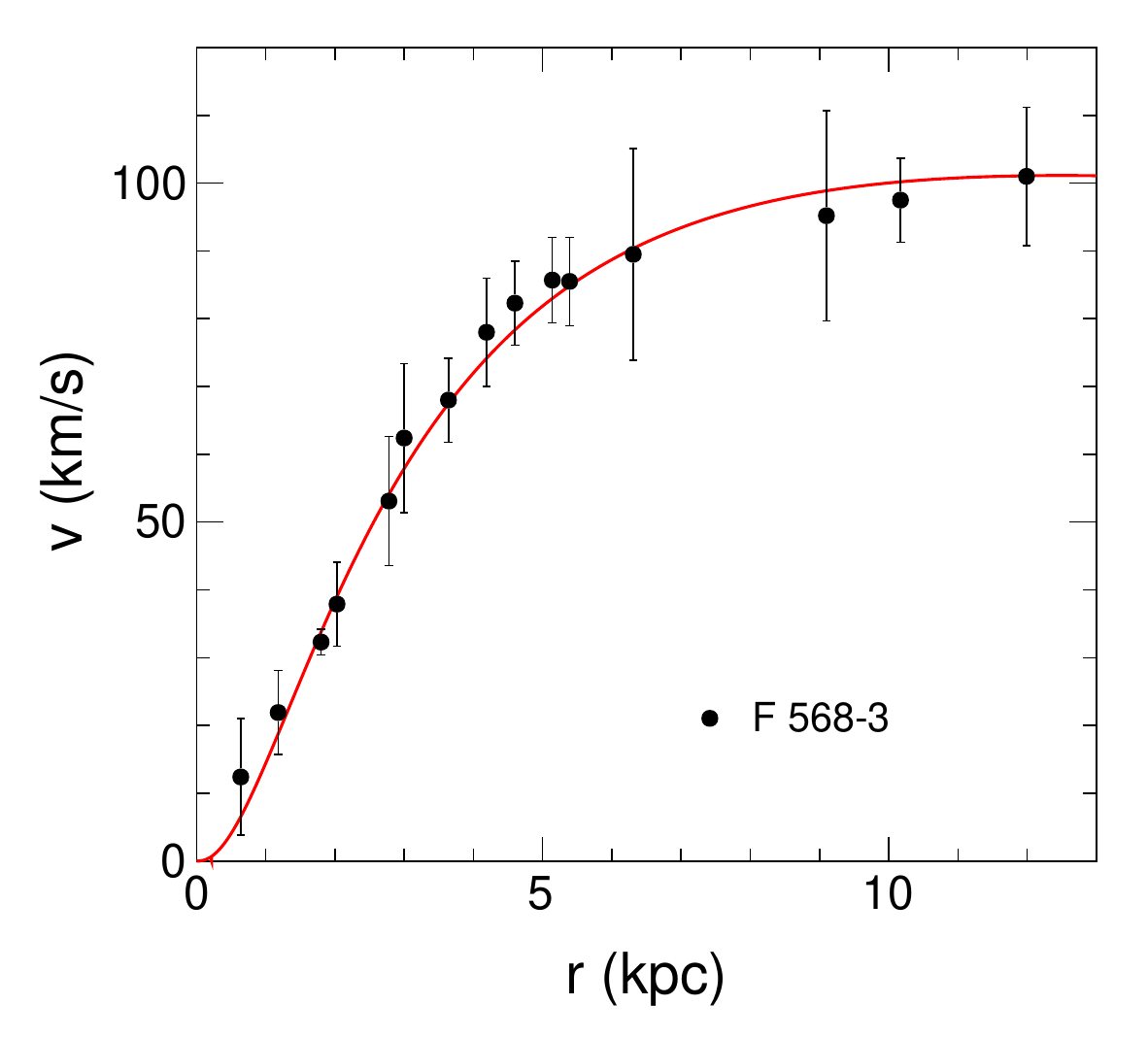}}\vspace{0.2cm}
	\centerline{
		\includegraphics[scale = 0.26]{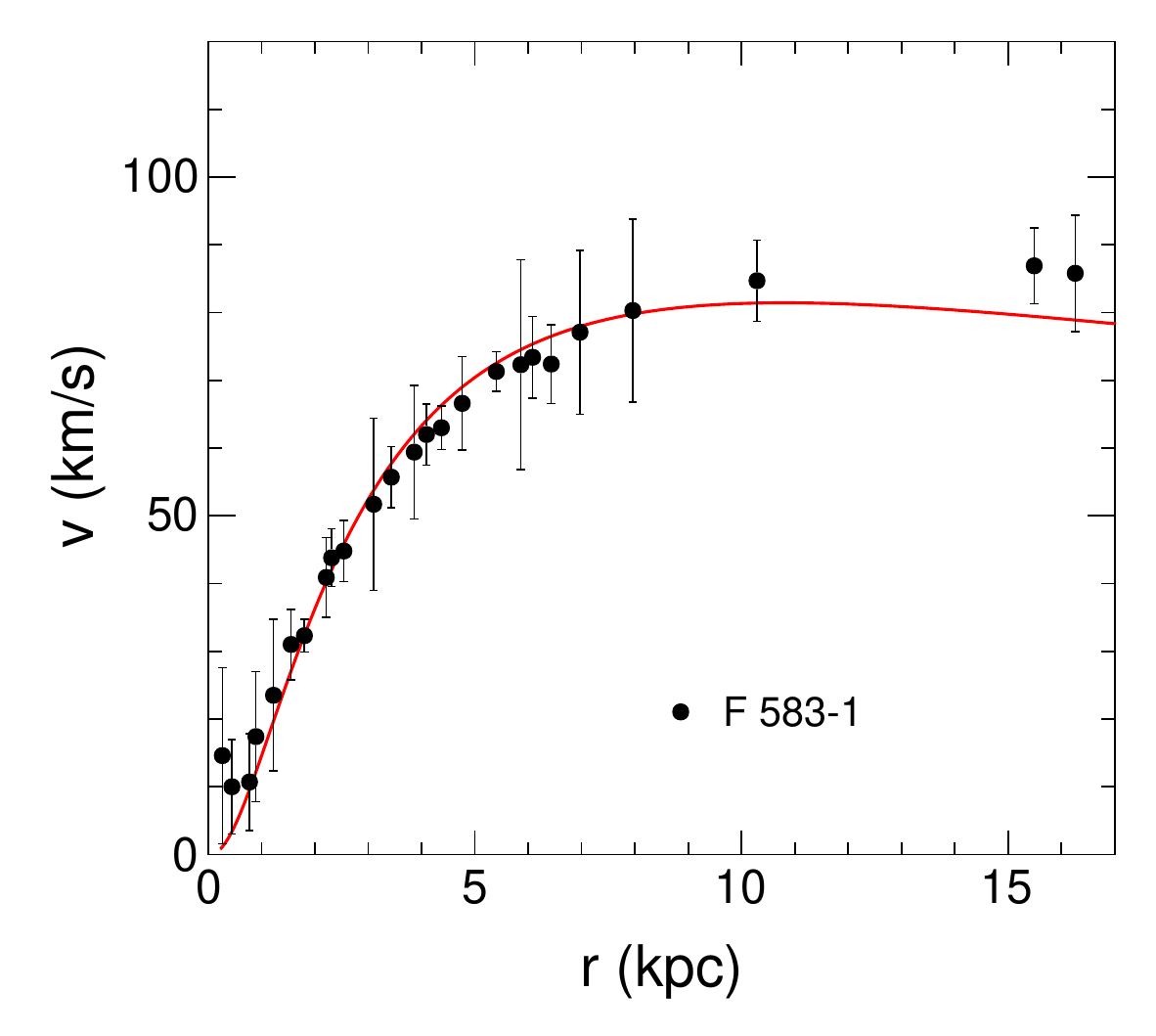}\hspace{0.3cm}
		\includegraphics[scale = 0.26]{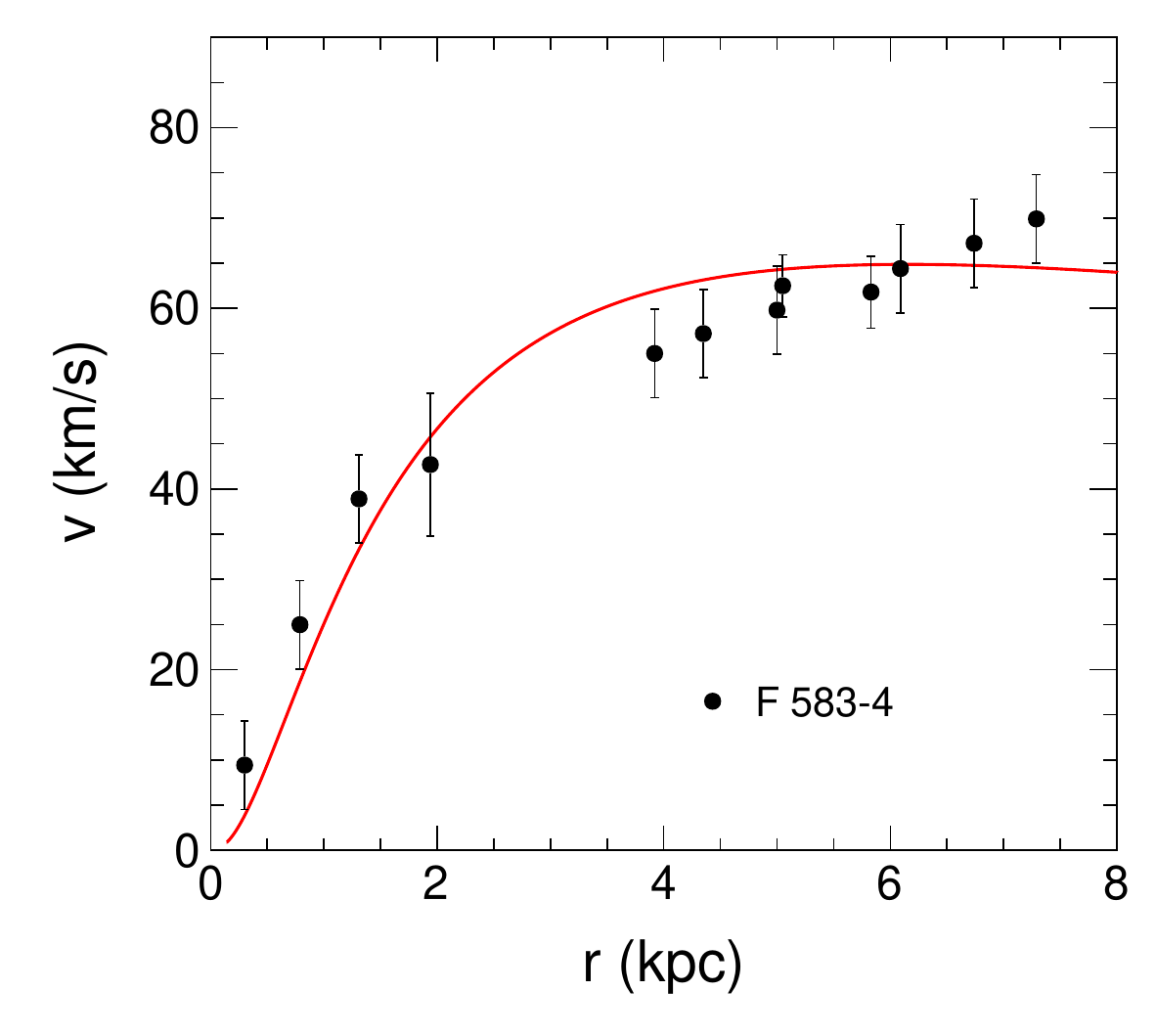}\hspace{0.3cm}
		\includegraphics[scale = 0.26]{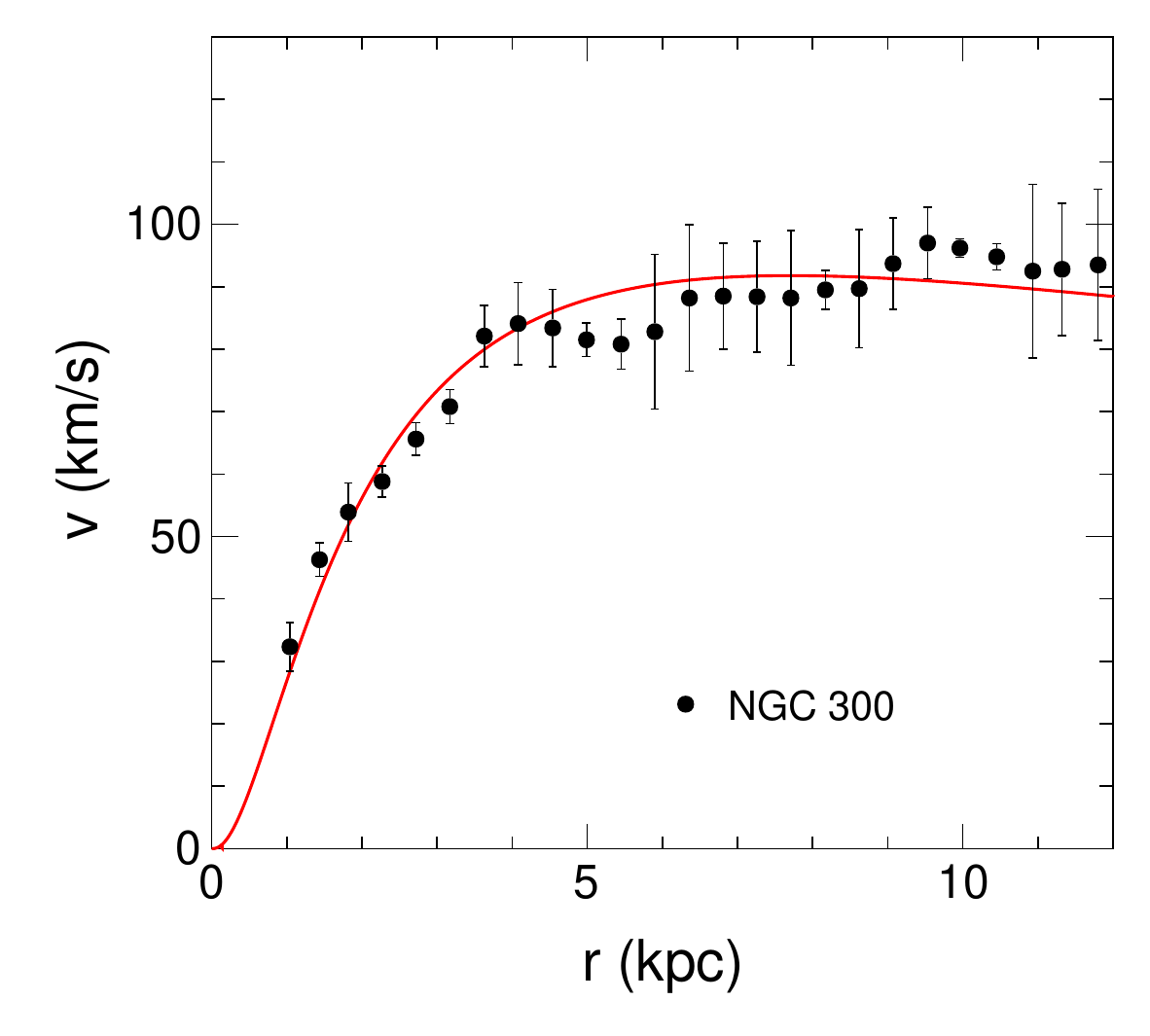}}\vspace{0.2cm}
	\centerline{
		\includegraphics[scale = 0.26]{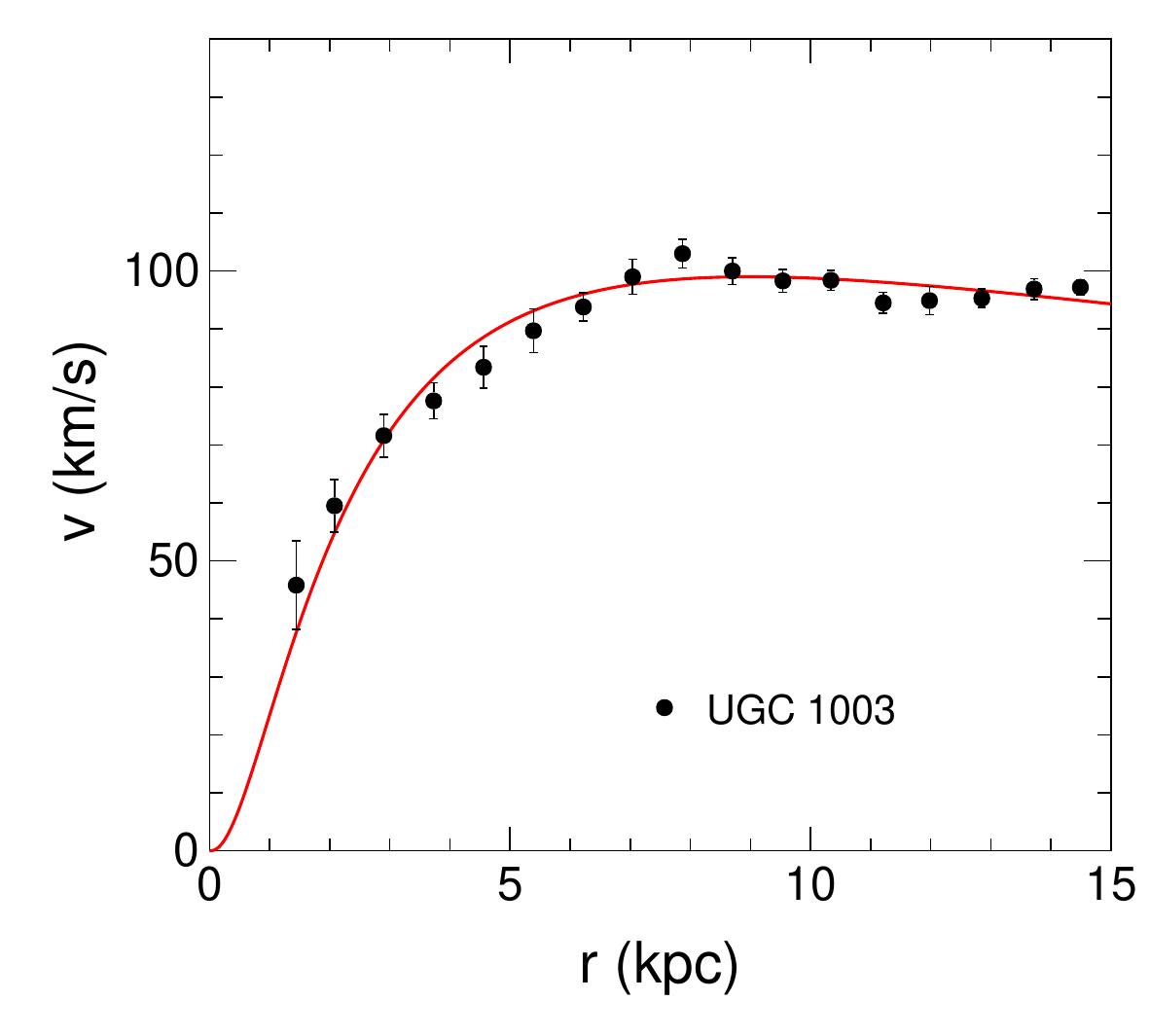}\hspace{0.3cm}
		\includegraphics[scale = 0.26]{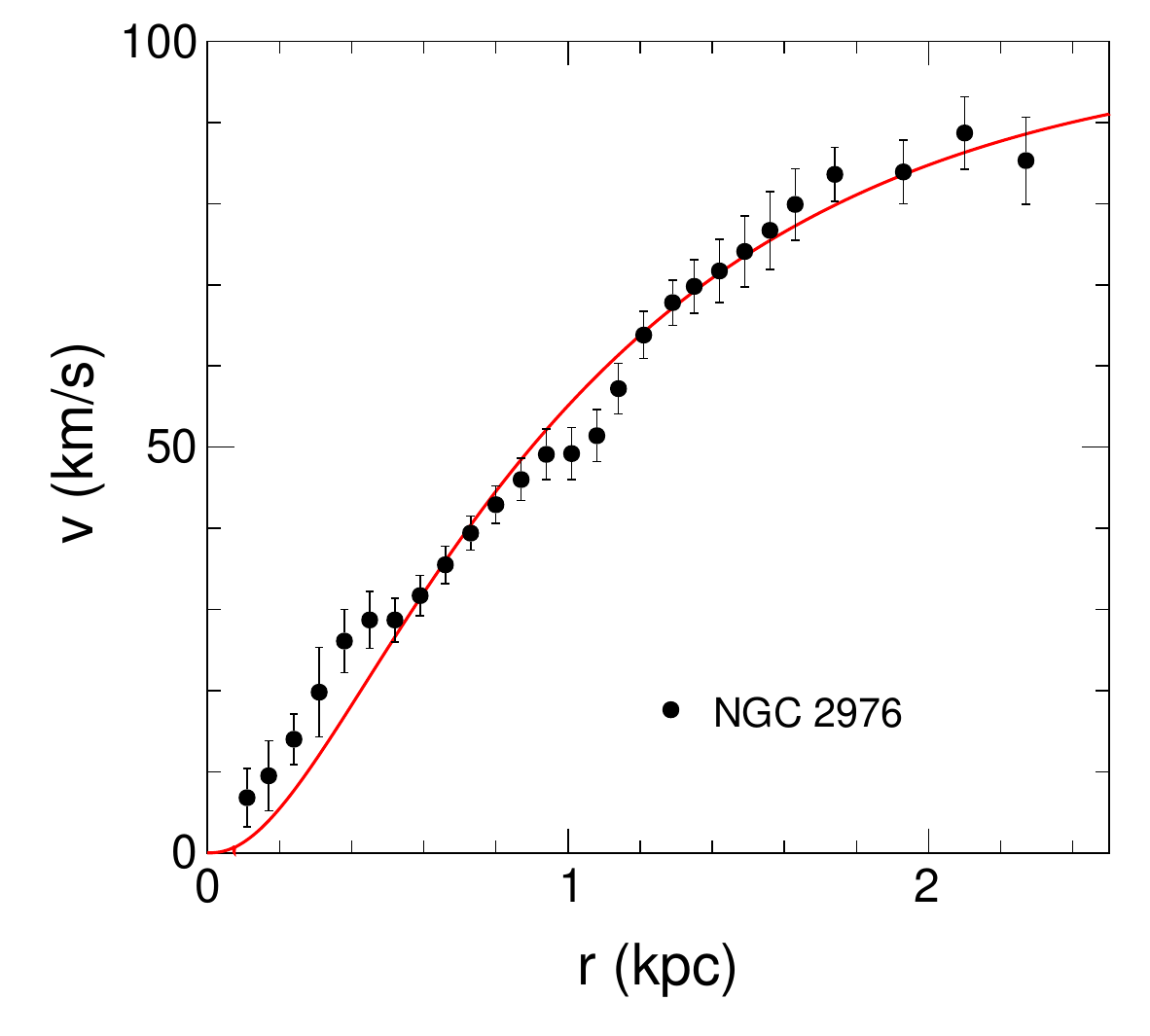}\hspace{0.3cm}
		\includegraphics[scale = 0.26]{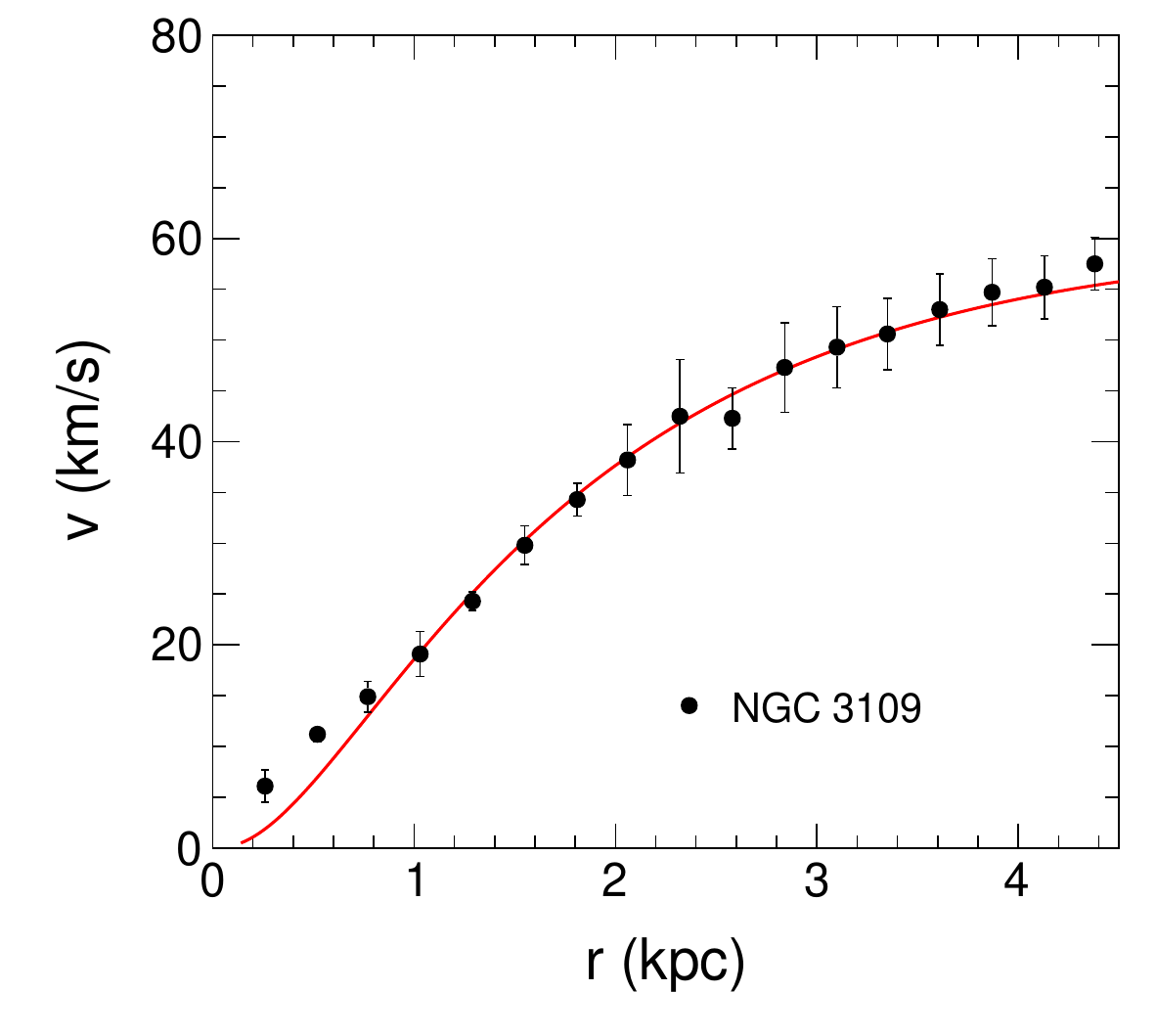}}\vspace{0.2cm}
	\centerline{
		\includegraphics[scale = 0.26]{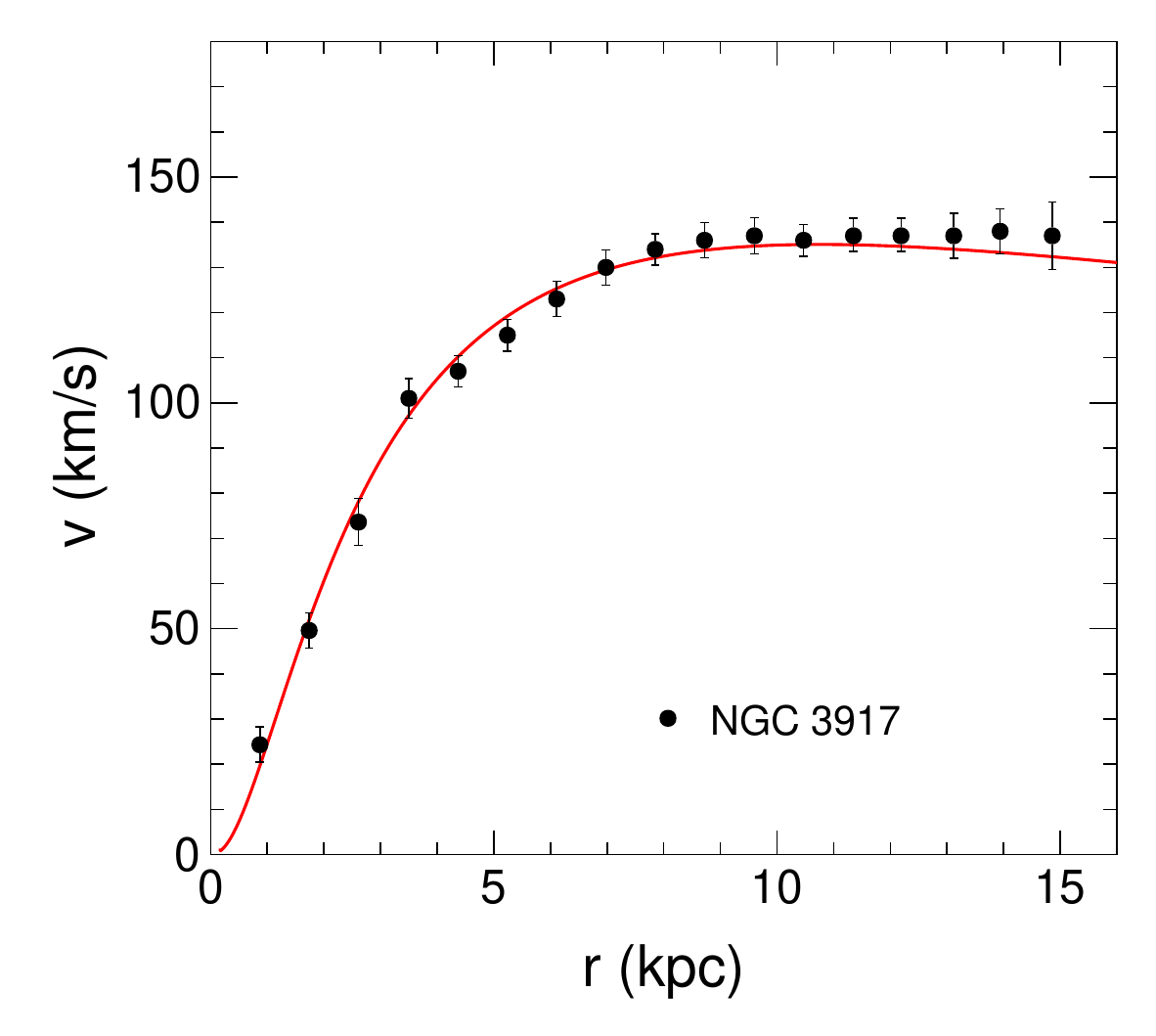}\hspace{0.3cm}
		\includegraphics[scale = 0.26]{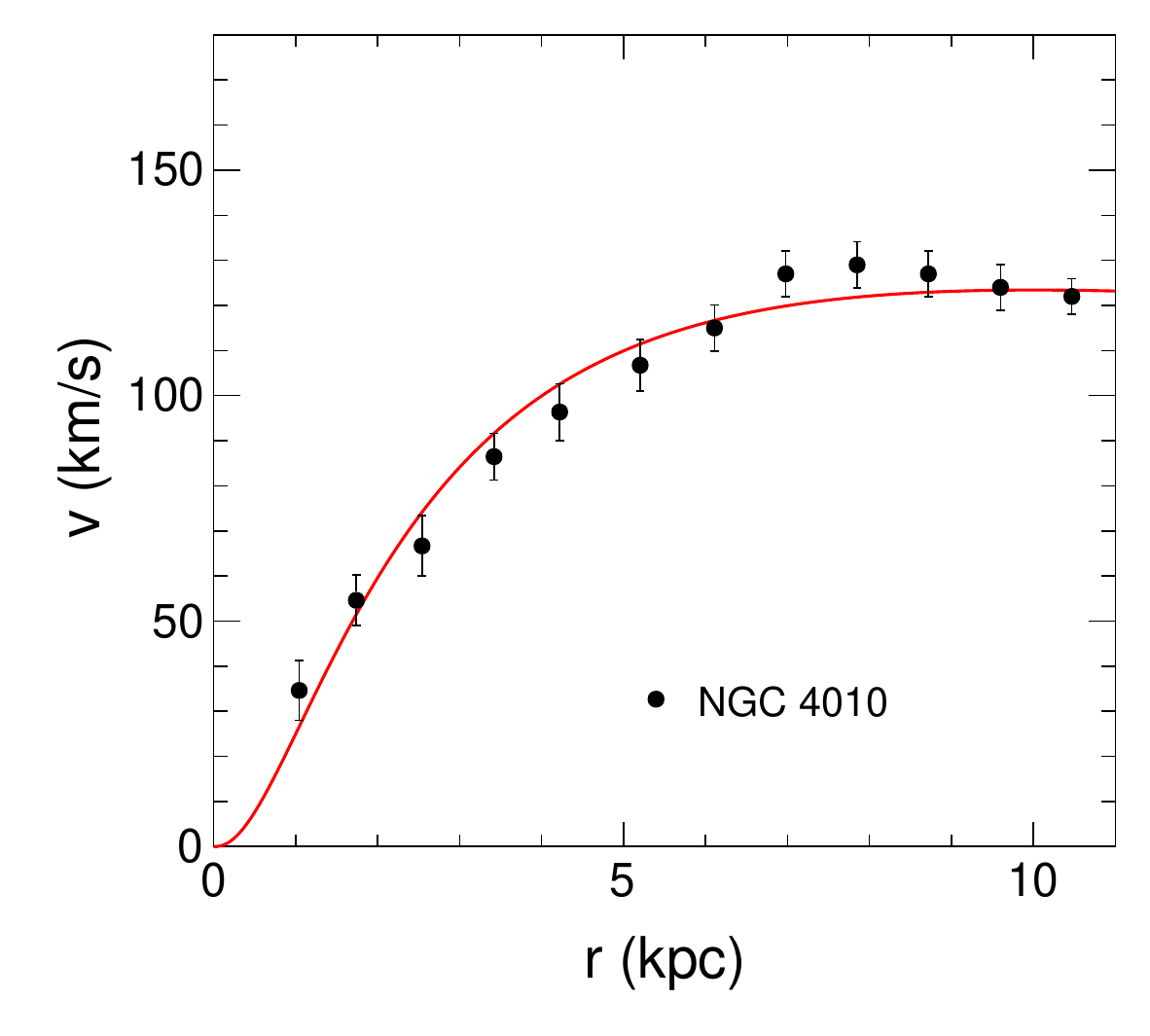}\hspace{0.3cm}
		\includegraphics[scale = 0.26]{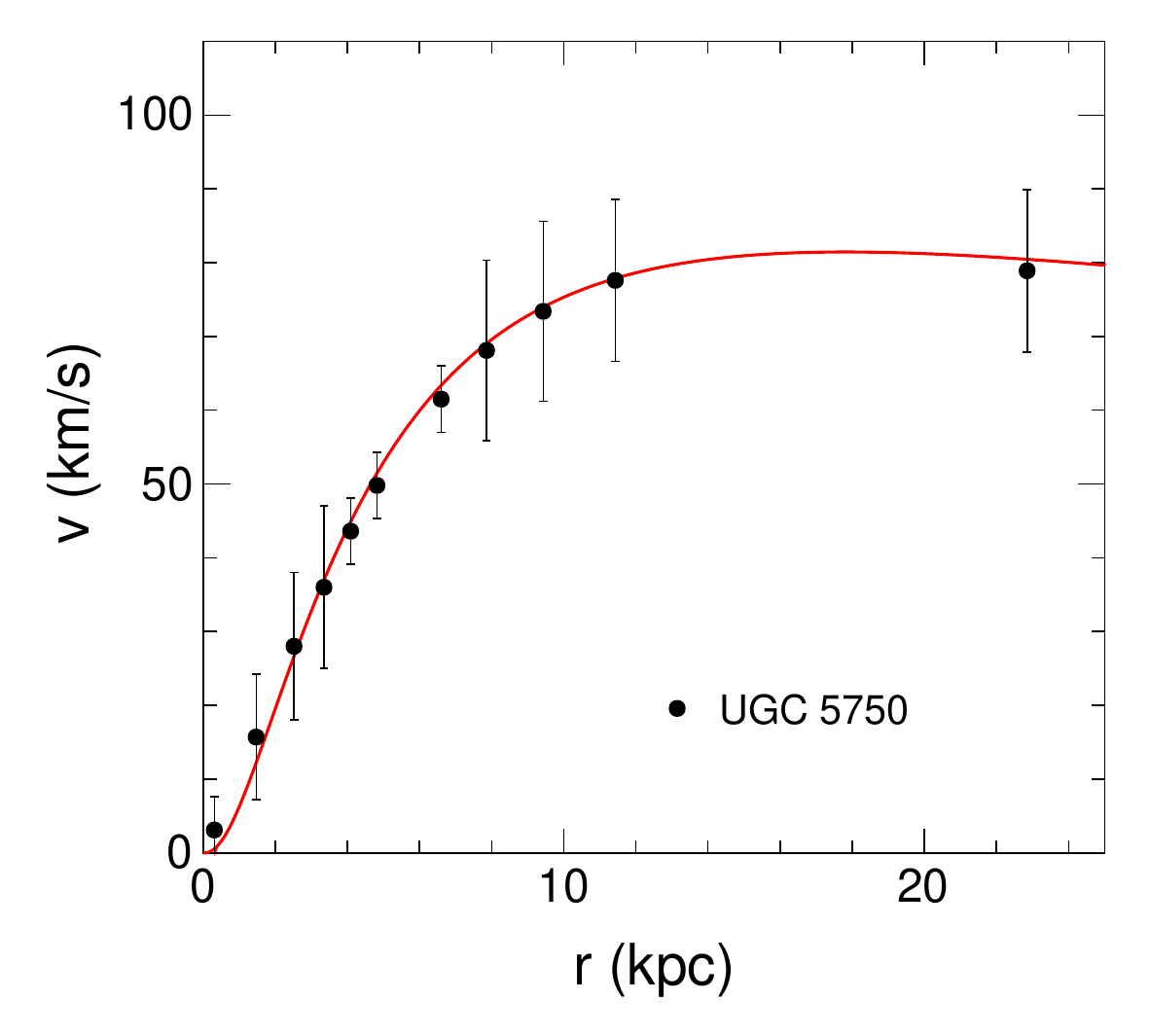}}\vspace{0.2cm}
	\centerline{
		\includegraphics[scale = 0.26]{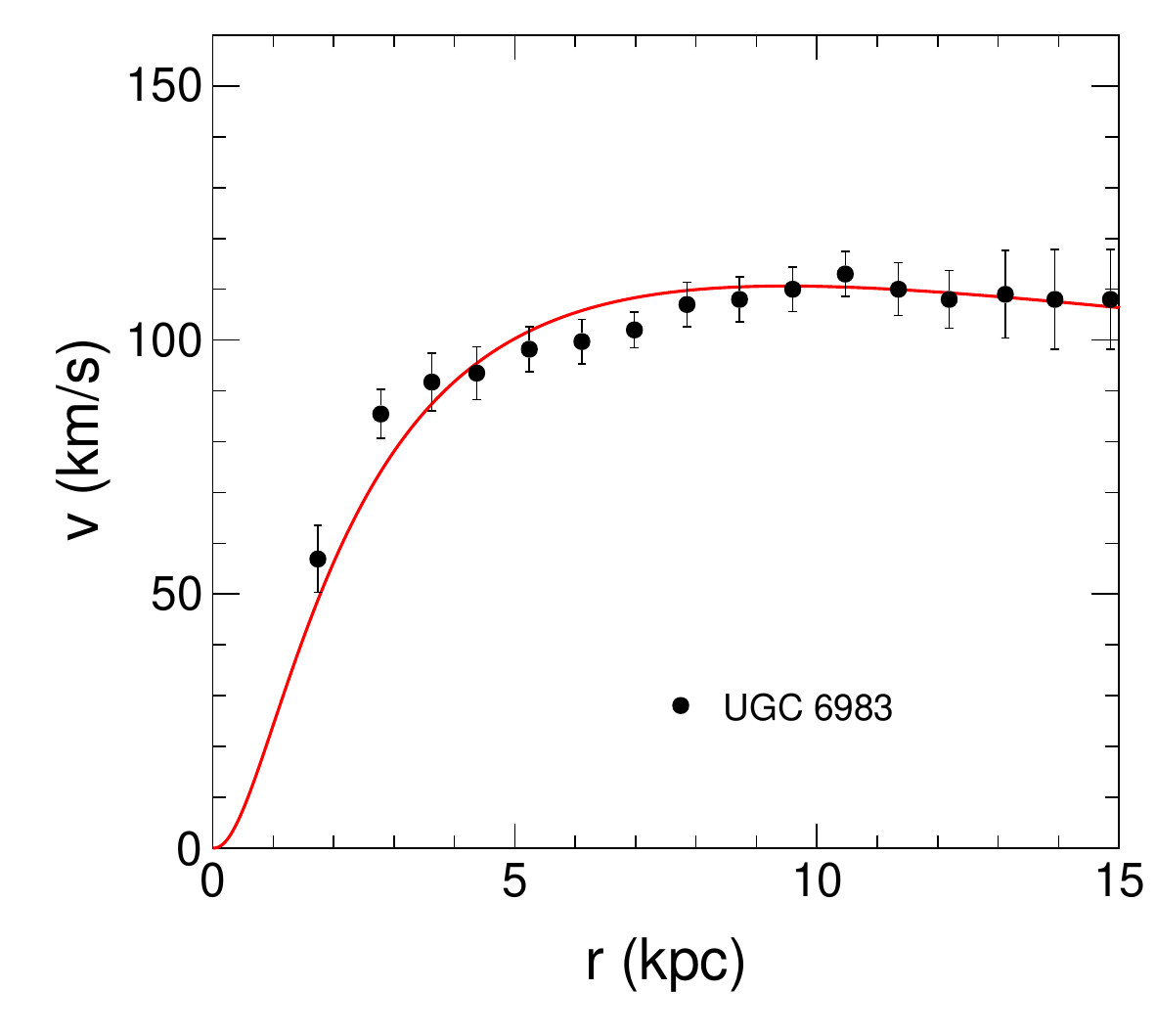}\hspace{0.3cm}
		\includegraphics[scale = 0.26]{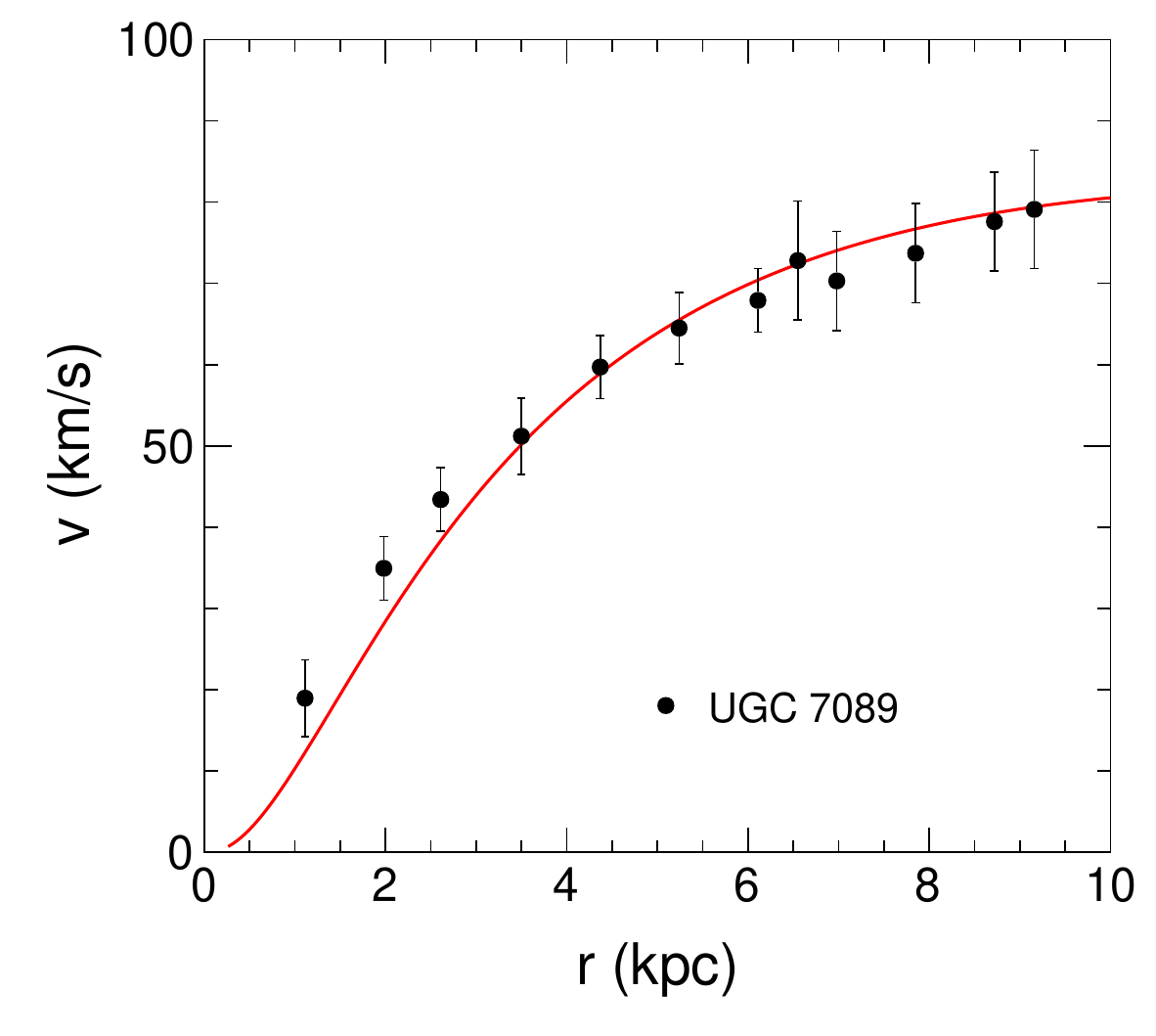}\hspace{0.3cm}
		\includegraphics[scale = 0.26]{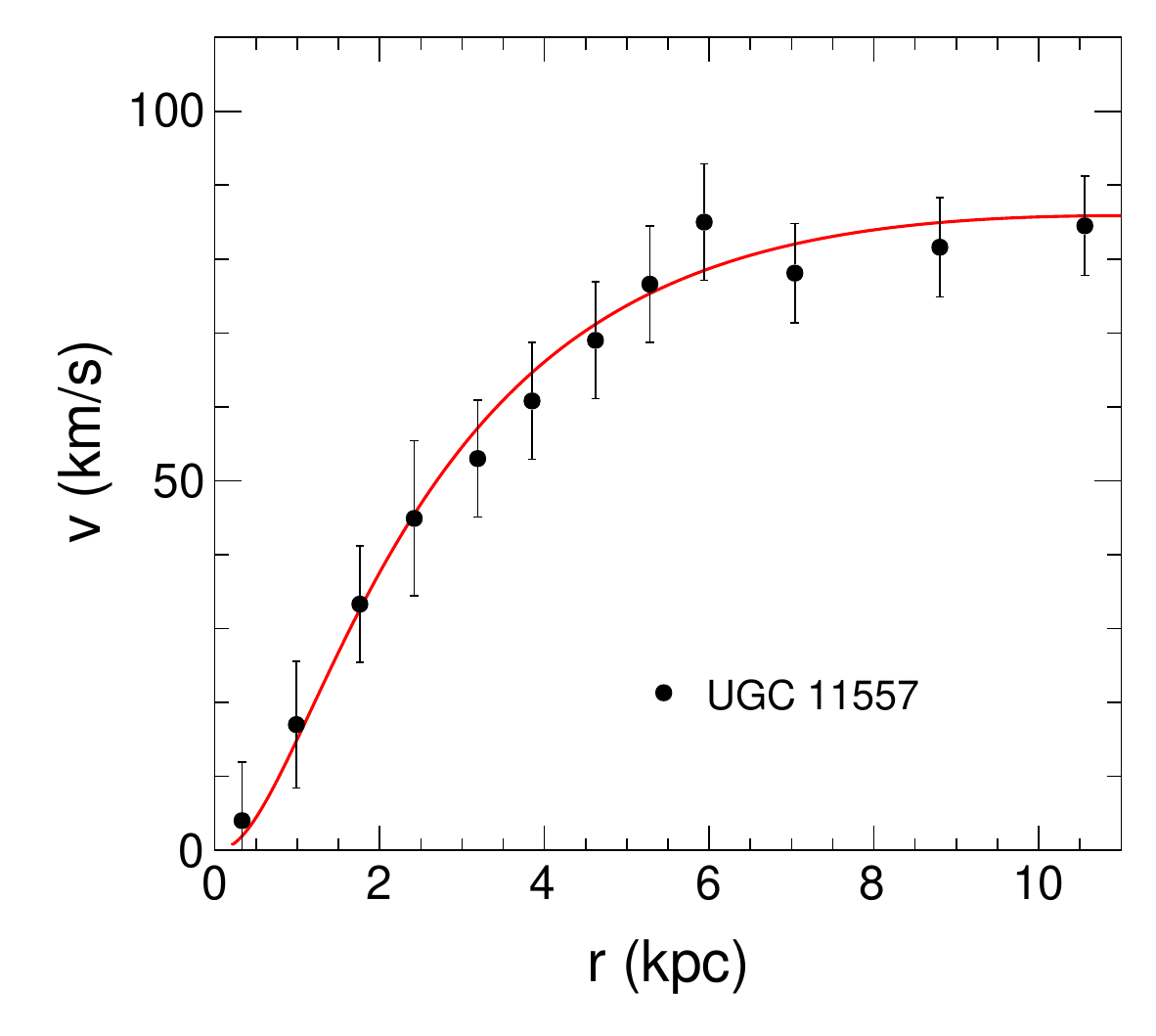}}\vspace{-0.2cm}
	\caption{Fitting of rotation curves generated from the $f(R)$ gravity model 
		\eqref{eq33} for rotation velocities of a sample of 15 LSB \\galaxies with their 
		quoted errors. The data points are observational values of rotational 
		velocities extracted from Ref.~\cite{2016_Lelli}.}
	\label{fig7}
\end{figure}

\subsection{Fitting of rotation curves and estimation of parameters}
The rotational velocity of a particle rotating around the galactic center
have been derived in the last subsection based on a recently introduced new
$f(R)$ gravity model \eqref{eq33}. We generate rotation curves for the test 
particle according to equation \eqref{eq66} and intend to examine the 
influence of scalaron on rotational profiles to explain galactic dynamics in 
the absence of dark matter by fitting the theoretically 
predicted curves with the observational data of a sample of 37 galaxies 
including fifteen LSB, ten HSB and twelve dwarf galaxies~\cite{2016_Lelli}.

\begin{figure}[!h]
	\centerline{
		\includegraphics[scale = 0.26]{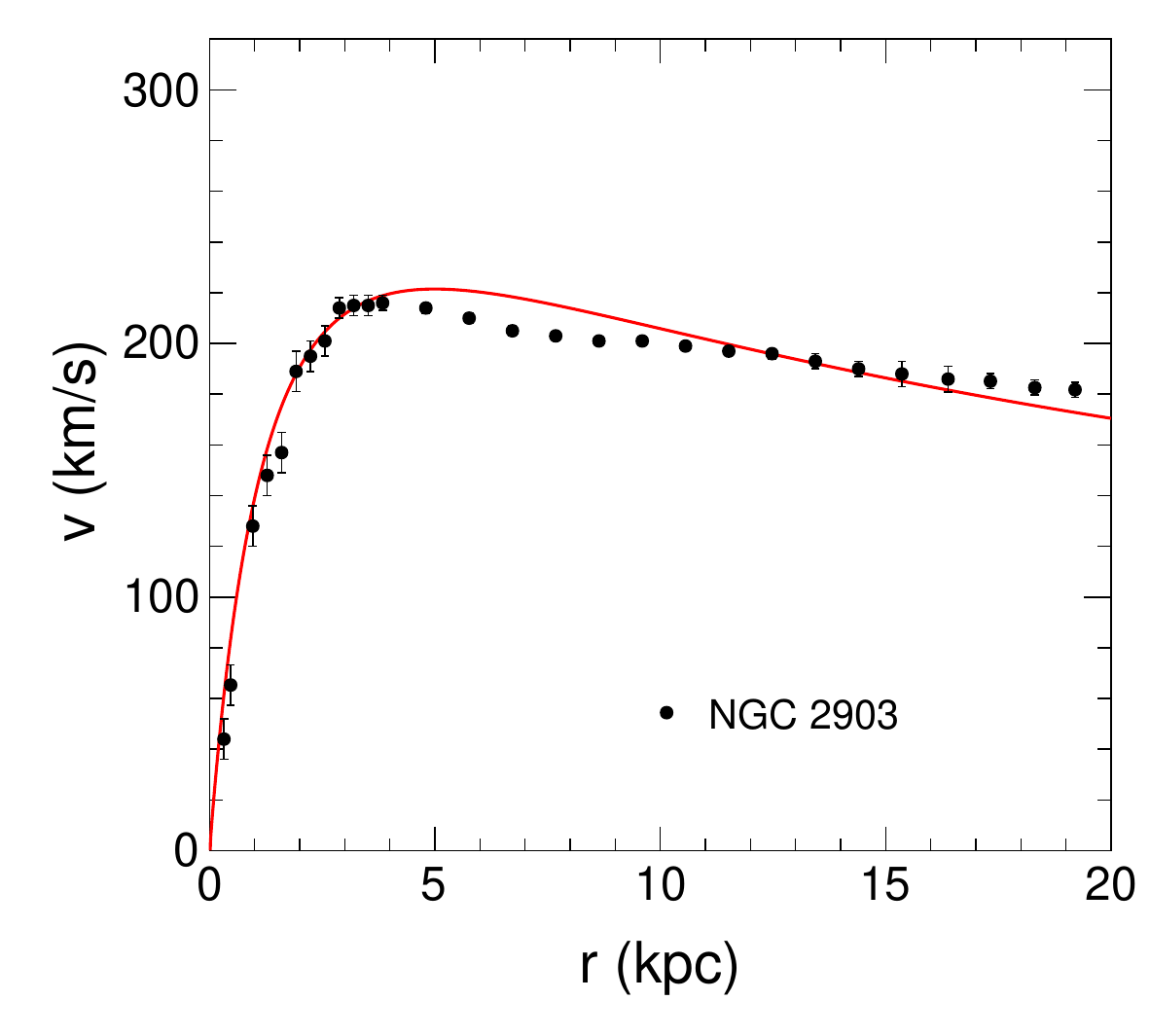}\hspace{0.3cm}
		\includegraphics[scale = 0.26]{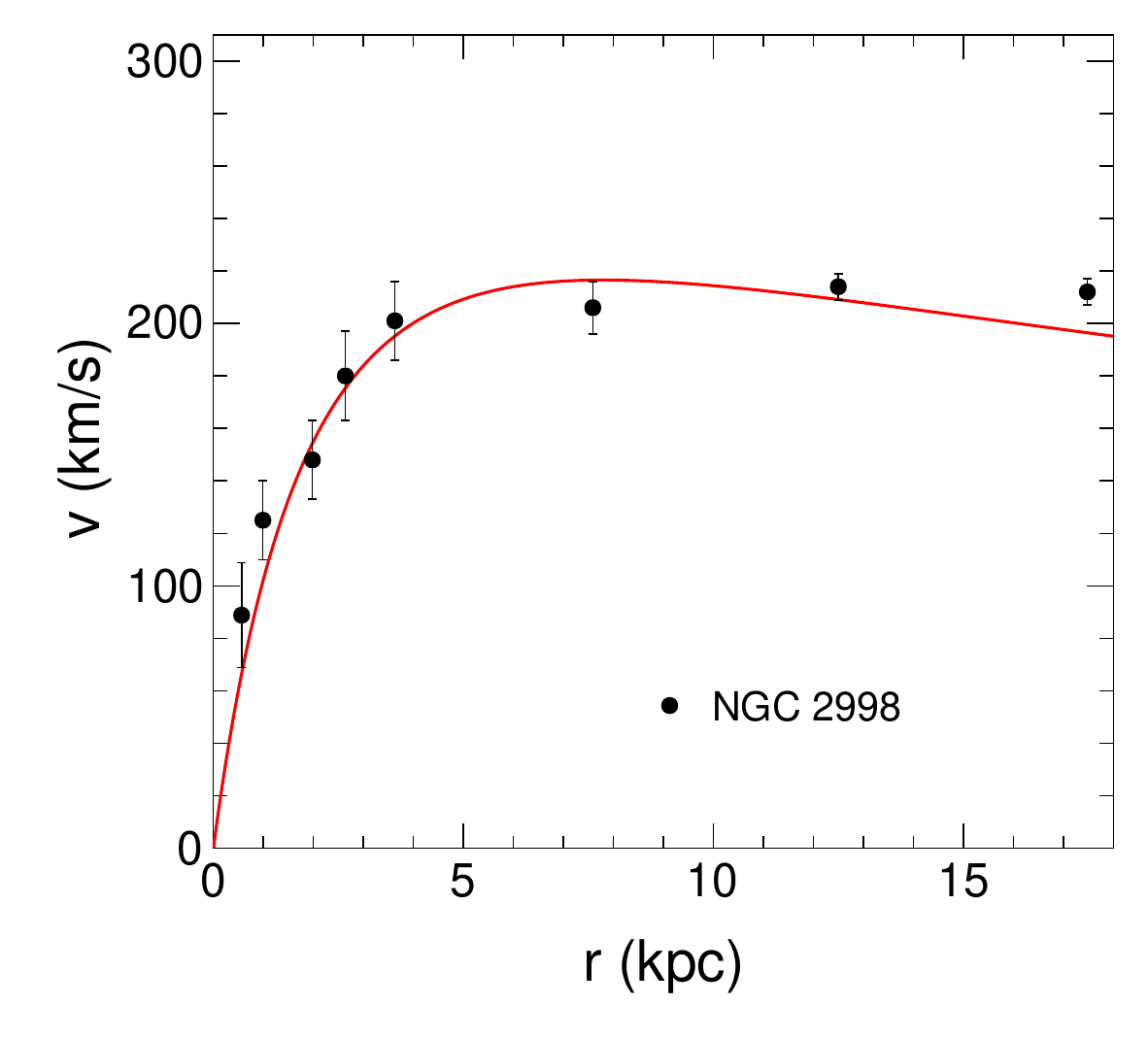}\hspace{0.3cm}
		\includegraphics[scale = 0.26]{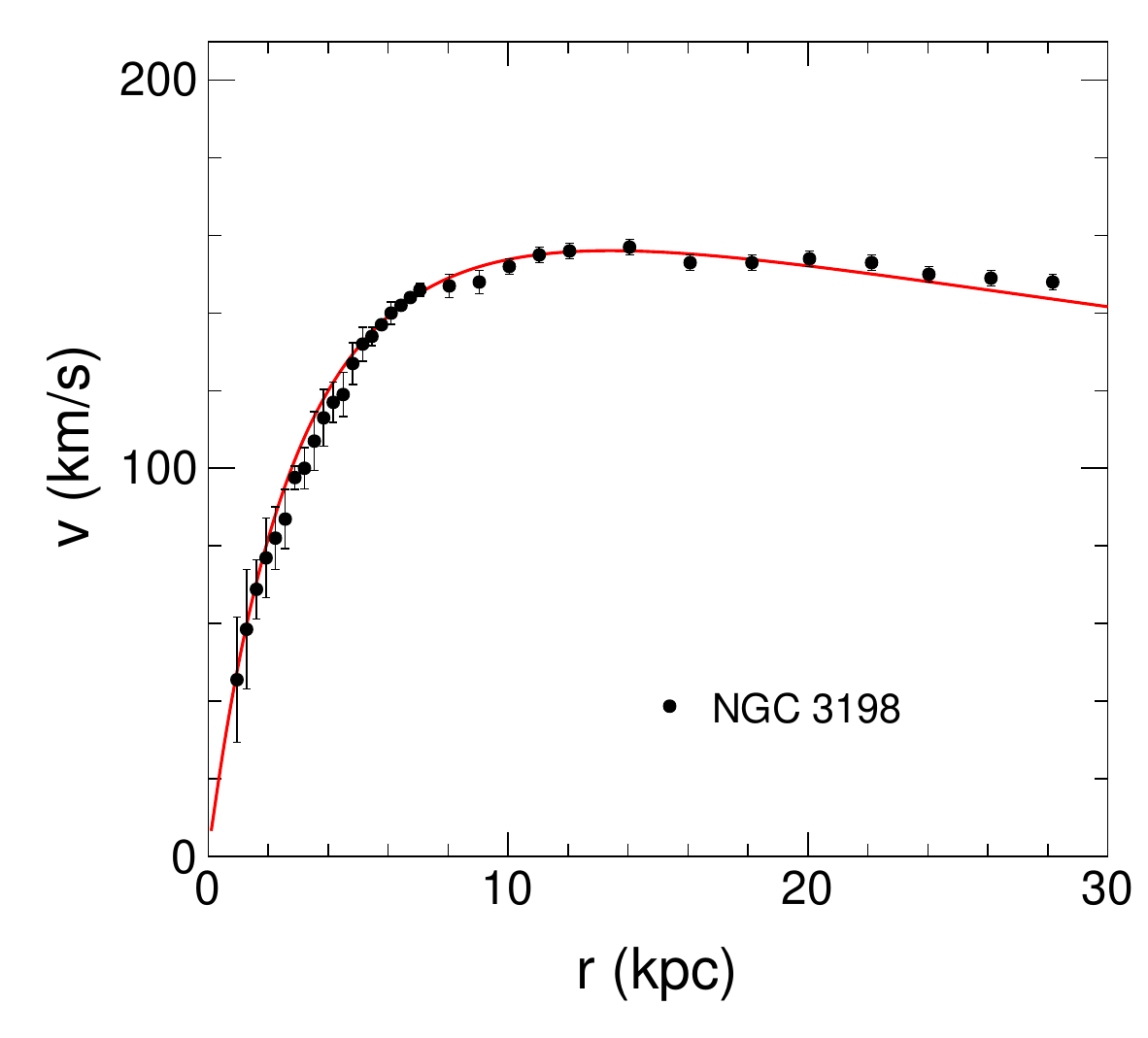}}\vspace{0.2cm}
	\centerline{
		\includegraphics[scale = 0.26]{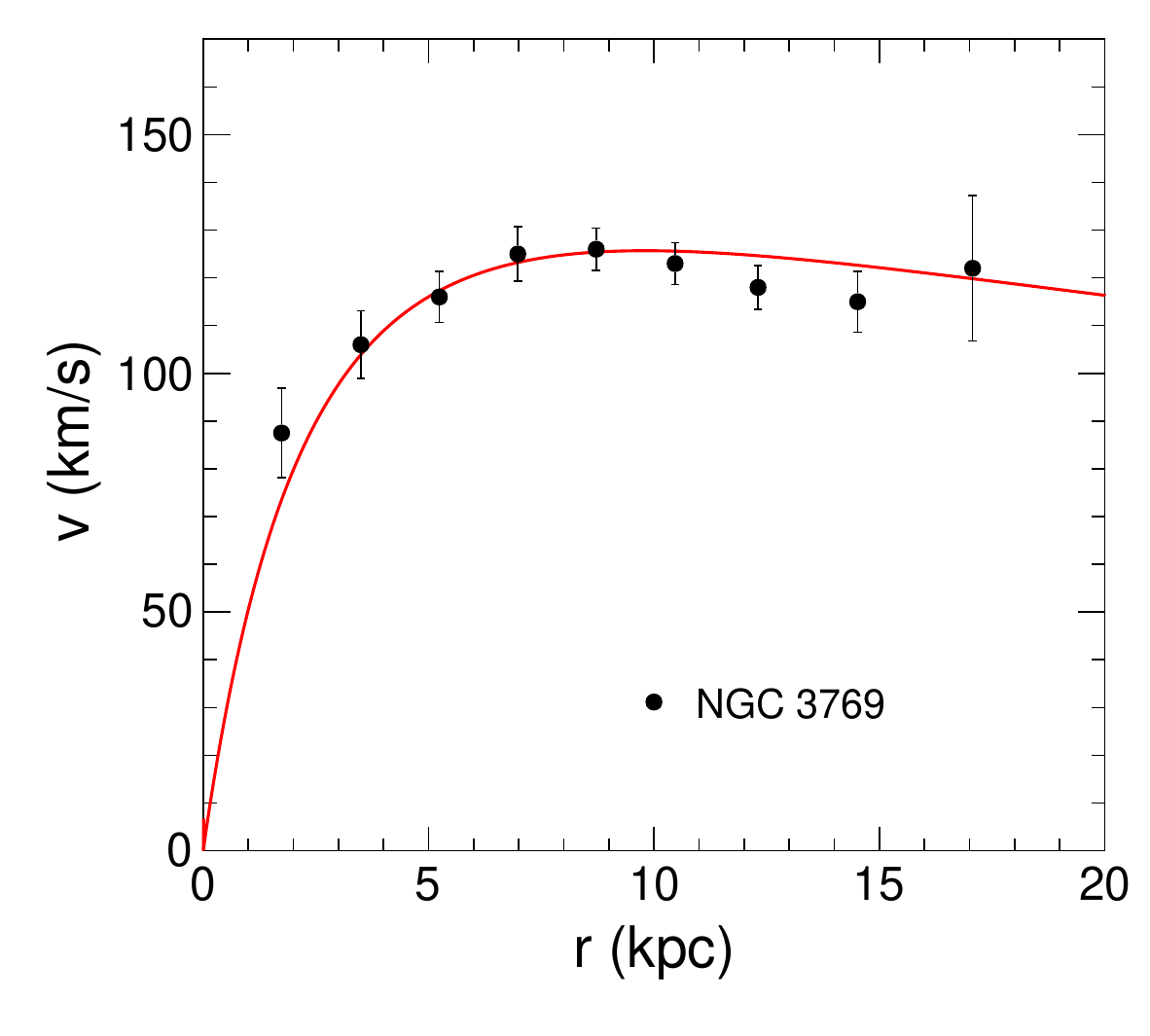}\hspace{0.3cm}
		\includegraphics[scale = 0.26]{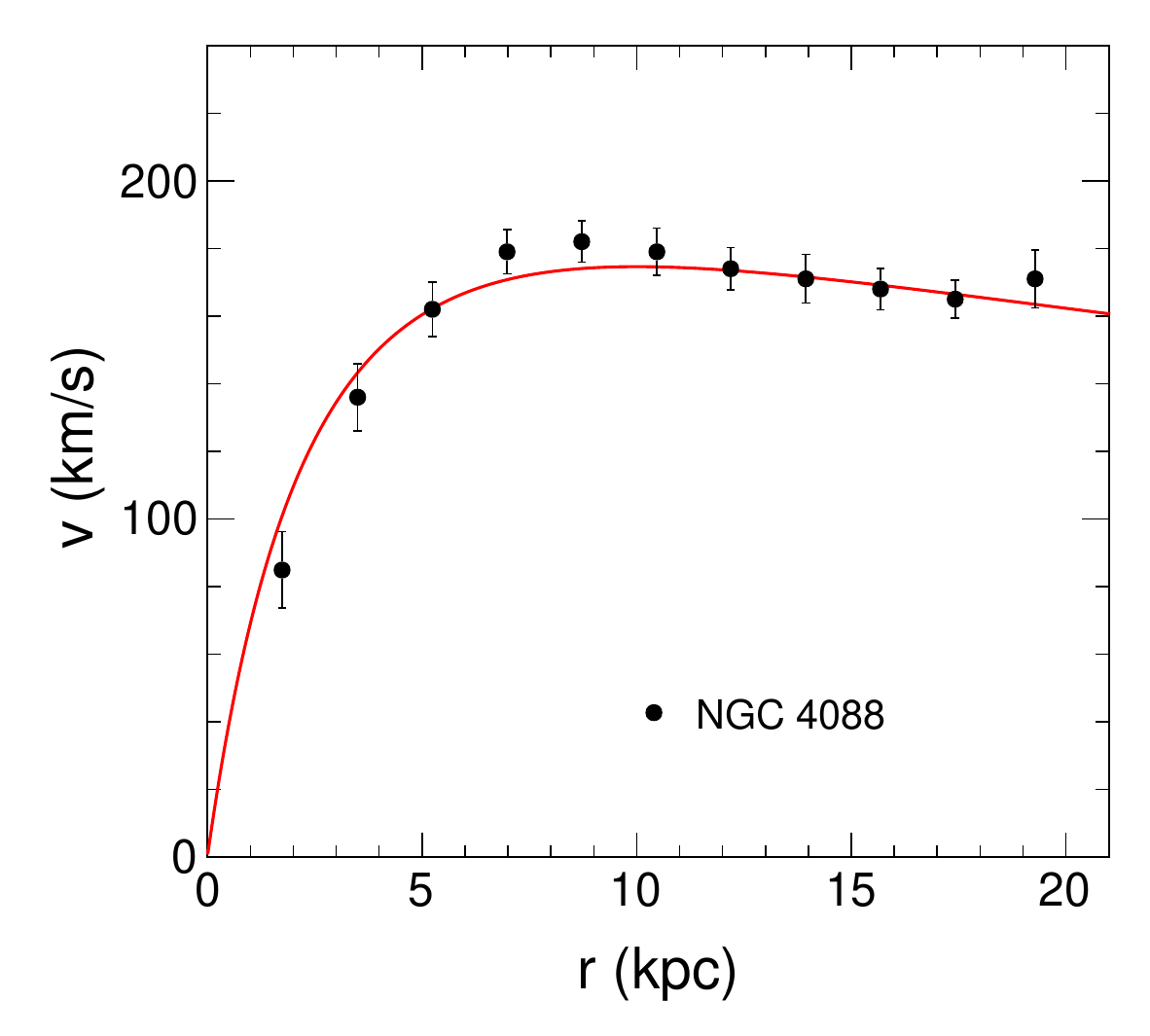}\hspace{0.3cm}
		\includegraphics[scale = 0.26]{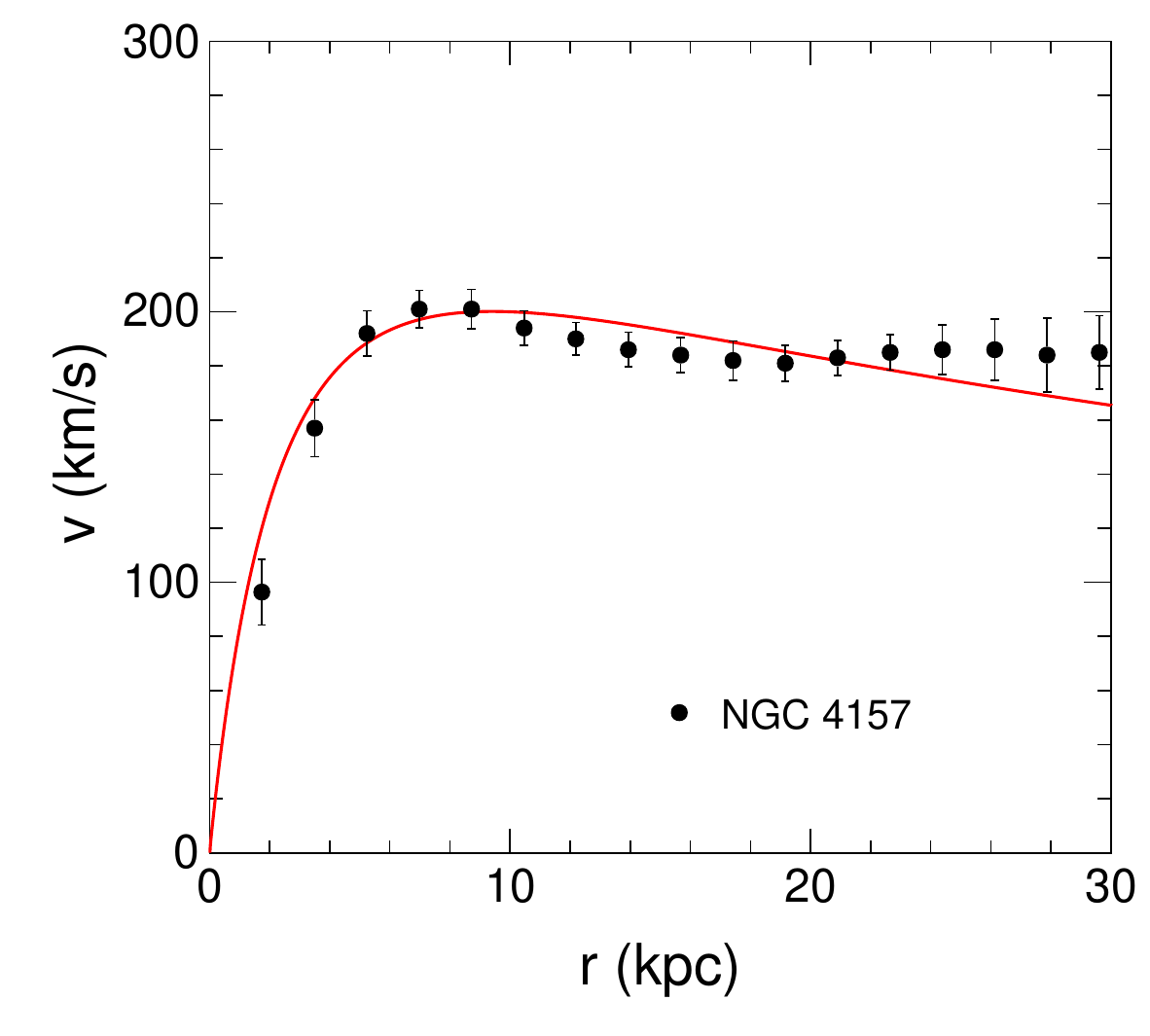}}\vspace{0.2cm}
	\centerline{
		\includegraphics[scale = 0.26]{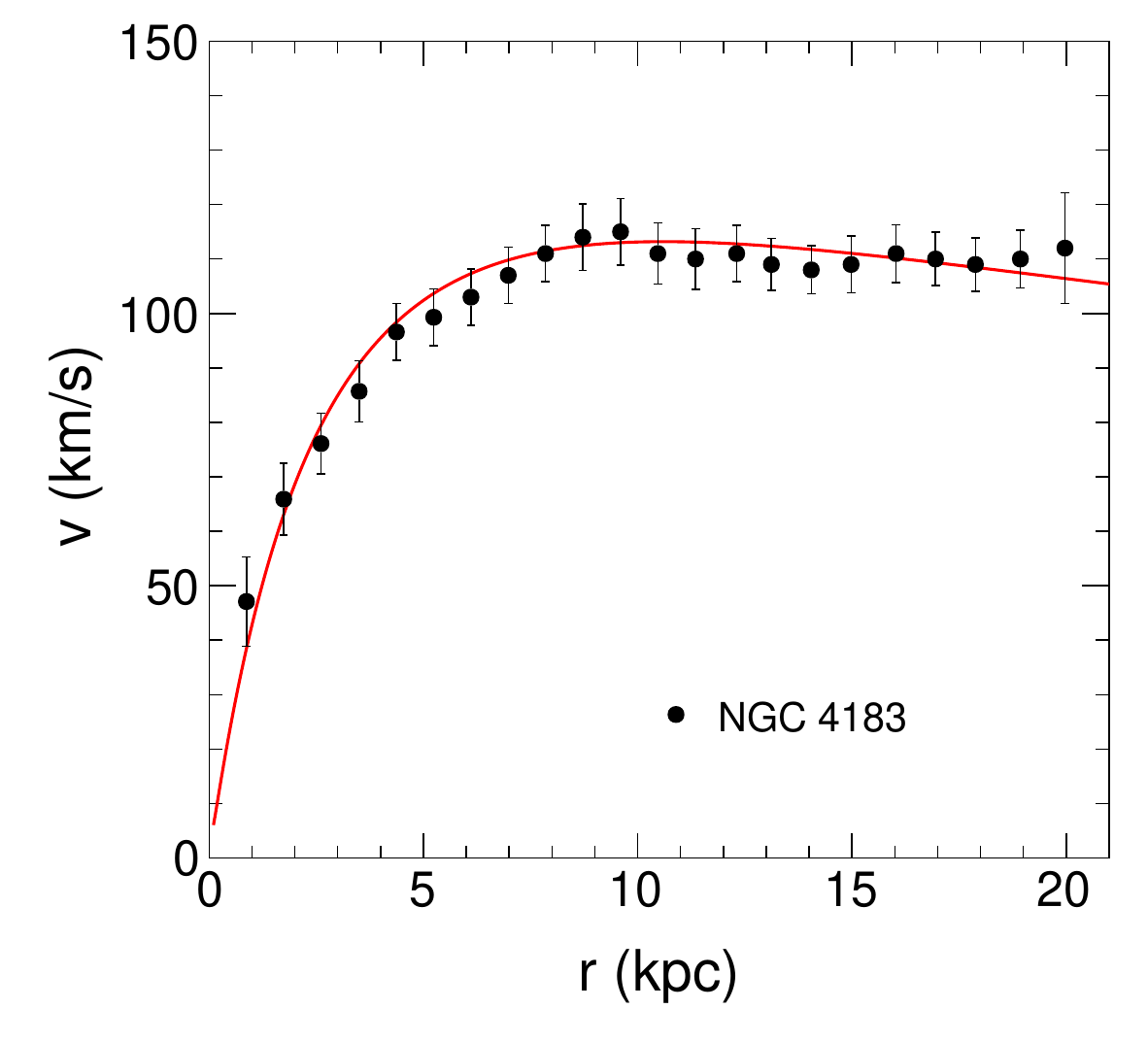}\hspace{0.3cm}
		\includegraphics[scale = 0.26]{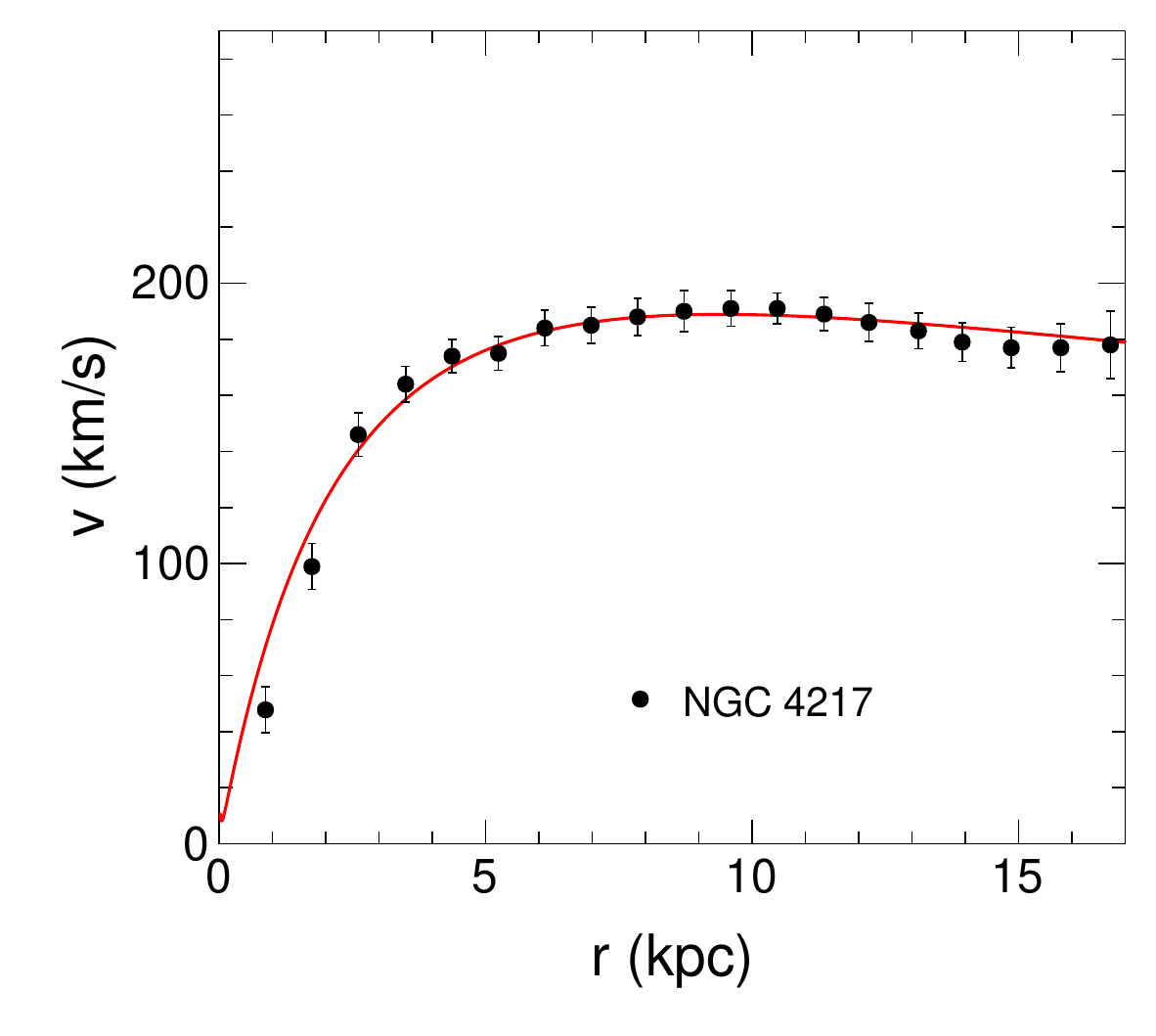}\hspace{0.3cm}
		\includegraphics[scale = 0.26]{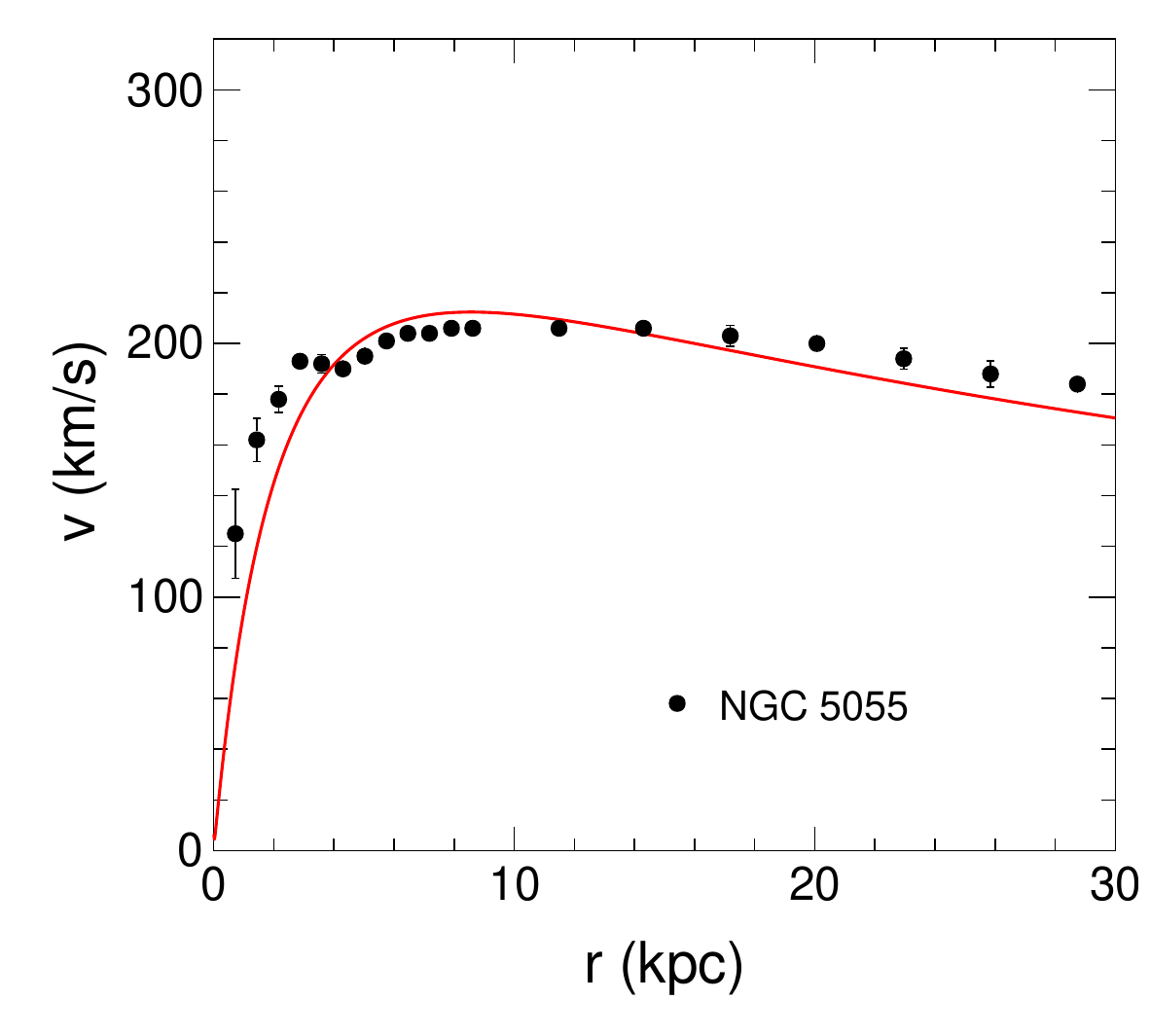}}\vspace{0.2cm}
	\centerline{
		\includegraphics[scale = 0.26]{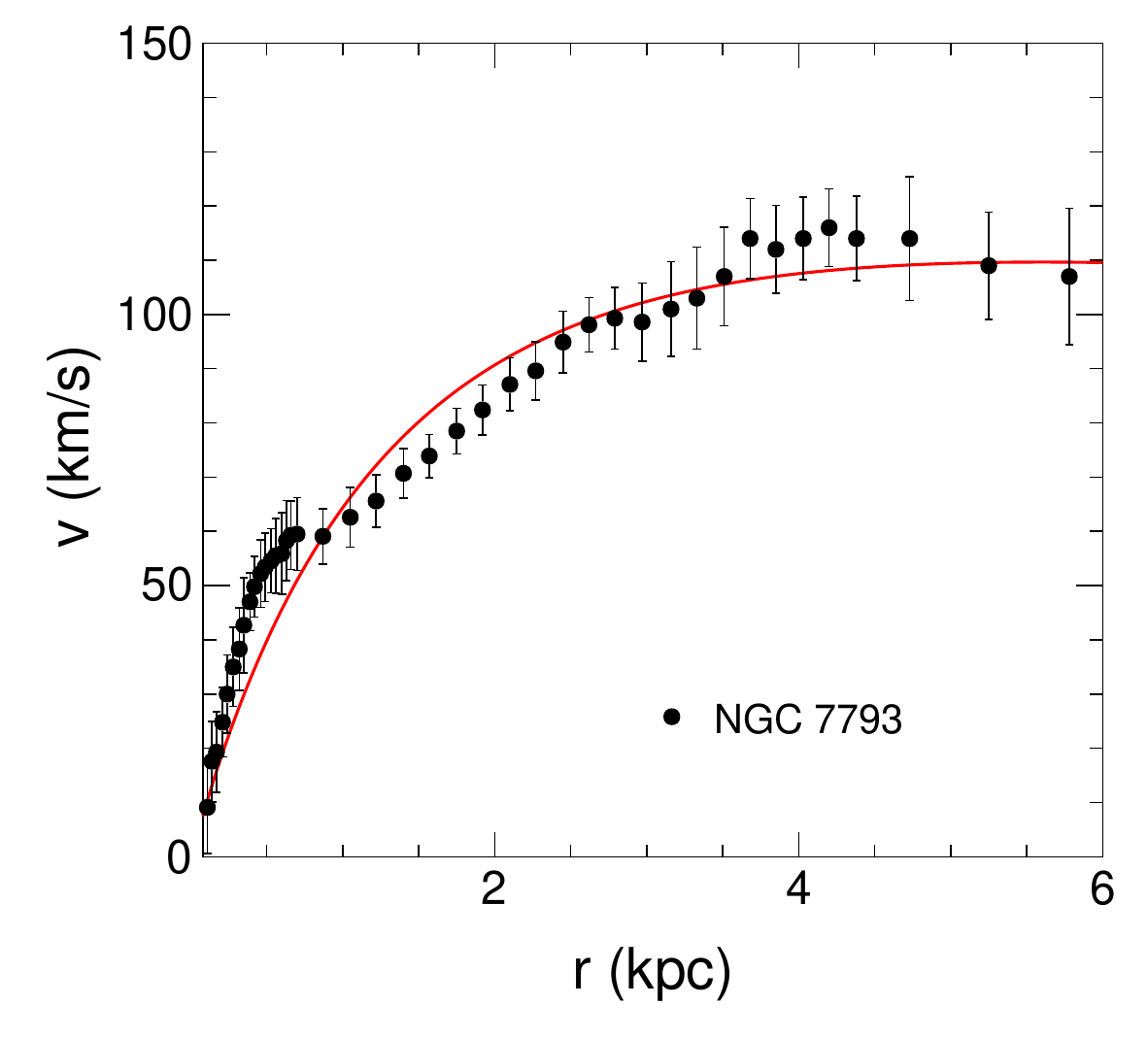}}\vspace{-0.2cm}
	\caption{Fitting of rotation curves generated from the $f(R)$ gravity 
		model \eqref{eq33} for rotation velocities of a sample of 10 HSB galaxies with 
		their quoted errors. The data points are observational values of rotational 
		velocities extracted from Ref.~\cite{2016_Lelli}.}
	\label{fig8}
\end{figure}
It is noteworthy that LSB galaxies are slowly evolving objects that have DM 
domination in all radii. They live in less dense, extremely extended DM halos 
and are found to be more isolated relative to HSB galaxies
\cite{1997_Bothun,1997_Blok}. Refs.~\cite{2002_blok,1996_blok,2017_Karukes} 
have analyzed that the rotation curves of LSB and Dwarf galaxies are slowly 
rising, shallower at small radii. Contrary to this, an HSB galaxy has a higher 
central mass density that leads to a steeply rising rotation curve in the inner 
region followed by a relatively flat outer part~\cite{2014_McGaugh}. Taking 
note of these results, rotation curves for all these three types of galaxies 
are produced and fitted by considering a predetermined set of model parameters 
$\alpha=0.0005$ and $R/R_c=2$ from Ref.~\cite{2020_dhruba} for all the 
samples. Also, another model parameter $\beta$ is set to $0.0001$ for each 
galaxy, and fitting of theoretical curves is carried out by constraining 
parameters $M_0$ and $r_c$ using the best-fitted values for each sample 
galaxy. As already mentioned the parameter $b$ determines the 
slope of the mass profile \eqref{eq67}. Higher values of $b$ produce a 
more slowly rising mass profile, whereas lower values yield a steeply rising 
mass profile leading respectively gradually and steeply rising rotation 
curves of galaxies. These two characteristics are indicative of LSB and Dwarf 
galaxies, and HSB galaxies, respectively. We have analyzed the value of $b$ by 
allowing it to be a free parameter in the fitting of the predicted rotation 
curves defined by equation \eqref{eq66} and observed that $b=1$ provides a good fit for the steeply rising rotation curves of HSB galaxies, whereas 
$b=2$ tends to fit slowly rising shallower at small radii curves of the LSB and 
Dwarf galaxies. Hence, similar to Refs.~\cite{2006_brown,2018_sporea} the 
value of the parameter $b$ is taken as $1$ for HSB and it is $2$ for 
both LSB and Dwarf galaxies. 
After fitting a notable agreement between predictions and actual observations 
is observed for all samples. These perfectly matched rotation curves with 
observational data, extracted from Ref.~\cite{2016_Lelli}, estimate the values 
of $M_0$ and $r_c$  well and are obtained as per the characteristics of 
different classes of galaxies. The LSB galaxies can
be fitted from a maximum radial distance $2.5$ kpc to $22.5$ kpc. Similarly, 
HSB galaxies' maximum fitted radial distance varies from $6$ kpc to $30$ kpc 
and in the case of dwarf galaxies it is from $6$ kpc to $12$ kpc. In 
Tables~\ref{table1}, \ref{table2} and \ref{table3}, the successfully fitted 
values of $M_0$ and $r_c$  are listed together with some other galactic
parameters extracted from Ref.~\cite{2012_mann} and reduced $\chi^2$ values of fitting for different 
galaxies. From the tables, it is seen that for HSB galaxies total mass $M_0$,
in the solar mass unit (here we consider $M_0=10^{10}M_{\odot}$), is large and 
in the case of LSB and dwarf galaxies this value is small.

\begin{figure}[!h]
	\centerline{
		\includegraphics[scale = 0.26]{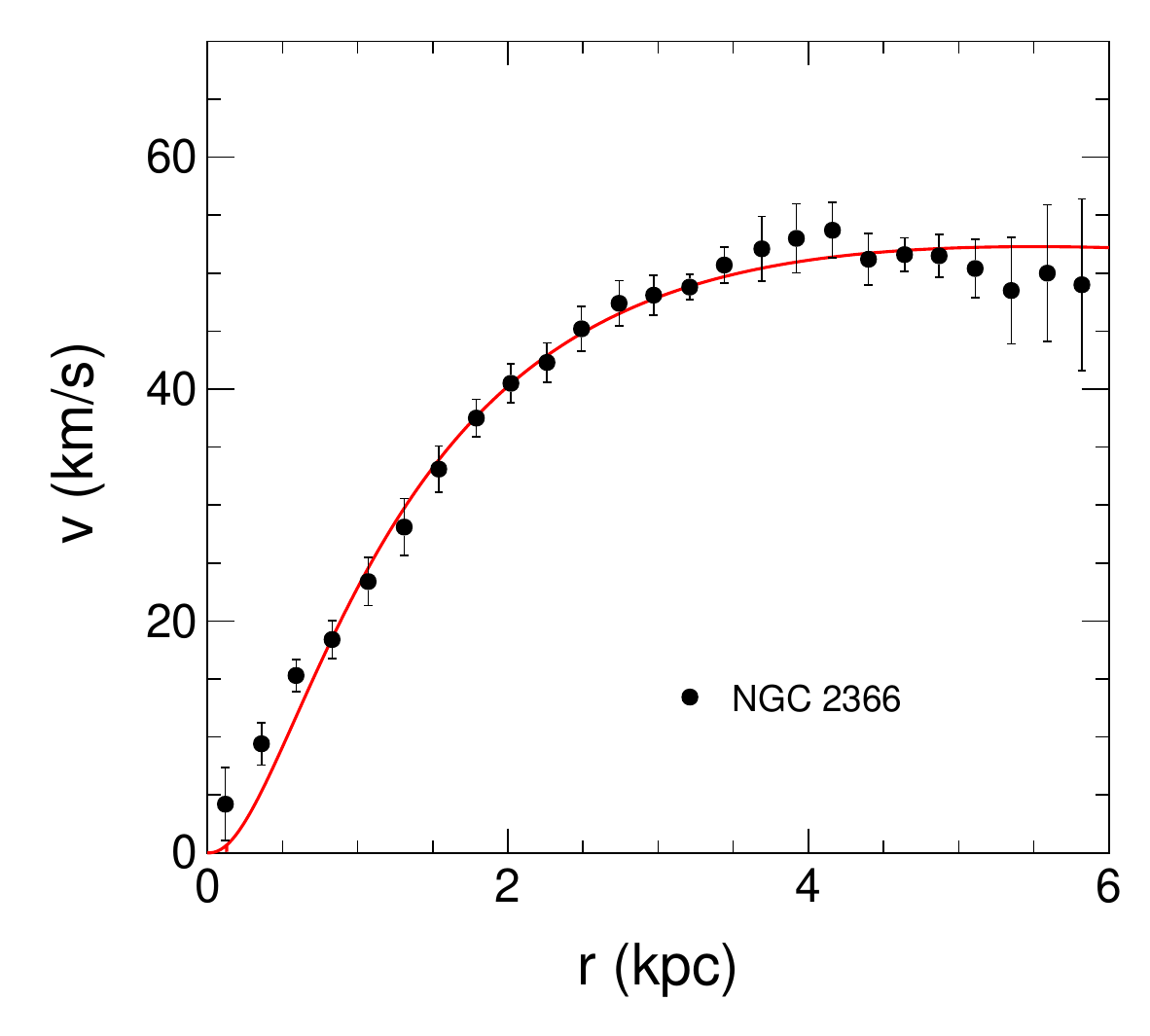}\hspace{0.3cm}
		\includegraphics[scale = 0.26]{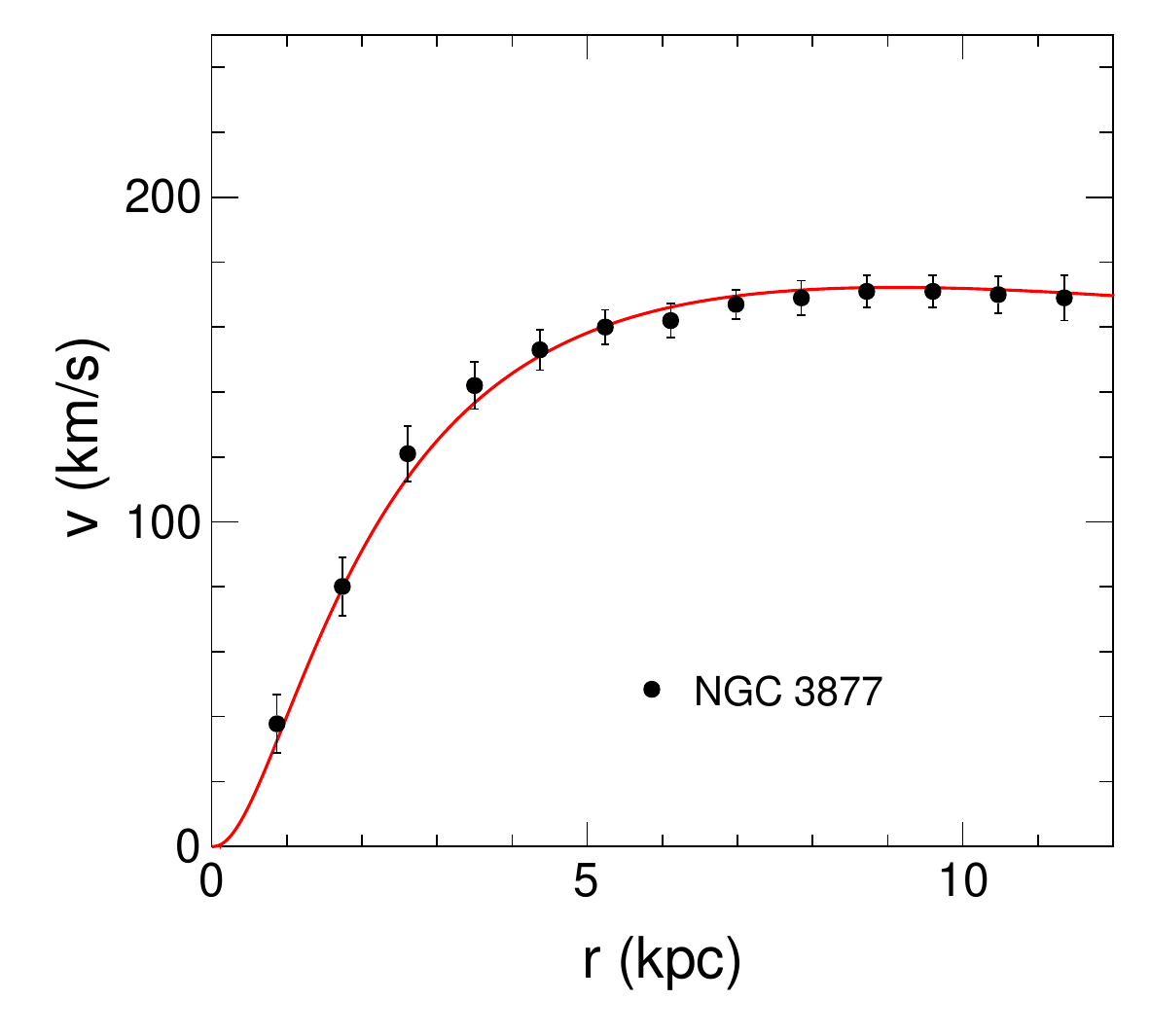}\hspace{0.3cm}
		\includegraphics[scale = 0.26]{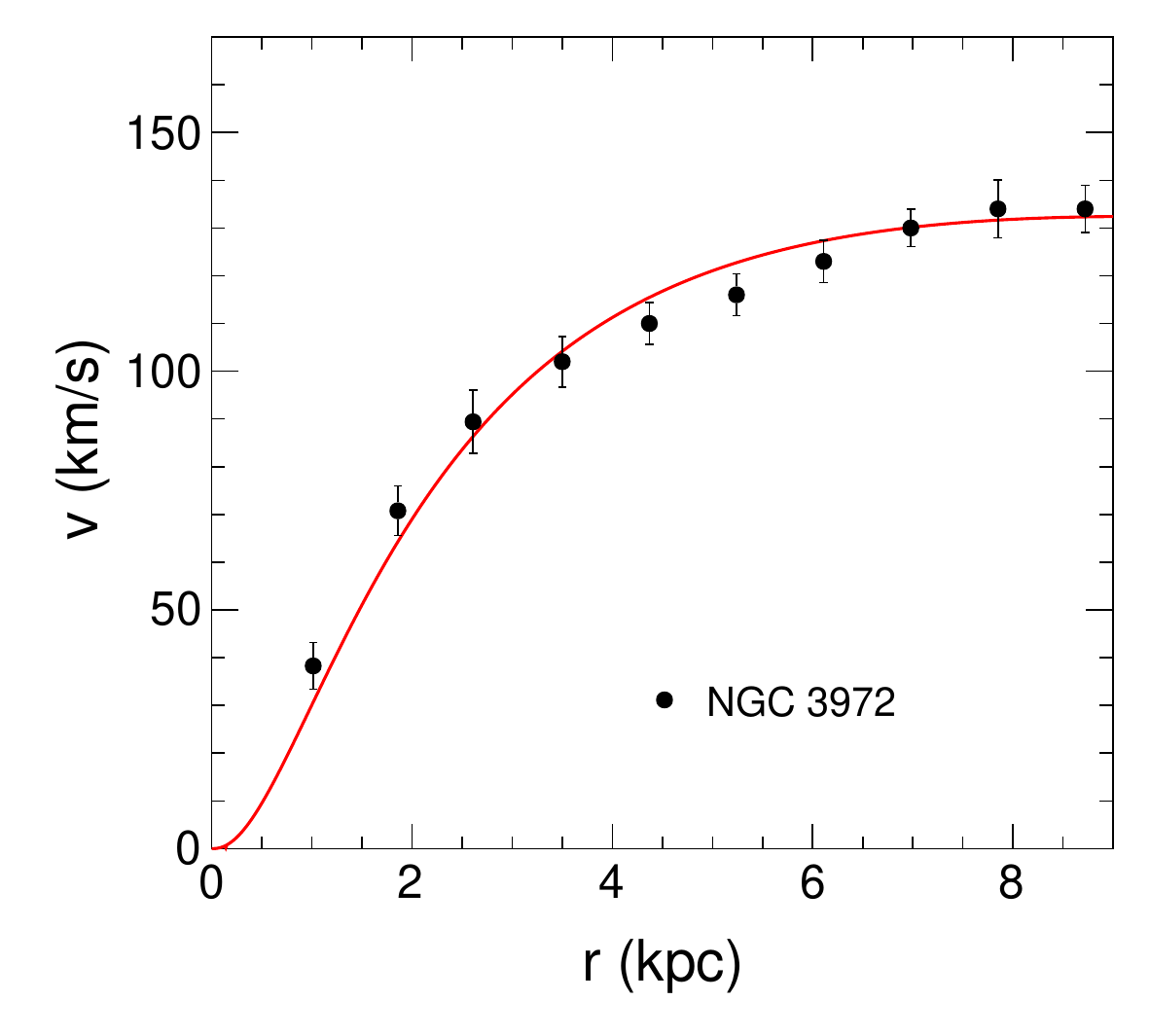}}\vspace{0.3cm}
	\centerline{
		\includegraphics[scale = 0.26]{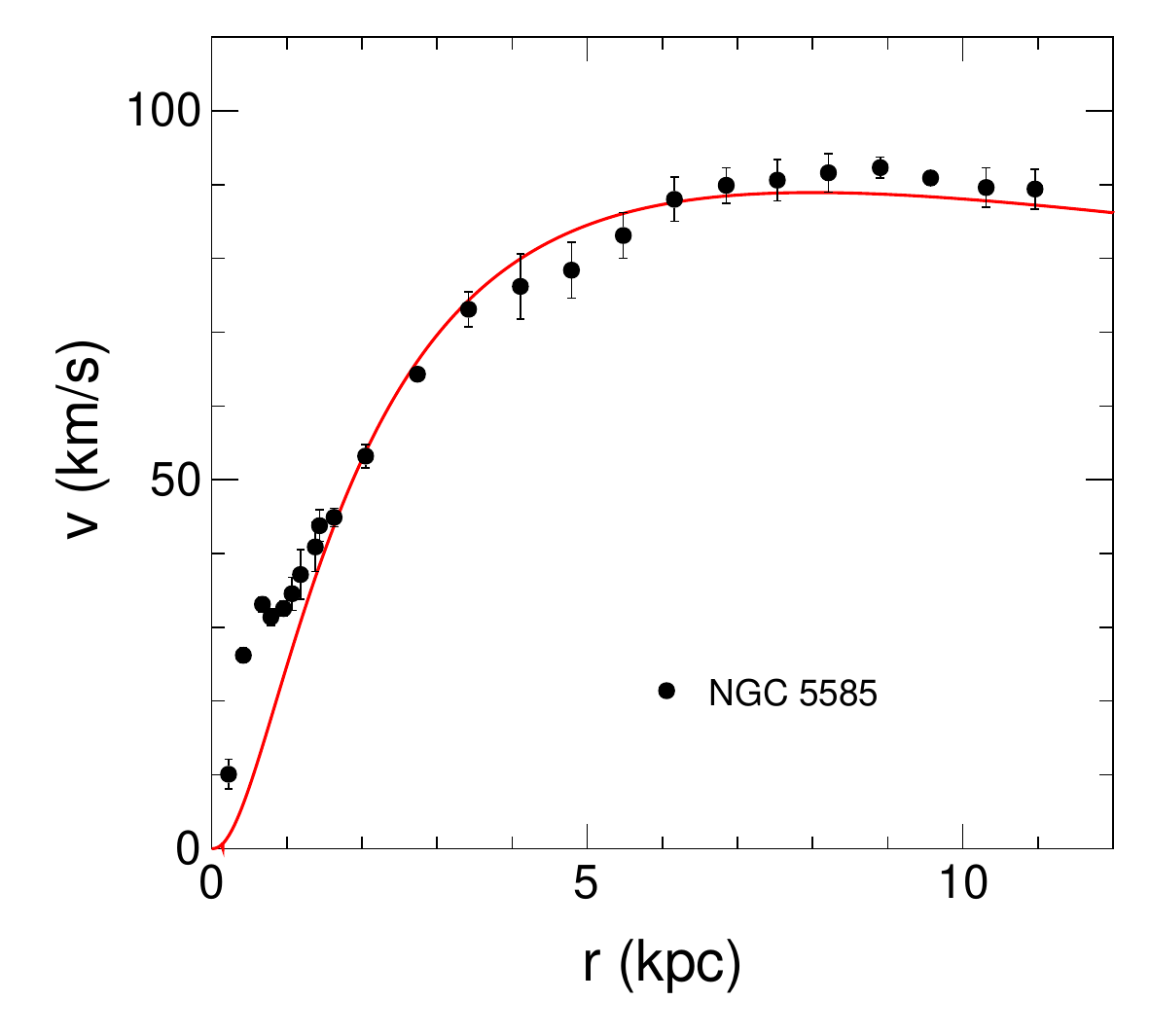}\hspace{0.3cm}
		\includegraphics[scale = 0.26]{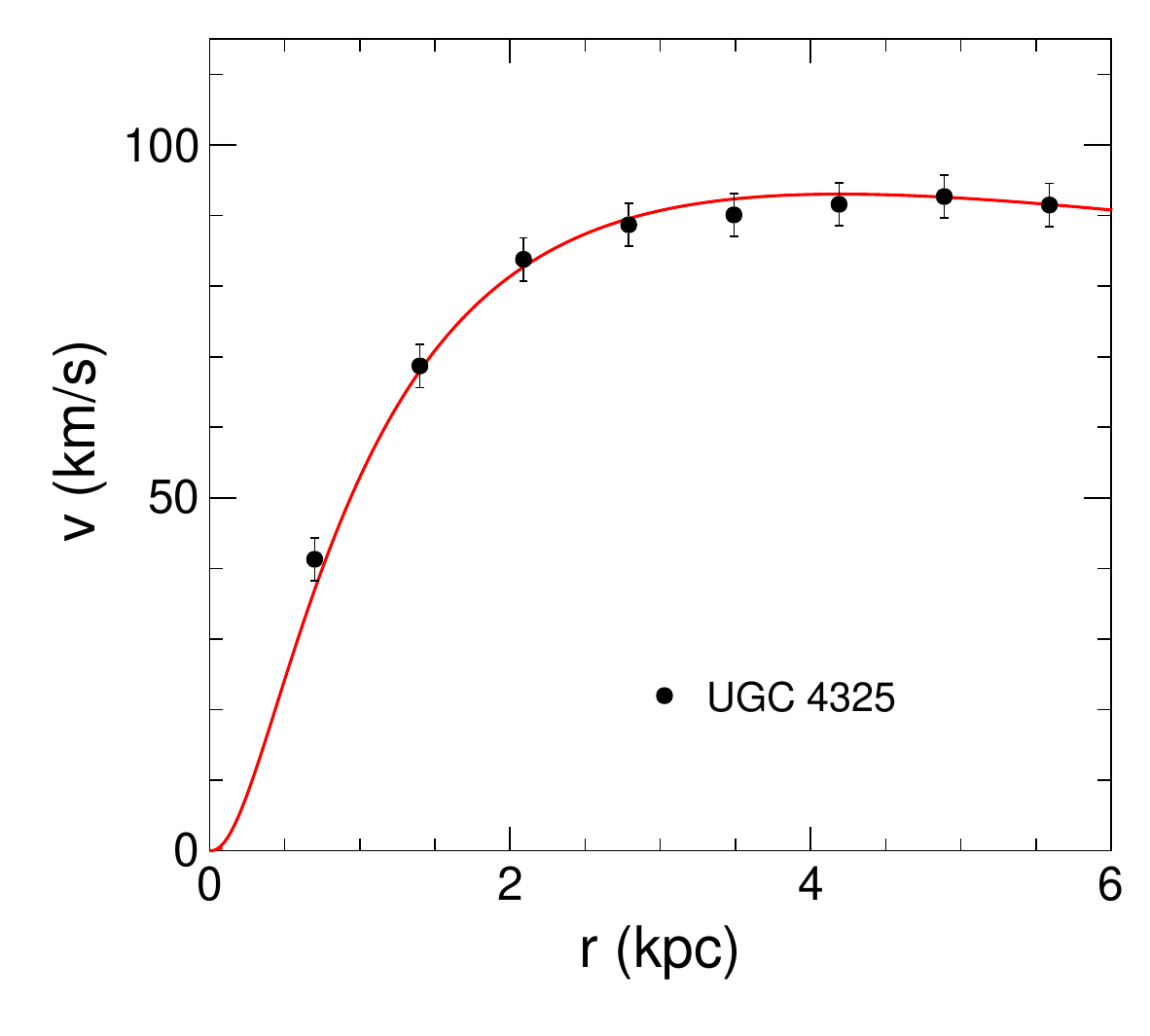}\hspace{0.3cm}
		\includegraphics[scale = 0.26]{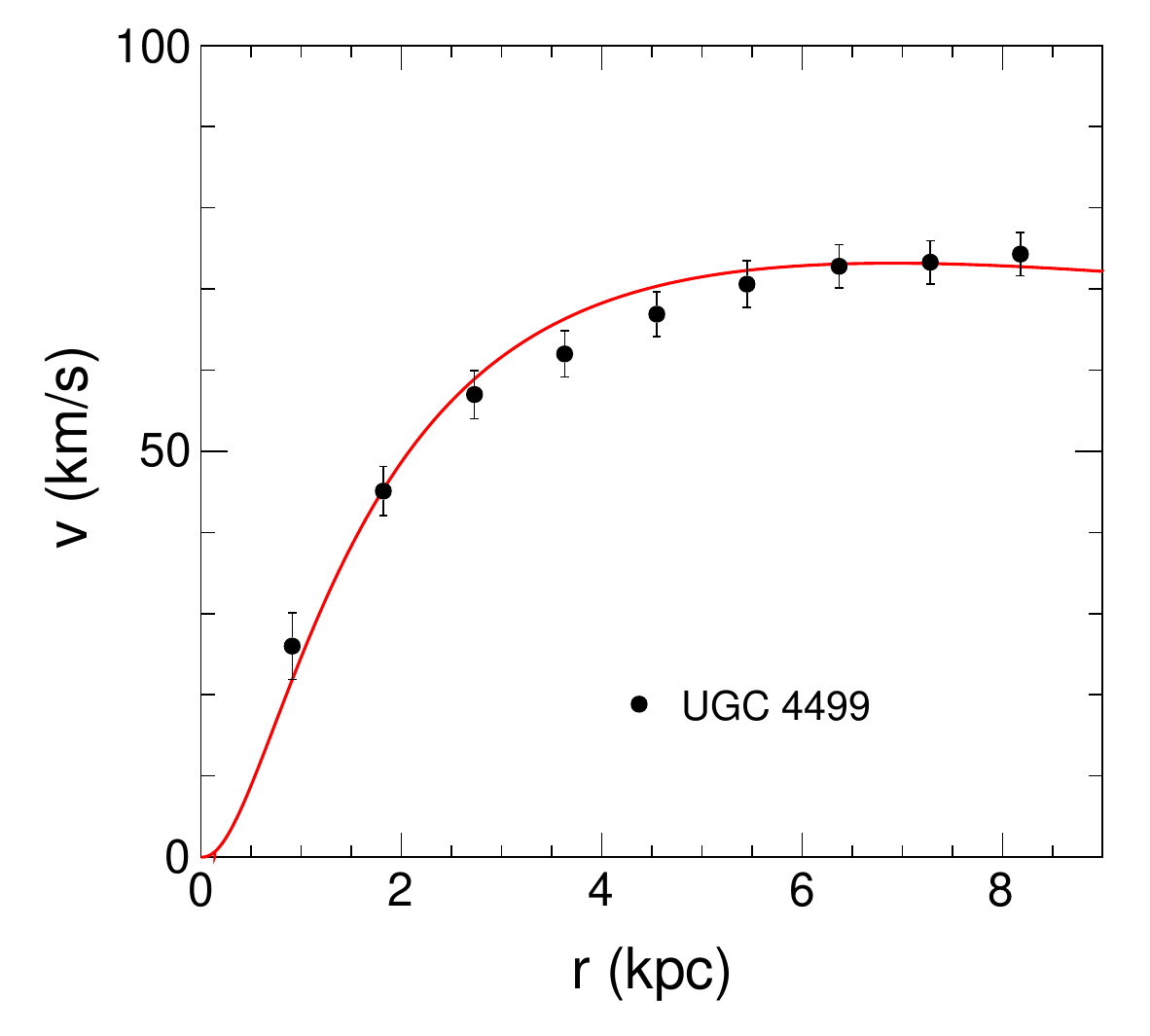}}\vspace{0.3cm}
	\centerline{
		\includegraphics[scale = 0.26]{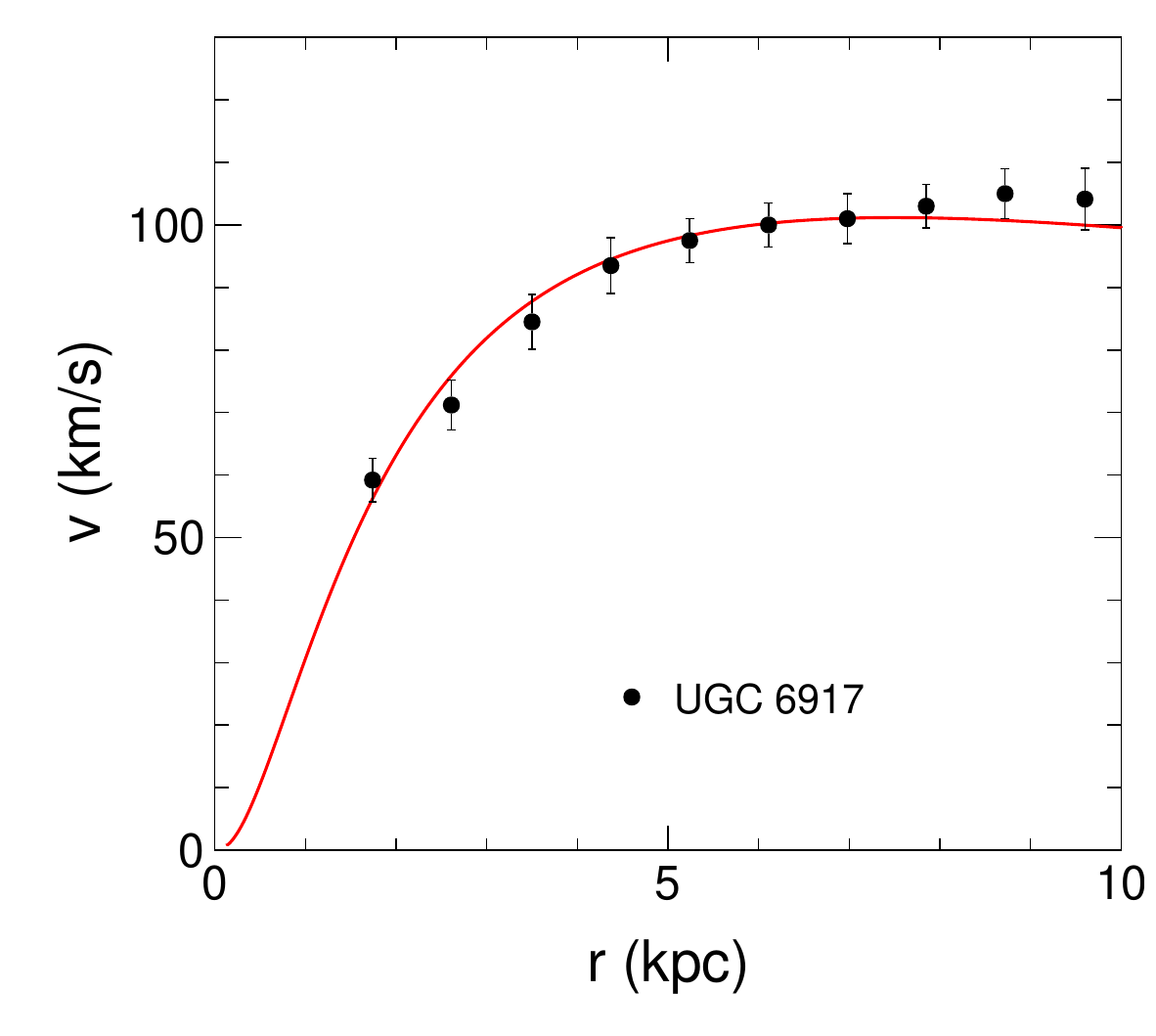}\hspace{0.3cm}
		\includegraphics[scale = 0.26]{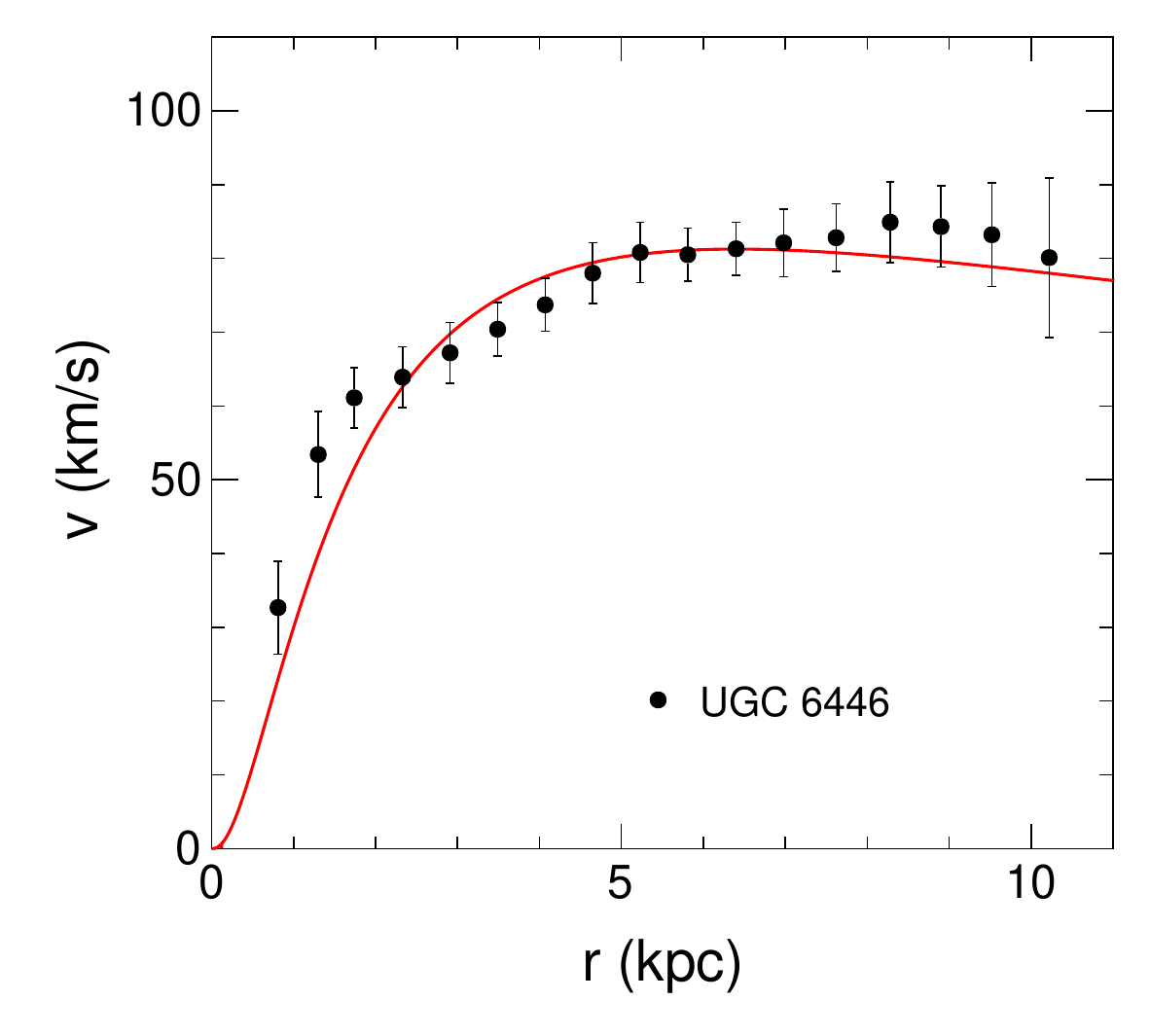}\hspace{0.3cm}
		\includegraphics[scale = 0.26]{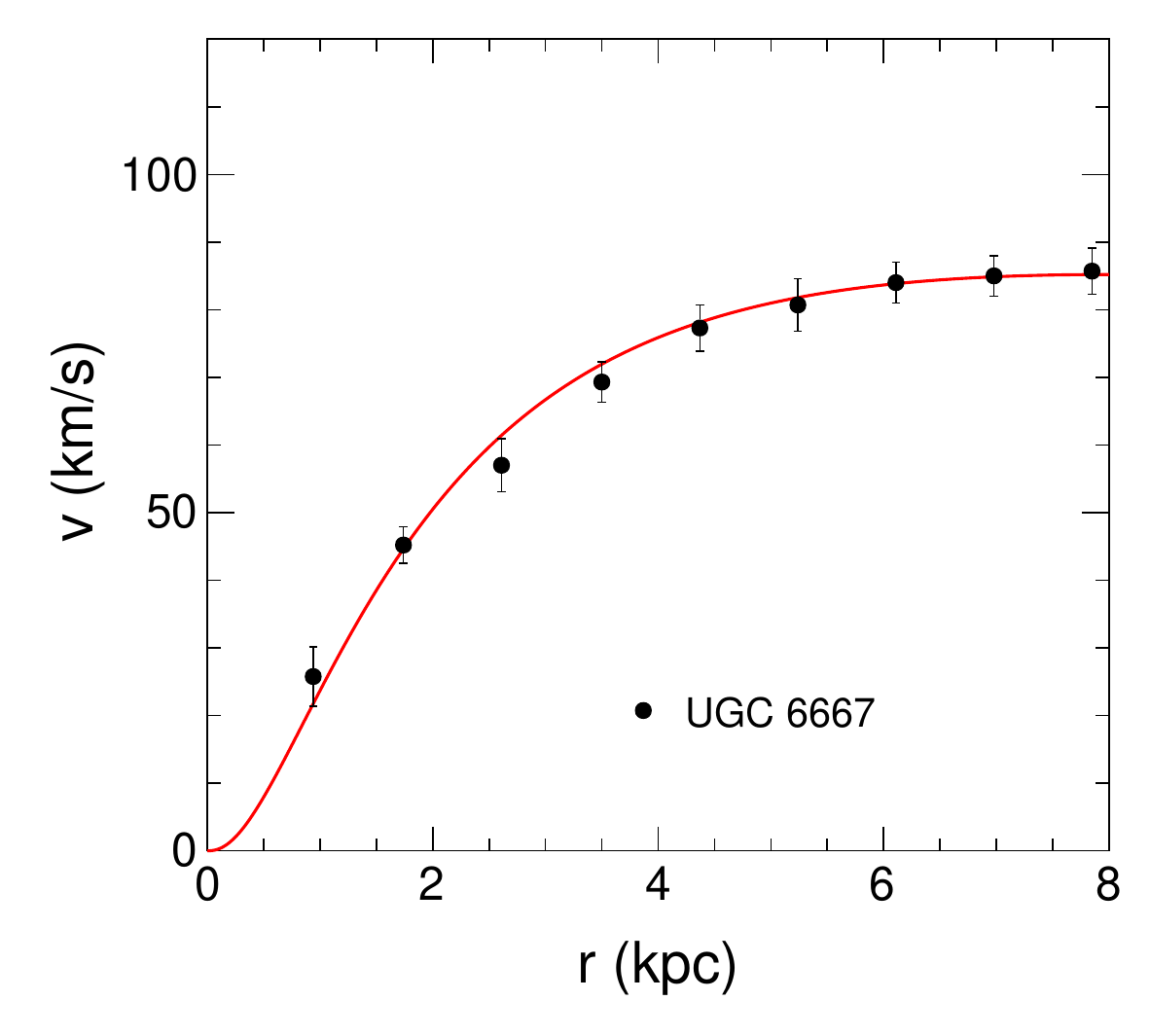}}\vspace{0.3cm}
	\centerline{
		\includegraphics[scale = 0.26]{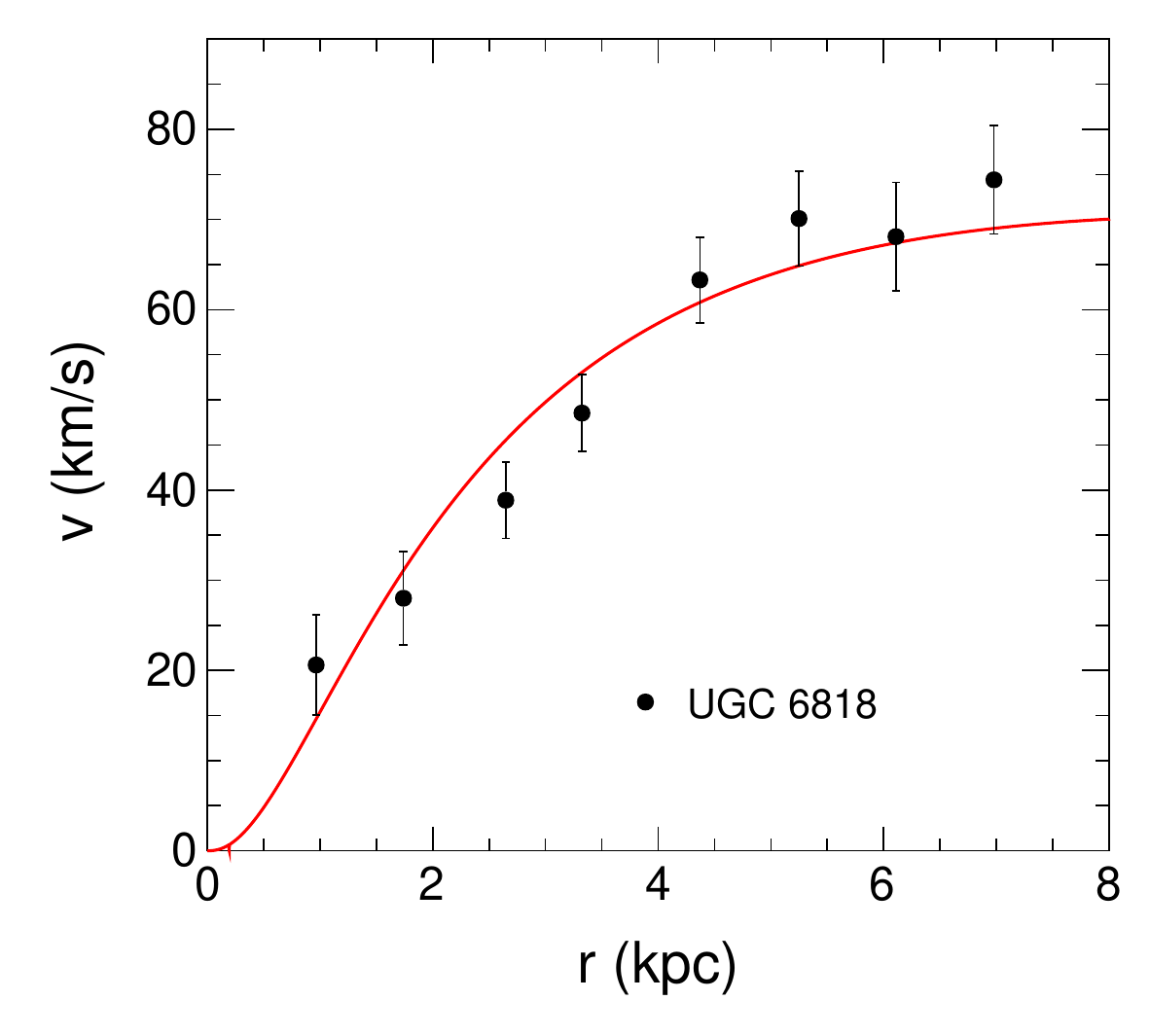}\hspace{0.3cm}
		\includegraphics[scale = 0.26]{ugc6917.pdf}\hspace{0.3cm}
		\includegraphics[scale = 0.26]{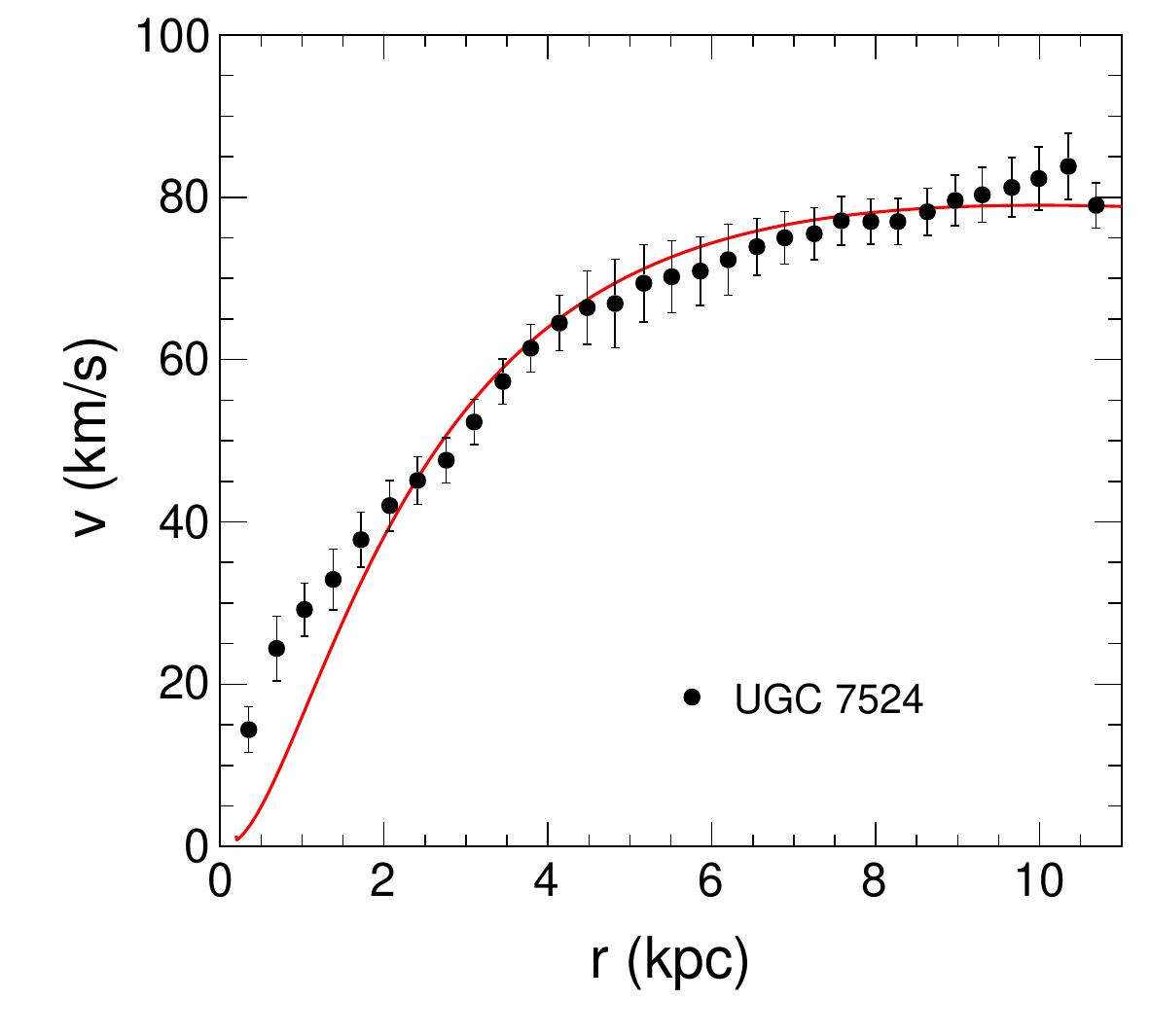}}\vspace{-0.2cm}
	\caption{Fitting of rotation curves generated from the $f(R)$ gravity model 
		\eqref{eq33} for rotation velocities of a sample of 12 dwarf galaxies with 
		their quoted errors. The data points are observational values of rotational 
		velocities extracted from Ref.~\cite{2016_Lelli}.}
	\label{fig9}
\end{figure}

\begin{figure}[!h]
	\centerline{
\includegraphics[scale = 0.35]{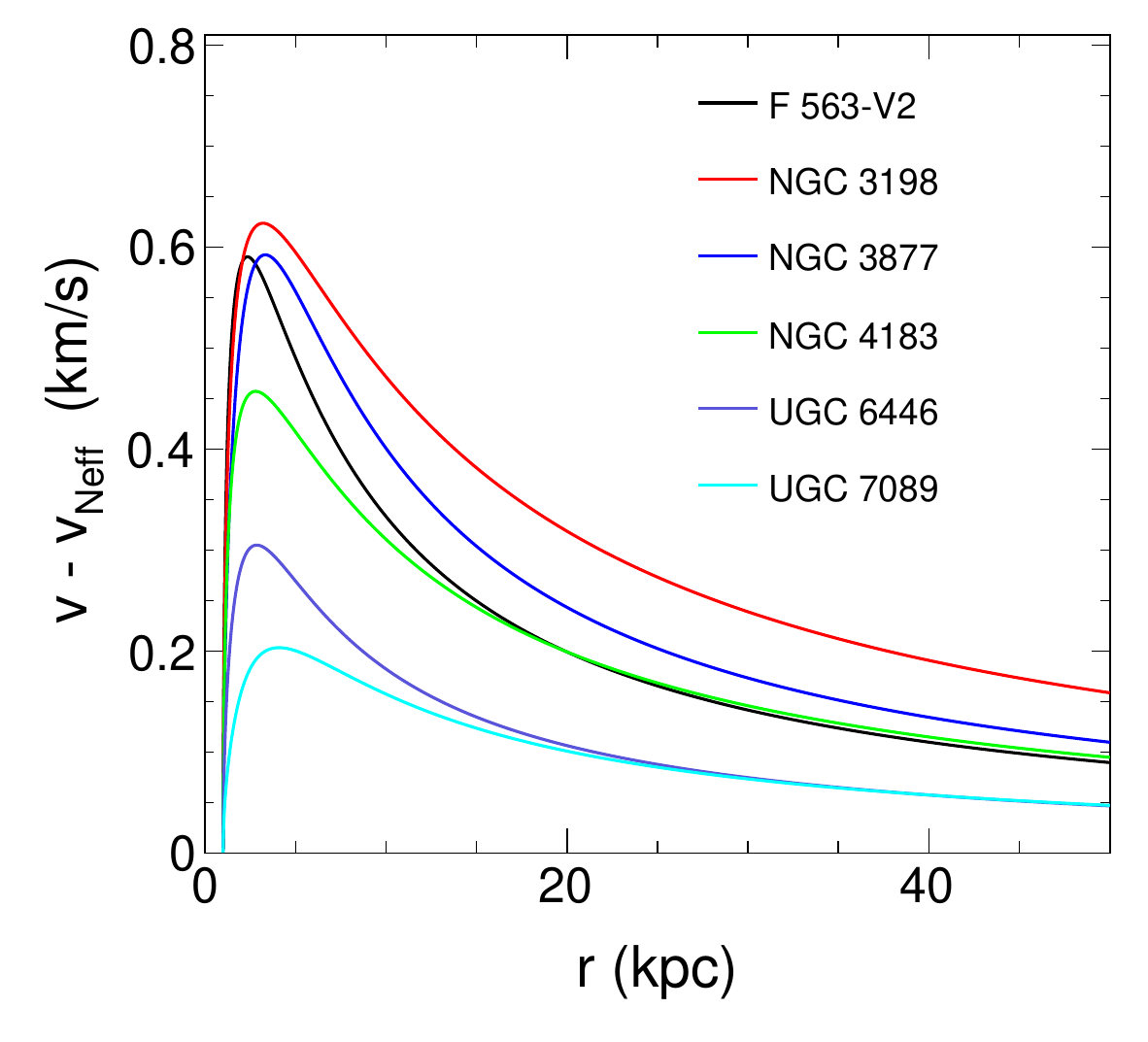}}
\vspace{-0.2cm}	
\caption{Difference of velocity $V$ and effective Newtonian velocity $V_{Neff}$ 
	i.e., $(V-V_{Neff})$ against radial distance $r$. $V$ is derived from 
	equation \eqref{eq66} and $V_{Neff} = \sqrt{GM(r)/r}.$
\label{fig10}}
\end{figure}

The behavior of galactic rotation curves of 37 sample galaxies are presented 
in Figs.~\ref{fig7}, \ref{fig8} and \ref{fig9} respectively. Furthermore, we 
are endeavoring to understand the effectiveness of the chameleon field on the 
dynamics of galaxies. For this purpose, we plot the difference between the 
velocity determined by equation \eqref{eq66} and the effective Newtonian velocity 
$\sqrt{GM(r)/r}$ as a function of radial distance $r$ for 6 sample galaxies taking 
two from each category LSB, HSB and Dwarf (see Fig.~\ref{fig10}) utilizing 
the values of model parameters that have already been taken into account and 
the same values of $M_0$ and $r_c$ obtained from Tables~\ref{table1}, 
\ref{table2} and \ref{table3} of respective galaxies. It is observed from 
Fig.~\ref{fig10} that the difference between these two velocities is not 
noticeably large. This result suggests a marginal impact of the chameleon 
field on galactic rotation curves. Contrarily, the major contribution to rotation curves is found due to the effective Newtonian velocity term which depends not only on the distance $r$ but also on the mass function $M(r)$ specified identically by the equation \eqref{eq67} and is connected to density profile \eqref{eq68}. Therefore, this velocity $v_{Neff}$ contributes significantly to account for the observed rotation curves, without the need for dark matter. As the chameleon mechanism acts as a screening mechanism, so in regions of high matter density, like galaxies, the chameleon field becomes effectively screened off, thereby its influence may reduce in the galactic environment. This screening mechanism ensures a minimal impact of the chameleon field on galactic dynamics. Thus, the behavior of rotation curves 
through modification of gravity with a scalar field attributing chameleonic 
nature can be illustrated using the newly introduced $f(R)$ model.

\section{Conclusions}
\label{sec5}
The DM issue is one of the greatest theoretical difficulties in modern 
physics in general and astrophysics in particular. Usually, the unique 
behavior of galactic rotation curves is interpreted by postulating the 
existence of this invisible matter distributed in a spherical halo around 
the galaxies. Of course, lack of acceptable evidences of such matter, 
modification of gravity is considered as a reasonable alternative for 
the explanation of observed galactic rotation curves. Looking into this, we 
thought about a popular and one of the simplest MTGs, the $f(R)$ gravity, for 
our study. Within the framework of this gravity, 
Refs.~\cite{2023_Nashiba,2018_Naik,2023_Martino} have implemented the theory to study galactic dynamics. Ref.~\cite{2023_Nashiba} employs the Palatini 
formalism along with the Weyl transformation and discusses rotation curves of 
different galaxies including an ultra diffuse galaxy. Ref.~\cite{2018_Naik} 
has studied the effect of chameleon $f(R)$ gravity on rotation curves and 
radial acceleration relations of disc galaxies. In our work utilizing the metric formalism of the theory we have investigated the impact of the scalaron on rotation curves and addressed singularity issue (in Appendix \ref{ap1}) faced by the $f(R)$ model considered. For this the field 
equations are derived first and then obtained the equation for the extra degree of freedom 
$f_R$ taking the trace of the field equation. 
To identify $f_R$ as a scalar field (scalaron),
we used conformal transformation, then tried to 
check the viability of a recently introduced $f(R)$ model, regarding the chameleonic behavior. For this relation between chameleon mass $m_{\phi}$ and local matter density $\rho$ is established in equation \eqref{eq39}. Fig.~\ref{fig1} illustrates this relationship and shows the growth of $m_{\phi}$ with $\kappa\rho$, indicating higher $m_{\phi}$ in regions of high density and lower in low density regions. The variation of chameleon mass is a consequence of the interaction of chameleon field with matter.

Then we aimed to investigate the effect of scalaron on the rotation curves of 
galaxies. For this, by considering a star as a test particle in a static 
spherically symmetric spacetime we have derived the rotation velocity equation \eqref{eq66}
of the particle moving around the galactic center in Einstein frame.
This equation demonstrates the influence of both effective Newtonian gravity and the chameleon force $F_{\phi}$ on the motion of the particle. The effective Newtonian gravity term in equation \eqref{eq66} arises according to equations \eqref{eq48} and \eqref{eq59}. Contribution of $F_{\phi}$ in the second term emerges from the variation of the chameleon field $\phi$ with  radial distance $r$, as described by equation \eqref{eq48} and \eqref{eq65}. To illustrate the galactic dynamics resulting from the combined effects of these forces, we have generated rotation curves for 37 galaxies using equation \eqref{eq66} to determine the rotational velocity of particles in stable circular orbits. To test the model we fit equation \eqref{eq66} with observational rotation curve data extracted from Ref.~\cite{2016_Lelli}. In all cases we observe a remarkable match between the model and the observations. Total mass 
$M_0$ and the core radius $r_c$ are determined for each galaxy from the 
fitting of predicted rotation curves to observed data of the sample galaxies 
by employing the $\chi^2$ minimization technique. The 
$\chi^2_{red}$ values are displayed in respective tables against each galaxy. 
It is seen that the $\chi^2_{red}$ values are not equivalent or close to $1$ 
relative to all the galaxies. In case of galaxies NGC 2366, NGC 2903, NGC 7793, 
UGC 4325, NGC 3877 and NGC 5585, its value is quite larger than $1$. Still, we 
can almost fit the curves successfully where the predicted curves exhibit a 
nice and good agreement with the observed curves as shown in respective 
figures. Moreover, in order to comprehend the extent of the impact of 
scalaron, we have depicted velocity difference curves of 
$6$ galaxies two each from the LSB, HSB and dwarf with respect to radial 
distance $r$ for similar values of model parameters $\alpha$, $\beta$ and of 
$M_0$ and $r_c$ estimated by the best-fitted curves of respective galaxies. Significantly, this plot does not suggest a notable influence of the scalaron on galactic dynamics. It is due to the screening characteristic of the chameleon mechanism for which in high matter density regions such as around massive objects like galaxies, the chameleon field becomes effectively screened resulting in a small acceleration of the field. For this, its influence is reduced on the motion of a particle around the center of a galaxy.

Hence, we may conclude that alternative of an explicit form of dark matter can be addressed by means of chameleon 
$f(R)$ gravity theory. However, in contrast to the approach centered on the single 
mass model, there are possibilities for comparative study by considering 
different mass profiles derived from the density distributions such as 
Navarro-Frenk-White (NFW), pseudo-isothermal (ISO)~\cite{1997_nava,2013_sofue,2008_block} etc.~with other convenient modified 
gravity models. We aim to develop the work further sustaining this viewpoint 
in the future.  

\section*{Acknowledgements} UDG is thankful to the Inter-University Centre for
Astronomy and Astrophysics (IUCAA), Pune, India for the Visiting
Associateship of the institute.\\

\appendix
\section{Singularity problem and its correction}
\label{ap1}

\begin{figure}[!h]
	\centerline{
		\includegraphics[scale = 0.277]{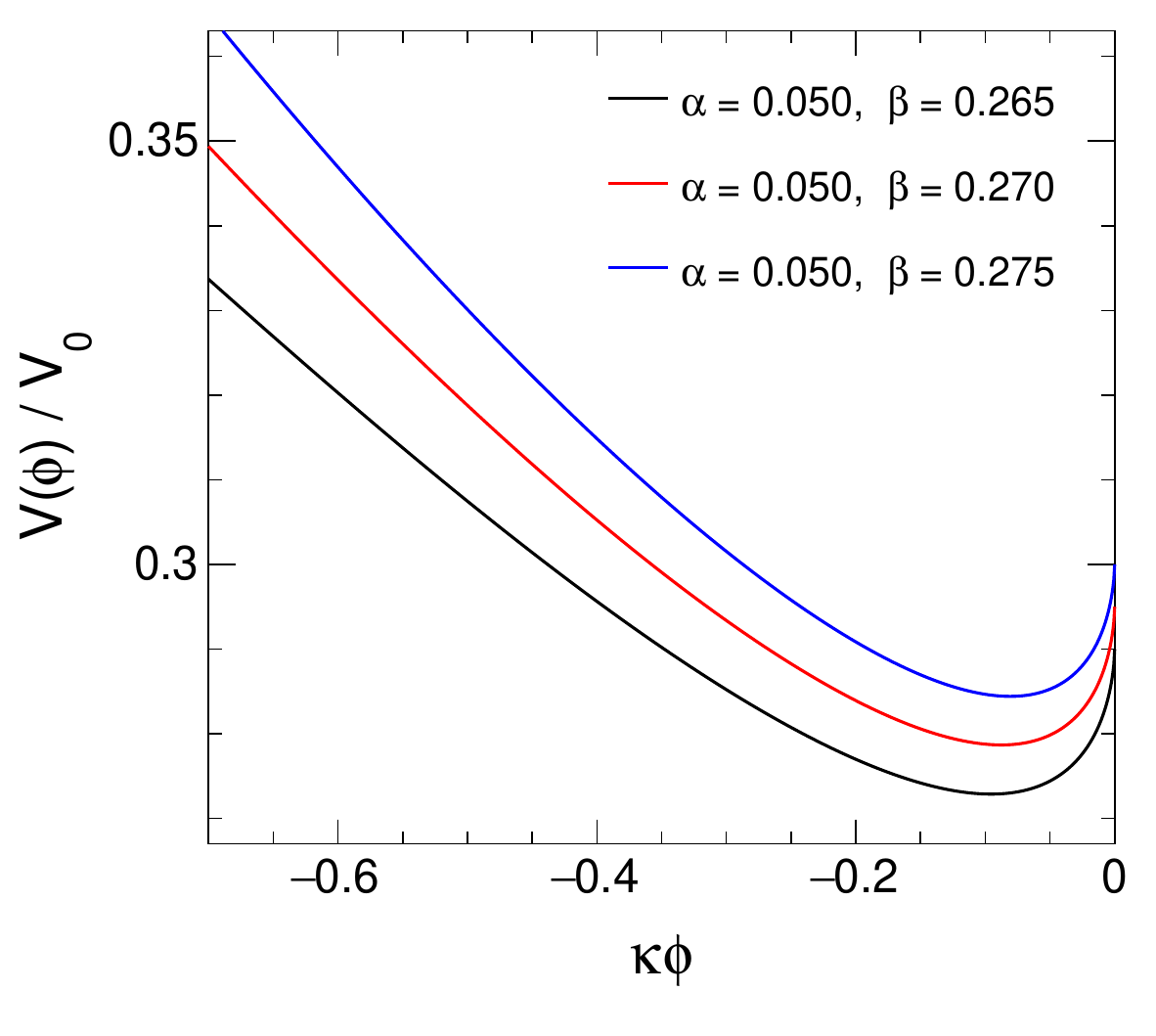}\hspace{1cm}
		\includegraphics[scale = 0.275]{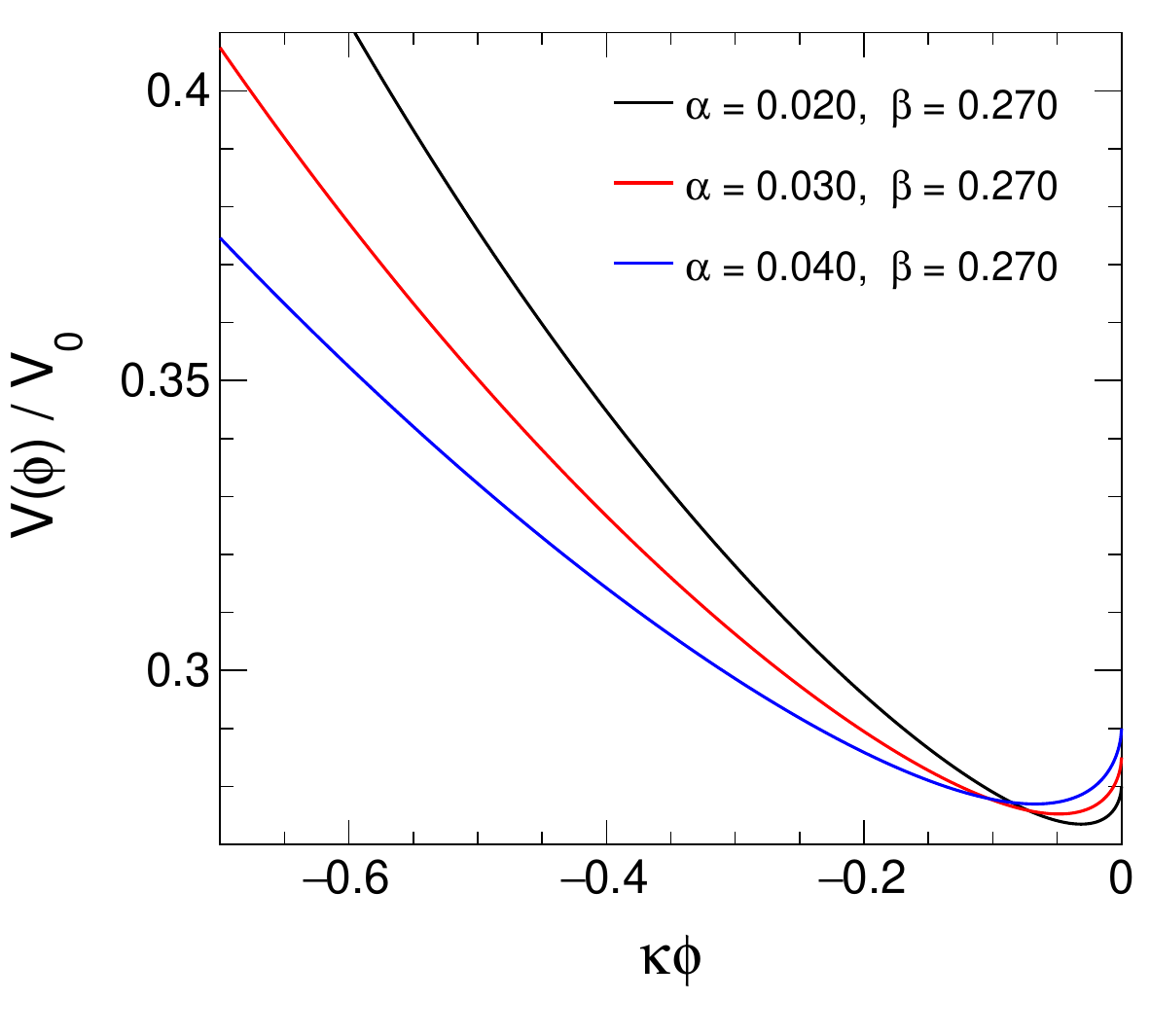}}\vspace{-0.2cm}
	\caption{Variation of $V(\phi)$ with respect to $\kappa\phi$ for different 
		values of model parameters $\alpha$ and $\beta$ with $R_c = 1$}.
	\label{fig2}
\end{figure}

To understand the 
singularity problem in the specific new $f(R)$ gravity model chosen for our 
study, initially, we take the field-dependent scalaron potential without 
matter contribution which is attained in the form:
\begin{equation} 
	\frac{V(\phi)}{V_0}=\left[\frac{\alpha}{2}+\beta-\left(\frac{2\alpha}{\pi}\right)^{1/3}\!\!\!\left(1-e^{\sqrt{\frac{2}{3}}\kappa\phi}\right)^{2/3} \right]e^{-2\sqrt{\frac{2}{3}}\kappa\phi},
	\label{eq69}
\end{equation}
where $V_0=R_c/2\kappa^2$ is the normalization factor that normalizes the 
potential $V(\phi)$. The relation between $V(\phi)$ and $\phi$ given in 
equation \eqref{eq69} is shown in Fig.~\ref{fig2}.

In the left panel, we depict an explicit behavior of the 
potential as a function of the field and see that the potential decreases 
with increasing $\phi$. It reaches a minimum at $\phi\sim -\ 0.1$ and then 
increases up to $\phi=0$. This behavior of the potential reflects the dynamics of the field $\phi$ indicating that the scalaron is rolling down in the  
slope of the potential, and potential minimum at $\phi\sim -\ 0.1$ suggests a stable equilibrium position of the field. The other two panels illustrate that the minimum of the potential moves towards the higher field value for larger values of 
$\alpha$ and $\beta$. Also, the potential minimum uplifts noticeably for 
higher values of $\alpha$ and $\beta$. 

Now, we consider the matter contribution to the potential as obtained 
	from equation \eqref{eq25} and plot the behavior of the effective 
potential in Fig.~\ref{fig3} against the scalaron field for a positive matter 
density using the normalized equation as given below:
\begin{equation} 
	\frac{V_{eff}(\phi)}{V_0}=\left[\frac{\alpha}{2}+\beta-\Big(\frac{2\alpha}{\pi}\Big)^{1/3}\Big(1-e^{\sqrt{\frac{2}{3}}\kappa\phi}\Big)^{2/3}\!\!+\frac{\rho\kappa^2}{2R_c}\,\right]e^{-2\sqrt{\frac{2}{3}}\kappa\phi}.
	\label{eq70}
\end{equation}
\begin{figure}[!h]
	\centerline{
		\includegraphics[scale = 0.27]{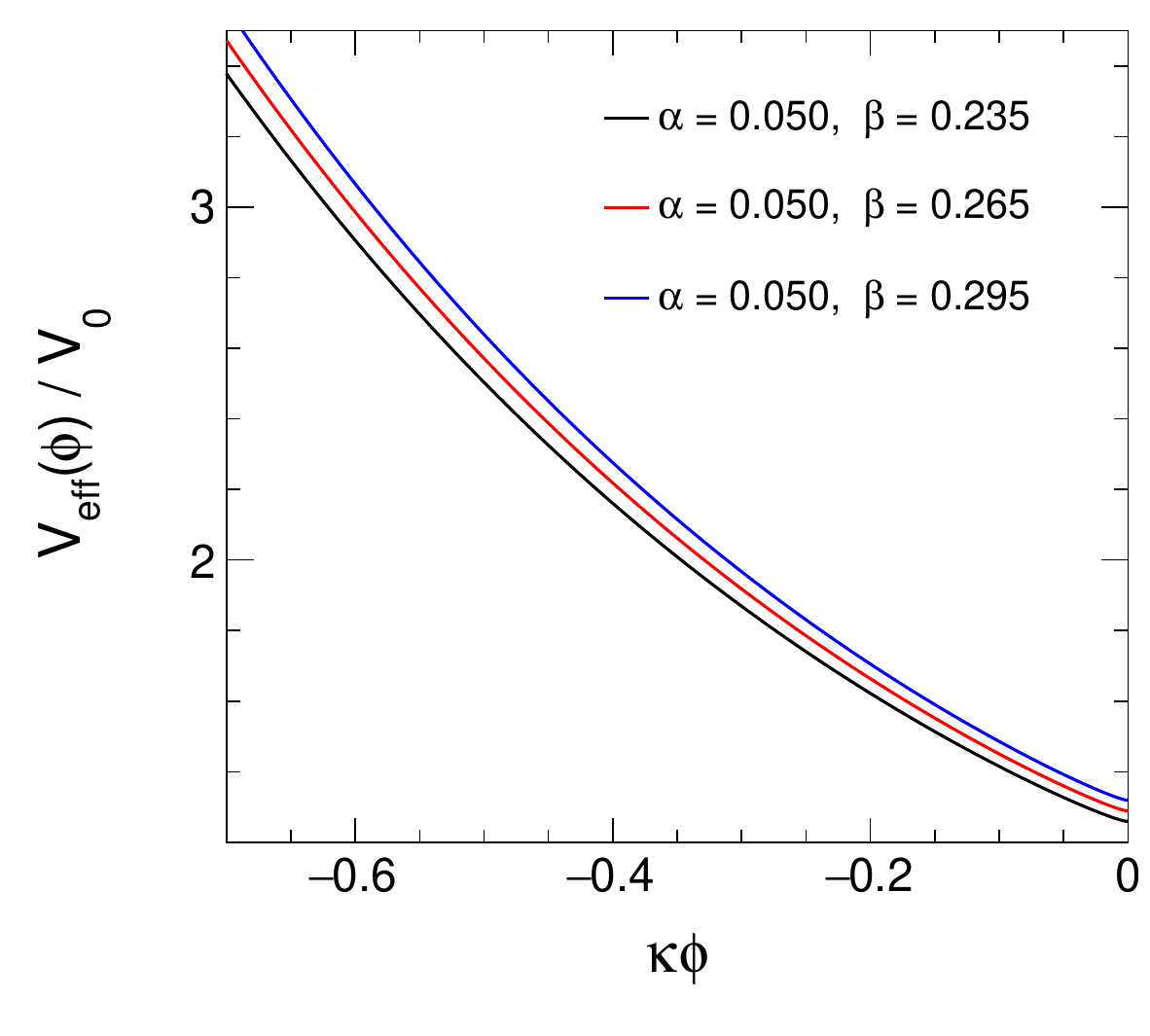}\hspace{1cm}
		\includegraphics[scale = 0.27]{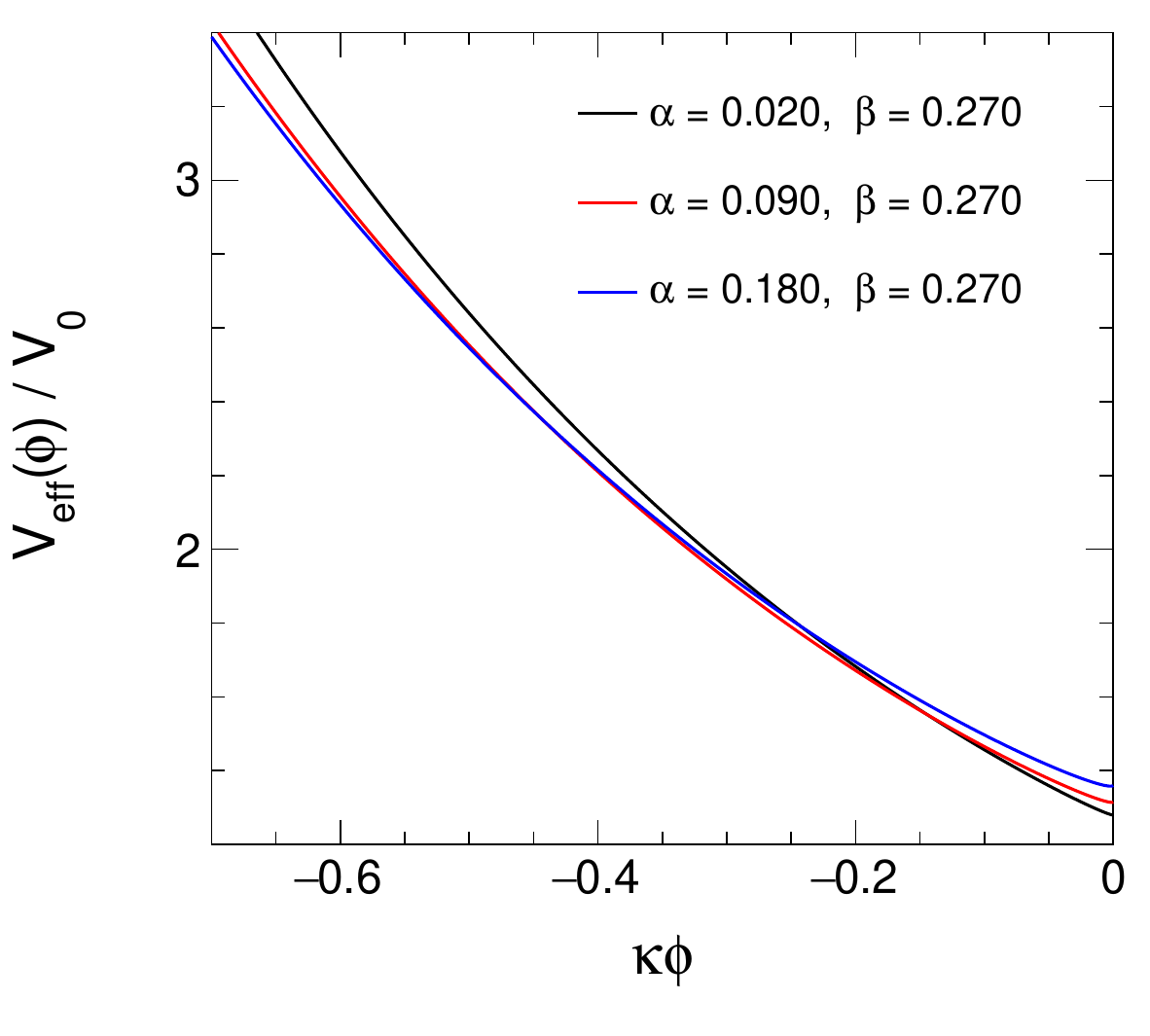}}\vspace{-0.2cm}
	\caption{Variation of effective potential $V_{eff}(\phi)$ as a function of 
		$\kappa\phi$ for different values of model parameters $\alpha$ and $\beta$
		with the matter density $\rho = \rho_c$, where $\rho_c$ is the critical
		matter density of the Universe and \textbf{$R_c = 1$}.}
	\label{fig3}
\end{figure}
One can see from Fig.~\ref{fig3} that the presence of matter  
flattened the minimum of the potential and shifted it very close to $\phi=0$. 
In fact, the minimum can smoothly correspond to zero of the scalaron field. 
This $\phi\rightarrow0$ correlates with the curvature singularity as hinted by 
equation \eqref{eq68} i.e., $R\rightarrow\infty$  when $\phi\rightarrow0$, can 
be easily obtainable in the presence of matter. Ref.~\cite{2008_frolov} 
suggested that the problem of curvature singularity is not unique to only a 
particular $f(R)$ gravity model, but it is endured by many infrared-modified 
$f(R)$ gravity models.

\begin{figure}[!h]
	\centerline{
		\includegraphics[scale = 0.30]{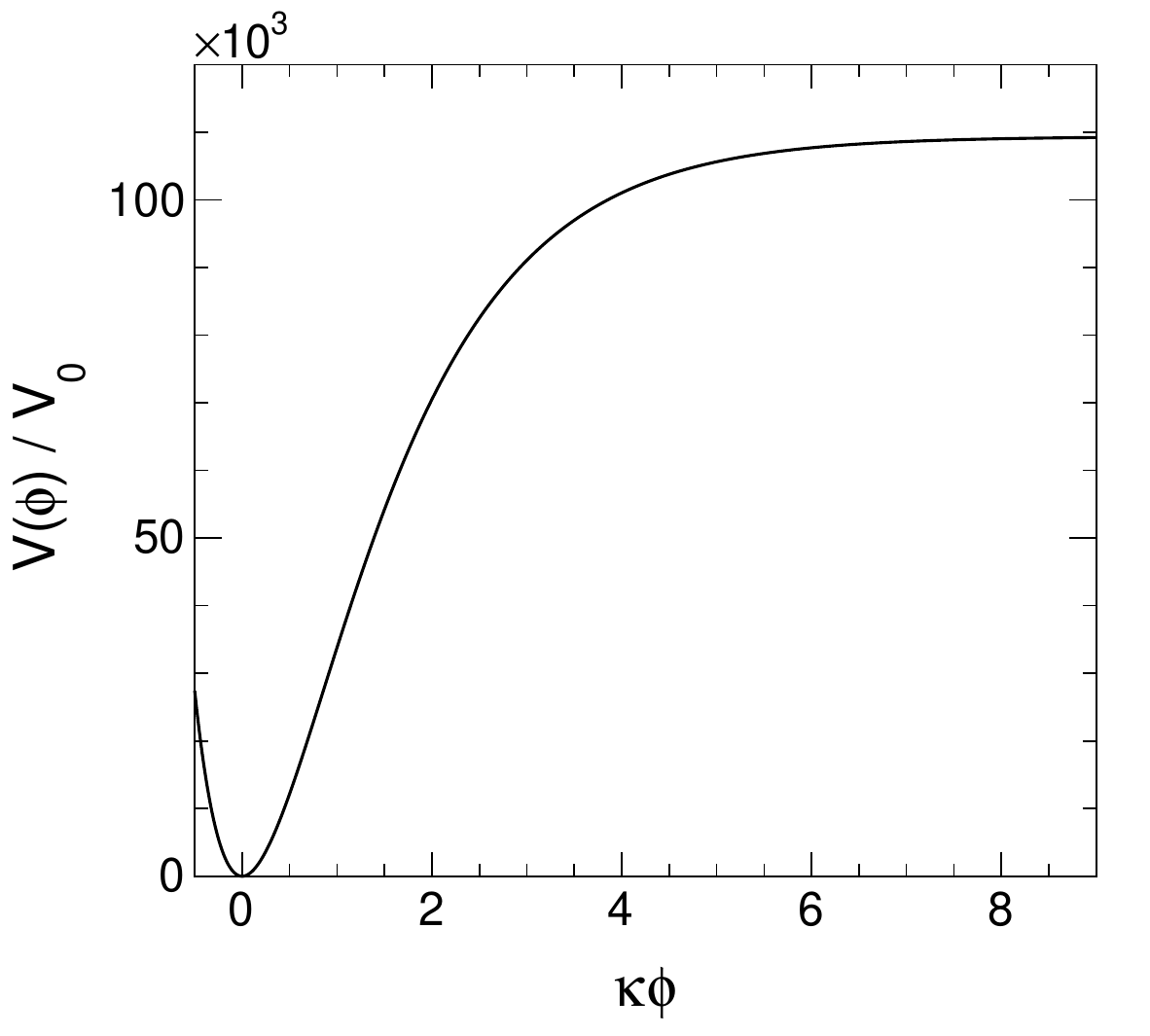}\hspace{1.0cm}
		\includegraphics[scale = 0.30]{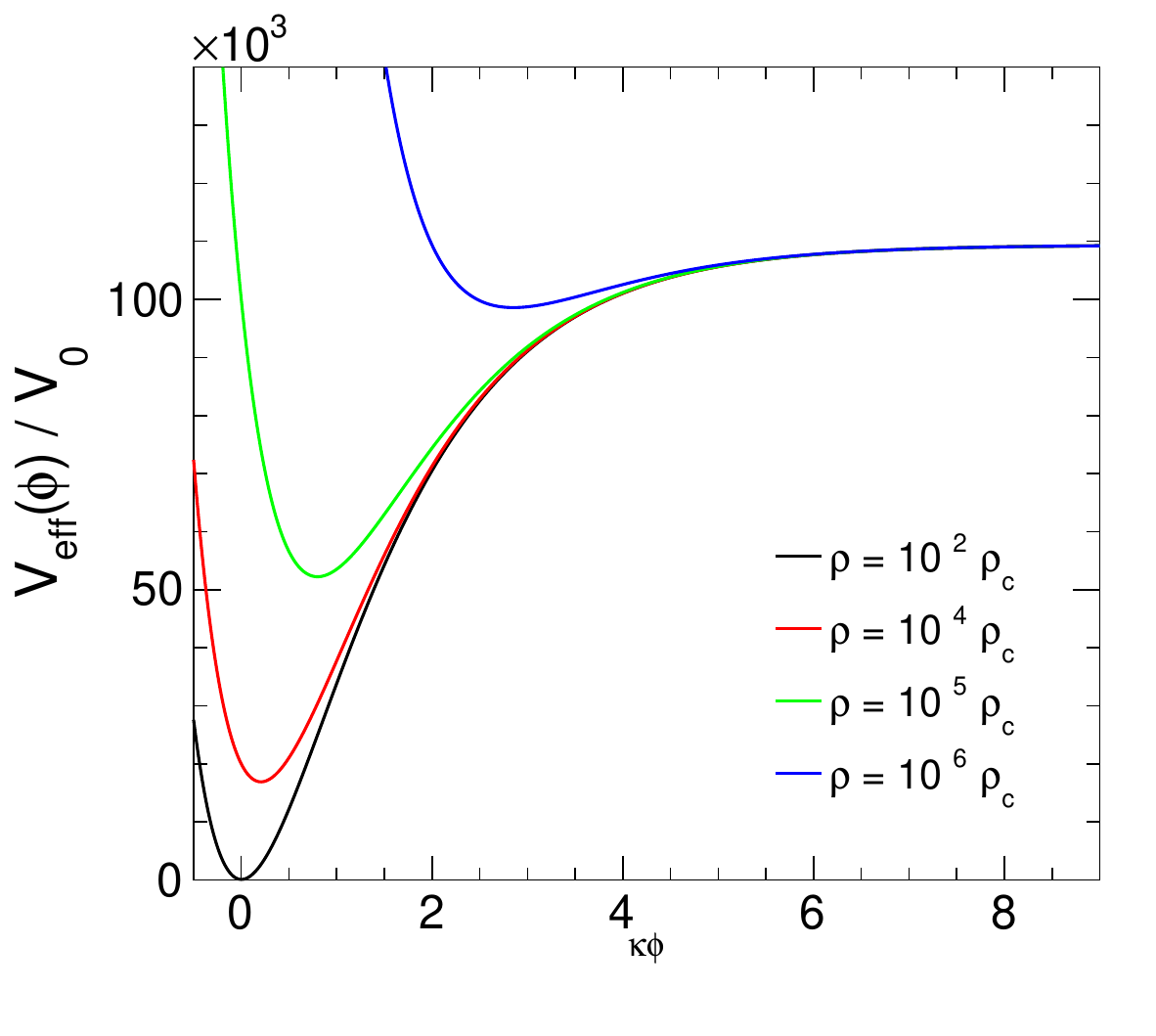}}
	\vspace{-0.2cm}
	\caption{Plot of $R^2$ corrected potential $V(\phi)$ as a function of 
		$\kappa\phi$ (left) and
		variation of $R^2$ corrected effective potential $V_{eff}(\phi)$ as a function 
		of $\kappa\phi$ for four positive values of 
		matter density (right), where $\rho_{c}$ is critical density of the Universe. Values of $\alpha$, $\beta$ and $R_c$ are taken as 0.05, 0.25 and 1 respectively for both the cases.}
	\label{fig4} 
\end{figure}

The above curvature singularity appeared in terms of the scalaron field 
potential in the large curvature regime can be removed by reforming the 
structure of the potential in this curvature regime.
This can be done by adding higher order correction term to the 
$f(R)$ model as follows~\cite{2021_Nashiba,2018_kat,2009_kobay,2008_Noji,2008_Dev,2008_Bamba}:
\begin{equation} 
	f(R)=R-\frac{\alpha}{\pi}\,R_c\,\cot^{-1}\!\left(\frac{R_c^2}{R^2}\right)-\beta R_c\left[1-e^{-\frac{R}{R_c}}\right]+\sigma R^2,
	\label{eq71}
\end{equation}
where $\sigma$ is a dimensional constant. In the higher curvature regime equation 
\eqref{eq71} may reduce to~\cite{2021_Nashiba,2018_kat}
\begin{equation} 
	f(R)\approx R-\frac{\alpha R_c}{2}-\beta R_c+\sigma R^2. 
	\label{eq72}
\end{equation}
Also, after introducing the correction term we obtain $f_R$ and the relation 
between curvature $R$ and the scalaron field through conformal transformation 
respectively as follows:
\begin{equation} 
	f_R(R) = 1-\frac{2\,\alpha R_c^3}{\pi R^3}-\beta\, e^{-\frac{R}{R_c}} + 2\sigma R, 
	\label{eq73}
\end{equation}
\begin{equation} 
	e^{\sqrt{\frac{2}{3}}\kappa\phi} = 1-\frac{2\,\alpha R_c^3}{\pi R^3}-\beta\, e^{-\frac{R}{R_c}} + 2\sigma R.
	\label{eq74}
\end{equation}
The noteworthy feature of this equation is that contrary to the resolution 
provided by the equation \eqref{eq68}, it results in $R\rightarrow\infty$ for 
$\phi\rightarrow \infty$. Equations \eqref{eq72}, \eqref{eq73} and \eqref{eq74} 
configure the $R^2$ corrected normalized field potential as 
\begin{equation} 
	\frac{V(\phi)}{V_0}=\left[\frac{\alpha}{2}+\beta\!+\frac{7}{64\sigma R_c}\big(e^{\sqrt{\frac{2}{3}}\kappa\phi}-1\big)^2\right]e^{-2\sqrt{\frac{2}{3}}\kappa\phi}.
	\label{eq75}
\end{equation}
The potential given by this equation \eqref{eq75} is plotted as a function of 
$\kappa\phi$ with $\sigma=10^{-6}R_c^{-1}$ GeV$^{-2}$ and for different values
of the parameters of the new $f(R)$ gravity model in the left panel 
of Fig.~\ref{fig4}. The figure shows that in the large curvature regime the potential is modified and attains a definite value. At $\phi=0$, 
$R$ takes a small value $\approx \left({{\alpha R_c^3}/{\pi\sigma}}\right)^{1/4}$ instead of infinity as expected.  

Then, by considering the matter contribution, the equation for the $R^2$ 
corrected effective potential is derived and obtained in the following 
normalized form:
\begin{equation} 
	\frac{V_{eff}(\phi)}{V_0}=\left[\frac{\alpha}{2}+\beta\!+\frac{7}{64\sigma R_c}\big(e^{\sqrt{\frac{2}{3}}\kappa\phi}-1\big)^2+\frac{\rho\kappa^2}{2R_c}\right]e^{-2\sqrt{\frac{2}{3}}\kappa\phi}.
	\label{eq76}
\end{equation}
The variation of this effective potential \eqref{eq76} is depicted in the 
right panel of Fig.~\ref{fig4} for four matter distributions. It exhibits 
that the effective potential achieves specific values in the high matter density regions (large curvature regions). Moreover, it is seen 
that the minimum of effective potential gradually becomes shallower as its 
position rises up and moves towards higher field values with the increasing 
matter densities, and finally it goes up to the plateau for the very high 
density environment. Thus, there is no possibility to have the 
curvature singularity after the $R^2$ correction in our working model.

\begin{figure}[!h]
	\centerline{
		\includegraphics[scale = 0.34]{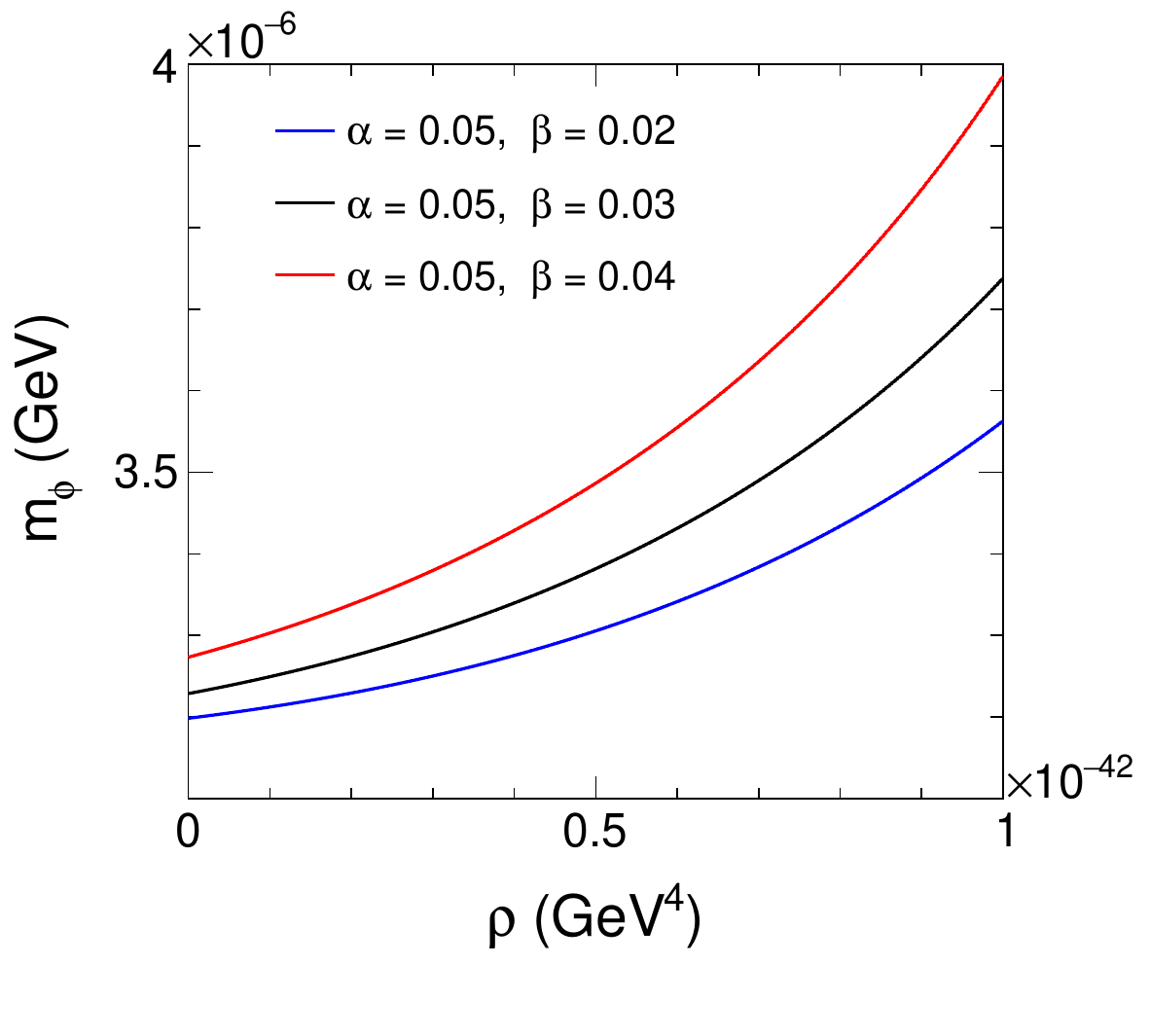} \hspace{1.0cm}
		\includegraphics[scale = 0.34]{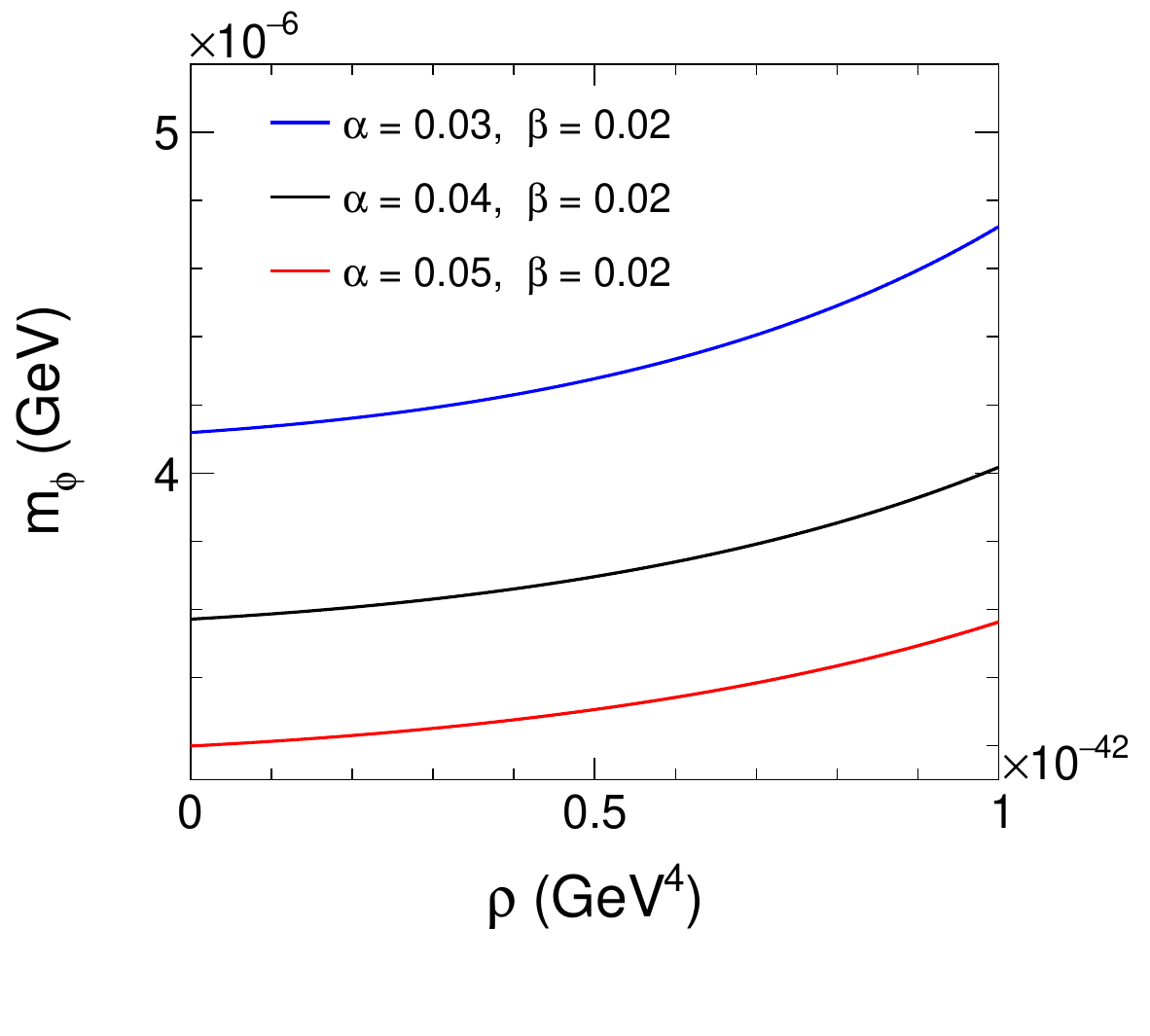}}
	\vspace{-0.5cm}
	\caption{Relations between scalaron mass and matter density for three sets of 
		model parameters $\alpha$ and $\beta$ after singularity correction. The left plot is for different values of $\beta$ by fixing $\alpha$ to 0.05 and the right plot is for different values of $\alpha$ by fixing $\beta$ to a value 0.02. In both cases with $R_c$ is set to 1.}
	\label{fig5}
\end{figure}

Here, it is to be noted that we realized very high scalaron mass in 
high-density regions according to equation \eqref{eq39} (see Fig.~\ref{fig1}) 
basically due to the singularity problem. Now, the singularity correction in 
the model modifies mass relation \eqref{eq39} into the following form provided $R_0$ is as given by equation \eqref{eq38}:
\begin{align} 
	m_\phi^2=\frac{\pi R_0^4}{3}\left[R_c\left(6\alpha R_c^4  + \pi R_0^4\,\big(\beta\, e^{-R_0/R_c} + 2 \sigma R_c\big)\right)^{-1} -  \left(\pi R_0^3\, \big(1- \beta\,e^{-R_0/R_c} + 2 \sigma R_0\big) - 2\alpha R_c^3 \right)^{-1}\right].		
	\label{eq77}
\end{align}
In order to compare the mass represented by the equation \eqref{eq77} to the 
previous one predicted by equation \eqref{eq39}, we present Fig.~\ref{fig5} 
which illustrates the change in the mass of scalaron with the density of the 
environment for three different sets of model parameters $\alpha$ and $\beta$ 
as considered before the correction. The figure indicates a reasonably lighter 
scalaron for the same environment after the singularity correction of the 
model. Moreover, a notable decrease in mass is observed in both scenarios with varying $\alpha$ and $\beta$ values.

In a region where $R_c< R< 1/\sigma$, scalaron mass will be~\cite{2018_kat}
\begin{equation}
	m_\phi^2\approx\frac{1}{6\sigma}.
	\label{eq78}
\end{equation}
Equation \eqref{eq78} indicates that the scalaron mass becomes stable within a 
range of considerable degree of curvature and is given by $m_{\phi}\propto 1/\sigma$, so it is completely controlled by the parameter $\sigma$. On the other 
hand, in an extremely high curvature regime i.e., at $R>>R_c$\, (and $R>>\frac{1}{\sigma}$\,) scalaron's mass is obtained as ~\cite{2018_kat,2021_Nashiba}
\begin{equation}
	m_\phi^2\approx\frac{1}{6\sigma(1+2\sigma R)}.
	\label{eq79}
\end{equation}
It hints at the possibility that the scalaron experiences a reduction in mass 
within regions of high curvature. It is worth mentioning that according to the 
analysis of Ref.~\cite{2018_kat} the upper boundary for the scalaron mass must 
be $<\mathcal{O}(1)$ GeV. Our analysis yields an suitable result within this 
prescribed limit.

\begin{figure}[!h]
	\centerline{
		\includegraphics[scale = 0.30]{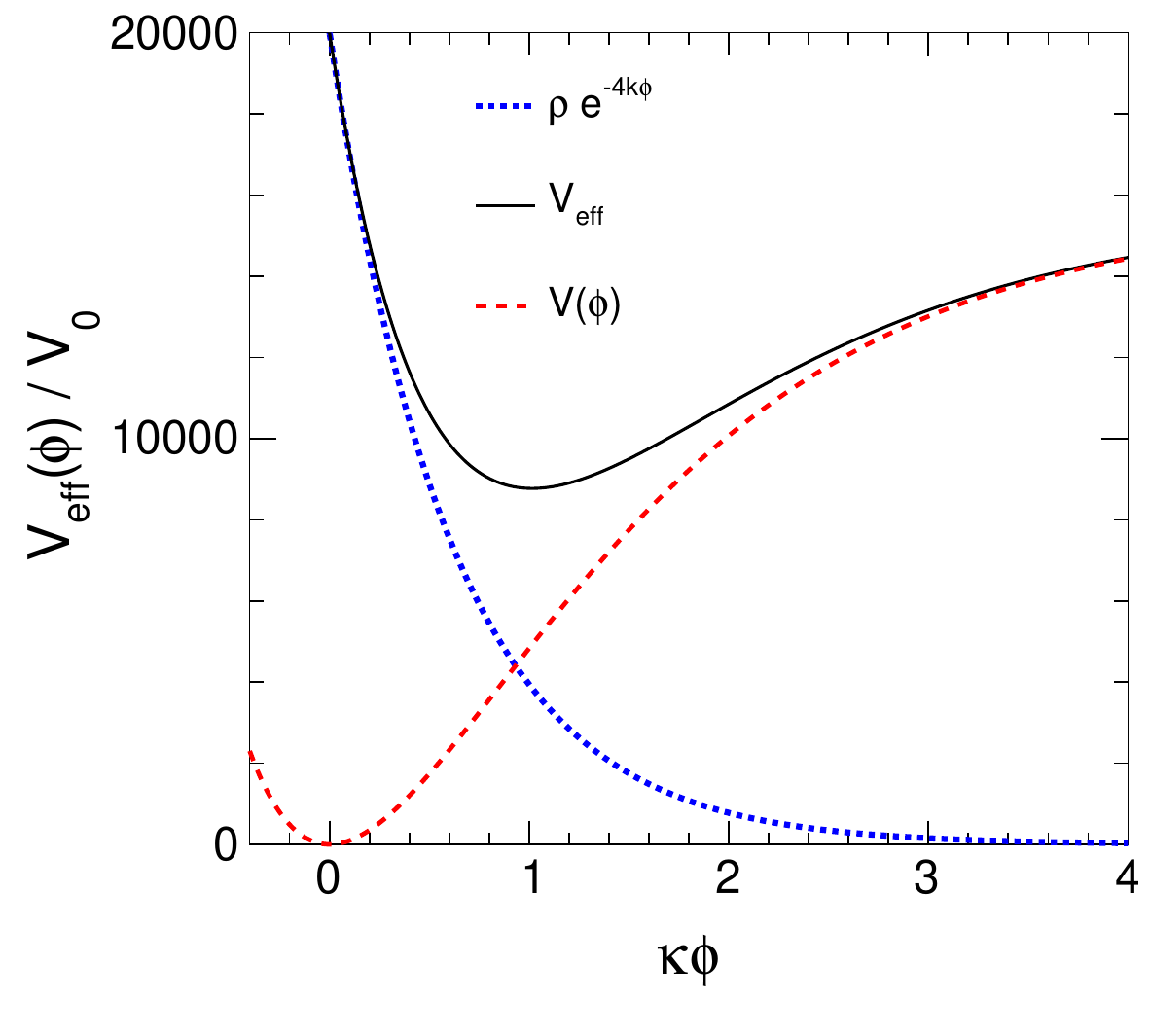}\hspace{1.0cm}
		\includegraphics[scale = 0.30]{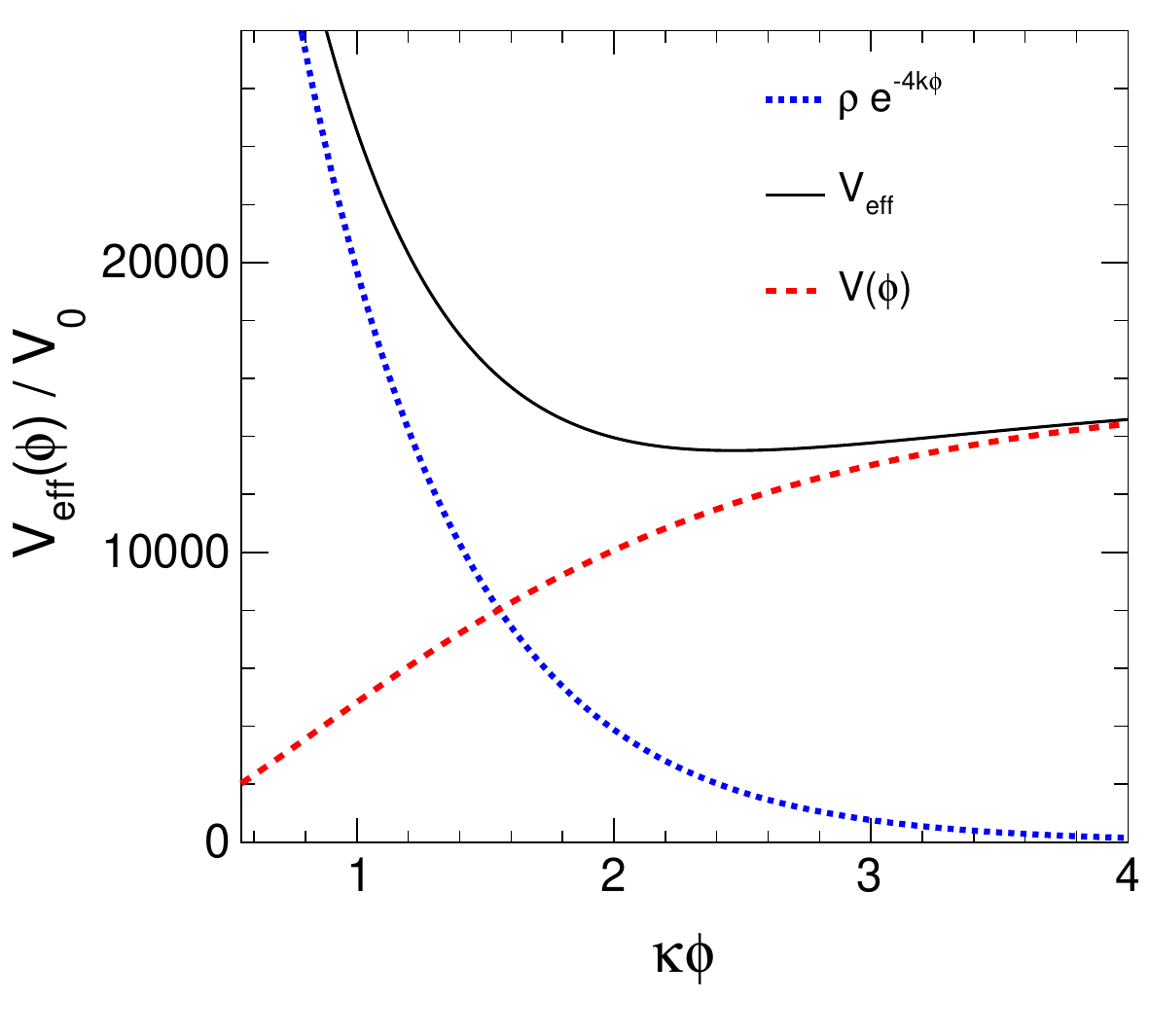}}
	\vspace{-0.2cm}
	\caption{$R^2$ corrected scalaron effective potential $V_{eff}(\phi)$ (black
		solid line) for the considered $f(R)$ gravity model with respect to two matter 
		distributions $\rho\sim10^4\rho_c$ (left) and $\rho\sim10^5\rho_c$ (right) 
		respectively with $R_c=1$. The red dashed lines show the contributions from the original 
		potential and the blue dotted lines represent the contributions of the 
		density-dependent term.}
	\label{fig6}
\end{figure}

Thus, after resolving this problem the 
scalaron can be made fairly lighter instead of extremely heavier. The cause of 
this attribution is seen clearly from Fig.~\ref{fig6} that the potential 
pushes the scalaron field to obtain larger values in denser environments and 
also, as mentioned already, the minimum of effective potential becomes 
shallower in the high-density region after fixing the singularity problem 
suffered by the model.
\vspace{0.5cm}\\
Units of $m_\phi$:
 1) Considering equation \eqref{eq39}, we proceed with the dimensional analysis as given below.
As previously mention, we have set $\hbar=c=1.$ i.e., we have chosen the 
natural unit system which ensures that all basic quantities namely mass, 
length, time, energy can be expressed in terms of single mass unit as 
$$\text{mass} = \text{length}^{-1} = \text{time}^{-1} = \text{energy},$$
where unit of mass, length, time and energy are GeV,\, Gev$^{-1},$\, Gev$^{-1}$\, and GeV respectively.
In our equation constants $\alpha,\, \beta,\, \pi\,$ have no 
dimensions and $\,e^{-\frac{R_0}{R_c}}$ is also a dimensionless term. Dimensions 
of other associated quantities are
$$[\rho] = [ML^{-3}]\,,\,\,\, [\kappa]= [M^{-1}L^3T^{-2}]\,,\,\,\,[\rho \kappa^2]=[T^{-2}]\,,\,\,\,[R_c]=[R_0]=[L^{-2}].$$
Therefore,
$$[{\left(\rho\kappa^2+\alpha R_c+2\beta R_c\right)^{\!4}}]=[L^{-8}]\,,\,\,\,\, \bigg[\frac{R_c}{6\,\alpha\,R_c^4 + \pi \beta\left(\rho\kappa^2+\alpha R_c+2\beta R_c \right)^{\!4}\,e^{-\frac{R_0}{R_c}}}\bigg]=[L^6],$$
$$\bigg[\frac{1}{\pi\left(\rho\kappa^2+\alpha R_c+2\beta R_c\right)^{\!3}[1 - \beta\,e^{-\frac{R_0}{R_c}}]-2\alpha\,R_c^3}\bigg]= [L^6].$$\\
Hence,
$$ \bigg[\frac{R_c}{6\,\alpha\,R_c^4 + \pi \beta\left(\rho\kappa^2+\alpha R_c+2\beta R_c \right)^{\!4}\,e^{-\frac{R_0}{R_c}}}-
\frac{1}{\pi\left(\rho\kappa^2+\alpha R_c+2\beta R_c\right)^{\!3}[1 - \beta\,e^{-\frac{R_0}{R_c}}]-2\alpha\,R_c^3}\bigg]=[L^6].$$ 
These lead the dimension of right hand side = $[L^{-8}] [L^6] = [L^{-2}]$.
Thus, $[m_\phi^2]=[L^{-2}].$
This gives $$[m_\phi]=[L^{-1}].$$
Hence, $m_\phi$ can be expressed in (GeV$^{-1})^{-1}$ = GeV unit.\\

2) Similarly in the case of equation \eqref{eq77} dimensions of respective terms are same as $[2\sigma R_c]$ has no dimension. Consequently, unit of mass is GeV in this case also.\\

\section{Derivation of equations \eqref{eq75} -- \eqref{eq79}:}
\label{ap2}
 After $R^2$ correction we obtain $f_{RR}$ along with equations \eqref{eq72} and \eqref{eq73} as follows:
\begin{equation}
	f_{RR}(R)=\frac{6\,\alpha\, R_c^3}{\pi\,R^4} + \frac{\beta}{R_c}\,e^{-\frac{R}{R_c}}+2\sigma.
	\label{eq80}
\end{equation}
Now, using equation \eqref{eq16} and \eqref{eq73} we may write
\begin{align*}
	e^{\sqrt{\frac{2}{3}}\kappa\phi}= 1-\frac{2\,\alpha R_c^3}{\pi R^3}-\beta\, e^{-\frac{R}{R_c}} + 2\sigma R,\\
	\Rightarrow 2\sigma R^4+R^3(1-\beta\,e^{-R/R_c}-e^{\sqrt{\frac{2}{3}}\kappa\phi})-\frac{2\,\alpha}{\pi} R_c^3= 0.
\end{align*}
As $e^{-R/R_c}<<1$ for $R>>R_c$,\, hence, we may simplify the LHS of above equation to
\begin{equation}
	R^4+R^3\,\left(\frac{1-e^{\sqrt{\frac{2}{3}}\kappa\phi}}{2\sigma}\right)-\frac{\alpha}{\sigma\pi} R_c^3= 0.
	\label{eq81}
\end{equation}
Solving this equation in WolframMathematica we find,
\begin{equation}
	R = \frac{1}{8 \sigma} \left( -1 + e^{\sqrt{\frac{2}{3}} k \phi} + 4 \sigma \sqrt{A - B + C} \right),
	\label{eq82}
\end{equation}
where
\begin{align*}
	A &= \frac{(-1 + e^{\sqrt{\frac{2}{3}} k \phi})^2}{16 \sigma^2},\\
	B &= \frac{8 \alpha R_c^3}{3^{1/3} \left( -9 \pi^2 (-1 + e^{\sqrt{\frac{2}{3}} k \phi})^2 \alpha R_c^3 + \sqrt{3} \sqrt{\pi^3 \alpha^2 R_c^6 \left(27 \pi(-1 + e^{\sqrt{\frac{2}{3}} k \phi})^4 + 4096 \alpha \sigma^3 R_c^3\right)} \right)^{1/3}},\\
	C &= \frac{1}{2 \cdot 3^{2/3} \pi \sigma} \left( -9 \pi^2 (-1 + e^{\sqrt{\frac{2}{3}} k \phi})^2 \alpha R_c^3 + \sqrt{3} \sqrt{\pi^3 \alpha^2 R_c^6 \left(27 \pi (-1 + e^{\sqrt{\frac{2}{3}} k \phi})^4 + 4096 \alpha \sigma^3 R_c^3\right)} \right)^{1/3}\!\!\!.
\end{align*}

It is seen that the solution \eqref{eq82} of equation \eqref{eq81} is a complicated one. To simplify our analysis, we ignore the second part of \eqref{eq82} as absence of this part does not change  $R$ effectively and then consider the solution as follows: 
\begin{equation}
	R\approx \frac{e^{\sqrt{\frac{2}{3}} \kappa \phi}-1}{8\sigma}.
	\label{eq83}
\end{equation}
Now, 
\begin{align*}
	Rf_{R}(R)-f(R)&= R\,e^{\sqrt{\frac{2}{3}}\kappa\phi}- R+\frac{\alpha R_c}{2}+\beta R_c-\sigma R^2\\
	&=R(e^{\sqrt{\frac{2}{3}}\kappa\phi}-1)+\frac{\alpha R_c}{2}+\beta R_c-\sigma R^2\\	
	&=\frac{\alpha R_c}{2}+\beta R_c+\frac{(e^{\sqrt{\frac{2}{3}}\kappa\phi}-1)^2}{8\sigma}- \frac{(e^{\sqrt{\frac{2}{3}}\kappa\phi}-1)^2}{64\sigma}\\
	&=\frac{\alpha R_c}{2}+\beta R_c+\frac{7}{64\sigma}\,(e^{\sqrt{\frac{2}{3}}\kappa\phi}-1)^2\\
	\Rightarrow\frac{Rf_{R}(R)-f(R)}{2\kappa^2f_R^2}&=\frac{R_c}{2\kappa^2f_R^2}\left[\frac{\alpha}{2}+\beta +\frac{7}{64\sigma R_c}\,(e^{\sqrt{\frac{2}{3}}\kappa\phi}-1)^2\right].\\
	\frac{V(\phi)}{V_0}&=\left[\frac{\alpha}{2}+\beta +\frac{7}{64\sigma R_c}\,(e^{\sqrt{\frac{2}{3}}\kappa\phi}-1)^2\right]\,e^{-2\sqrt{\frac{2}{3}}\kappa\phi},
\end{align*}
where $V_0=\frac{R_c}{2\kappa^2}$ is the normalization factor. Further,
\begin{align*}
	V_{eff} &= V(\phi)-\frac{1}{4}\,T_\mu^\mu\,\,e^{-2\sqrt{\frac{2}{3}}\kappa\phi}\\
	\Rightarrow\frac{V_{eff}(\phi)}{V_0} &= \frac{V(\phi)}{V_0}+\frac{1}{4V_{0}}\rho\,\,e^{-2\sqrt{\frac{2}{3}}\kappa\phi} \\
	\Rightarrow\frac{V_{eff}(\phi)}{V_0} & =\left[\frac{\alpha}{2}+\beta +\frac{7}{64\sigma R_c}\,(e^{\sqrt{\frac{2}{3}}\kappa\phi}-1)^2+ \frac{\rho \kappa^2}{2R_c}\right]\,e^{-2\sqrt{\frac{2}{3}}\kappa\phi}.
\end{align*}

Computation of $m_{\phi}$ after correction:
\begin{align*}
	m_\phi^2&=\frac{1}{3f_{RR}(R_0)}\left[1-\frac{R_0f_{RR}(R_0)}{f_R(R_0)}\right]\\
	&=\frac{1}{3}\left[\frac{1}{\frac{6\,\alpha\, R_c^3}{\pi\,R_0^4} + \frac{\beta}{R_c}\,e^{-\frac{R_0}{R_c}}+2\sigma}-\frac{R_0}{1-\frac{2\,\alpha R_c^3}{\pi R_0^3}-\beta\, e^{-\frac{R_0}{R_c}} + 2\sigma R_0}\right]\\
	&=\frac{1}{3}\left[\frac{\pi R_c R_0^4}{6\,\alpha\, R_c^4+\pi\beta\,R_0^4\,e^{-\frac{R_0}{R_c}}+2\sigma\pi R_c R_0^4}-\frac{\pi\,R_0^4}{\pi\,R_0^3-2\,\alpha R_c^3-\pi\beta\,R_0^3\,e^{-\frac{R_0}{R_c}}+2\sigma\pi R_0^4}\right] \\
	&=\frac{\pi R_c R_0^4}{3(6\,\alpha\, R_c^4+\pi\beta\,R_0^4\,e^{-\frac{R_0}{R_c}}+2\sigma\pi R_c R_0^4)}\left[1-\frac{6\,\alpha\, R_c^4+\pi\beta\,R_0^4\,e^{-\frac{R_0}{R_c}}+2\sigma\pi R_c R_0^4}{\pi R_c \,R_0^3-2\,\alpha R_c^4-\pi\beta\,R_c \,R_0^3\,e^{-\frac{R_0}{R_c}}+2\sigma\pi R_c\,R_0^4}\right]\\
	&=\frac{\pi R_c R_0^4}{3\times2\sigma \pi R_c R_0^4(1+\frac{\,3\alpha}{\sigma\pi}\frac{\,R_c^3}{R_0^4} +\frac{\beta}{2\sigma R_c}\,\,e^{-\frac{R_0}{R_c}})}\left[\,1-\frac{\,2\pi\sigma R_c R_0^4\,[1+\frac{3\,\alpha}{\sigma\pi}\,\frac{R_c^3}{R_0^4} +\frac{\beta}{2\sigma R_c}\,e^{-\frac{R_0}{R_c}}]}{2\pi\sigma R_c R_0^4[1+\frac{1}{2\sigma R_0}
		-\frac{\alpha}{\pi\sigma}\,\frac{R_c^3}{R_0^4}-\frac{\beta}{2\sigma R_0}\,e^{-\frac{R_0}{R_c}}]}\right]\\
	&=\frac{1}{6\sigma}\left[\frac{1}{(1+\frac{\,3\alpha}{\sigma\pi}\frac{\,R_c^3}{R_0^4} +\frac{\beta}{2\sigma R_c}\,\,e^{-\frac{R_0}{R_c}})}\right]\left[\,1-\frac{\,[1+\frac{3\,\alpha}{\sigma\pi}\,\frac{R_c^3}{R_0^4} +\frac{\beta}{2\sigma R_c}\,e^{-\frac{R_0}{R_c}}]}{[1+\frac{1}{2\sigma R_0}
		-\frac{\alpha}{\pi\sigma}\,\frac{R_c^3}{R_0^4}-\frac{\beta}{2\sigma R_0}\,e^{-\frac{R_0}{R_c}}]}\right].
\end{align*} 
In a region where $R_c< R < 1/\sigma$,\\
$\frac{1}{6\sigma}\approx 166667;\,\,\,
\bigg[\frac{1}{(1+\frac{\,3\alpha}{\sigma\pi}\frac{\,R_c^3}{R_0^4} +\frac{\beta}{2\sigma R_c}\,\,e^{-\frac{R_0}{R_c}})}\bigg]\approx1;$\,\, and $\bigg[\,1-\frac{\,[1+\frac{3\,\alpha}{\sigma\pi}\,\frac{R_c^3}{R_0^4} +\frac{\beta}{2\sigma R_c}\,e^{-\frac{R_0}{R_c}}]}{[1+\frac{1}{2\sigma R_0}
	-\frac{\alpha}{\pi\sigma}\,\frac{R_c^3}{R_0^4}-\frac{\beta}{2\sigma R_0}\,e^{-\frac{R_0}{R_c}}]}\bigg] \approx 1,$\\ So, we may write,
\begin{align*}
	m_\phi^2&\approx\frac{1}{6\sigma}.
\end{align*}
In higher curvature condition $R>R_c$ and $R>1/\sigma$\,
similar to previous case, 
$$\frac{1}{6\sigma}\approx 166667.$$ 
Further,
$$1+\frac{\,3\alpha}{\sigma\pi}\frac{\,R_c^3}{R_0^4} +\frac{\beta}{2\sigma R_c}\,\,e^{-\frac{R_0}{R_c}}\approx1,$$
In this case we find \, \,$\frac{1}{2\sigma R_0}<1$ instead of \, \,$\frac{1}{2\sigma R_0}>1$\,\, in previous case along with two negligible terms\\ $\frac{\alpha}{\pi\sigma}\,\frac{R_c^3}{R_0^4}$\,\, and \,$\frac{\beta}{2\sigma R_0}\,e^{-\frac{R_0}{R_c}}$.\,\, Also, $\bigg[\,1-\frac{\,[1+\frac{3\,\alpha}{\sigma\pi}\,\frac{R_c^3}{R_0^4} +\frac{\beta}{2\sigma R_c}\,e^{-\frac{R_0}{R_c}}]}{[1+\frac{1}{2\sigma R_0}
	-\frac{\alpha}{\pi\sigma}\,\frac{R_c^3}{R_0^4}-\frac{\beta}{2\sigma R_0}\,e^{-\frac{R_0}{R_c}}]}\bigg] \not\approx 1.$\\
This allows us to write the scalaron mass at $R=R_0$ as
\begin{align*}
	m_\phi^2&\approx\frac{1}{6\sigma}\left[1-\Big[1+\frac{1}{2\sigma R}\Big]^{-1} \right]\approx\frac{1}{6\sigma}\left[1-\frac{2\sigma R}{1+2\sigma R}\right]\approx\frac{1}{6\sigma}\left[\frac{1}{1+2\sigma R}\right].	
\end{align*}

\end{document}